\documentclass[preprint,
 amsmath,amssymb,
aps, physrev, superscriptaddress]{revtex4-2}
\usepackage{graphicx}
\usepackage{pgffor}
\usepackage{here}
\usepackage{booktabs}
\usepackage{newtxtext,newtxmath}
\usepackage{url}
\usepackage{caption}
\usepackage[english]{babel}
\usepackage{subcaption}
\newcommand{\argmax}{\mathop{\rm arg~max}\limits}

\begin{document}

%\preprint{APS/123-QED}

\title{Integrating Bayesian Spectral Deconvolution \\and Expert Scientific Reasoning for Robust Peak Estimation}% 

\newcommand{\affSDS}{Graduate School of Social Data Science, Hitotsubashi University, 2-1 Naka, Kunitachi, Tokyo 186-8601, Japan}
\newcommand{\affUTokyo}{Department of Advanced Materials Science, Graduate School of Frontier Sciences, The University of Tokyo, 5-1-5 Kashiwanoha, Kashiwa, Chiba 277-8561, Japan}
\newcommand{\affHIAS}{Hitotsubashi Institute for Advanced Study, Hitotsubashi University, 2-1 Naka, Kunitachi, Tokyo 186-8601, Japan}
\newcommand{\affCAL}{Department of Statistics and Applied Probability, University of California, Santa Barbara, Santa Barbara, CA 93106, USA}

\author{Hayato Okubo}\affiliation{\affSDS}
\author{Yoshifumi Amamoto}\affiliation{\affUTokyo}
\author{Toshimitsu Aritake}\affiliation{\affHIAS}
\author{Hiroyuki Kumazoe}\affiliation{\affSDS}
\author{Shiryu Nakano}\affiliation{\affSDS}
\author{Evan Jamison}\affiliation{\affCAL}
\author{Satoshi Tanaka}\affiliation{\affSDS}
\author{Yoh-ichi Mototake}\email{y.mototake@r.hit-u.ac.jp}\affiliation{\affSDS}

\date{\today}
\begin{abstract}
Spectral deconvolution is essential for extracting peak structures that encode material properties and chemical structures, but conventional automated methods often fail when spectra contain high-intensity noise or unknown background components. In practice, scientists rarely interpret spectra in isolation. Instead, they identify physically meaningful peaks by relating spectral structures to auxiliary information such as physical-property values, chemical structures, and trends across related measurements. 
Here, we propose a Bayesian framework that integrates spectral deconvolution with a model of expert scientific reasoning. In this work, expert scientific reasoning refers to the practice of evaluating candidate spectral structures by their consistency with independently measured physical-property values, rather than to manual expert intervention during inference. We formalize this reasoning as a physical-property regression layer, implemented using Gaussian process regression, and couple it with Bayesian spectral deconvolution. By averaging the physical-property likelihood over posterior predictive spectra inferred from Bayesian spectral deconvolution, the proposed method selects spectral models according to the consistency between inferred spectral structures and physical-property information. 
We validate the framework using synthetic spectra with high-intensity noise or unknown backgrounds and infrared spectra of poly(lactic acid). The method recovers physically meaningful peak structures that conventional Bayesian spectral deconvolution misses or misidentifies from spectra alone, including weak peaks in poly(lactic acid) IR spectra related to measured degradation rates. These results demonstrate that integrating expert scientific reasoning with Bayesian spectral deconvolution enables robust peak estimation under conditions where spectrum-only inference is unreliable.
\end{abstract}

\maketitle

\section{Introduction}
Spectral deconvolution estimates the number, position, and variance of peaks by regressing the spectra obtained by irradiating materials with X-rays or visible light as a sum of basis functions. This method is crucial for evaluating material properties and chemical structures.
Mass spectrometry\cite{harrison2018chemical} is used to determine the molecular weight, whereas infrared (IR) spectroscopy \cite{Fan2018,doi:10.1021/acs.chemrev.6b00448} and Raman spectroscopy\cite{kneipp1999ultrasensitive} are used to investigate the physical properties and chemical structure of synthesized compounds. Spectral deconvolution enables the identification of the molecular weight and partial structures.
Molecular bonding can be evaluated using the nuclear magnetic resonance spectra\cite{nmr}, whereas X-ray diffraction (XRD) determines the lattice constants and atomic arrangements in crystals\cite{XRD} by obtaining and deconvolving the spectra.
The properties and chemical structures of materials are determined by measuring the spectra and deconvoluting them into their components.
The automation of processes such as material synthesis and spectral measurement has led to an increase in data requiring spectral deconvolution \cite{HTPE}. Despite these advancements, spectral analysis continues to depend to some extent on expert scientific interpretation \cite{Baer2019Practical,Near2021Preprocessing}. The significance of spectral deconvolution, which seeks to promote objectivity and automation within an information science framework, is therefore being increasingly acknowledged.
\par
Existing spectral deconvolution methods may struggle to determine spectral models, such as the number of peaks or the basis function model, for spectral data with complex peak shapes and overlapping peaks.
Estimating the number and shape of peaks is a fundamental requirement for achieving spectral deconvolution.
To improve this capability, Bayesian spectral deconvolution, which applies Bayesian inference to the problem, has been proposed.
Bayesian spectral deconvolution constructs a statistical model by applying Bayesian inference to a regression model that represents the spectrum as a sum of basis functions. This approach enables the estimation of spectral deconvolution model parameters, including the optimal number of peaks required to represent the spectral data~\cite{nagata201282}.
Undesirable components such as high-intensity noise or substance-derived background often appear in spectral data owing to instrument errors, signal processing, or the measured sample and are problematic for analysis~\cite{noise,XPS_bacground}.
In fields lacking a well-defined physical model, the background model may not be well defined, complicating the spectral deconvolution analysis in some cases \cite {Diehl_2024}.
Model selection for various analyses, such as peak-number estimation, becomes difficult when the spectral data contain high-intensity noise or background~\cite{Macaro2014} in conventional Bayesian spectral deconvolution.
Thus, existing methods have the drawback that spectral deconvolution becomes difficult when noise dominates or when the background component model is unknown.\par
However, researchers may achieve spectral deconvolution by applying domain-specific knowledge, even in scenarios where noise is prevalent or the background component model is not defined.
Scientists do not rely solely on the spectral data to analyze the spectra. Rather, they analyze the spectra in the context of the physical properties and chemical structure of the material and variations observed in other measurements collected during their analysis.
Correlations between the changes in material properties or chemical structure often result in shifts in the spectral peaks. These shifts help in identifying the key candidate peaks for analysis and highlight the spectral regions warranting further investigation.
Scientists have achieved spectral deconvolution by leveraging the insights gained from such knowledge, enabling them to extract peak structures buried in noise or appearing against unknown backgrounds \cite{spectrum_cao}.
Consequently, peak-number estimation and other spectral model selections are achievable even for complex spectral shapes\cite{kan2}.
The analytical approach used by scientists to focus on the important spectral regions based on the physical property values is crucial for analyzing spectra dominated by noise or where background components are not modeled.
\par
We propose a spectral deconvolution framework that models the process followed by scientists, namely, using material properties and structures to identify the key regions, enabling robust peak estimation even in the presence of dominant noise or unknown backgrounds.
Several methods based on such scientific practices have already been developed.
A Bayesian analysis method for X-ray absorption near-edge structure spectra has been proposed, which improves the accuracy by assigning different model structures or prior distributions to each region, based on the knowledge that the peak-shape measurements differ across the energy regions \cite{kashiwamura2022}.
Other studies have applied regression modeling to the relationship between the physical property values and peak structures derived from spectral deconvolution\cite{HAN2020106846}.
However, these approaches merely provide a framework in which spectral data analysis and knowledge extraction from auxiliary measurement data, such as physical properties, are conducted separately.
Integrating these two steps is expected to automate spectral deconvolution while enabling more comprehensive knowledge extraction of the relationship between the physical properties and spectra.
In this context, our proposed framework comprehensively models the analytical process followed by scientists for deconvoluting complex spectra, as follows.
First, we employ the Bayesian spectral deconvolution model mentioned above to model the spectral data.
Next, we model the practice of examining the relationship between peak structures derived from deconvolution and the physical properties or chemical structures as Gaussian process regression.
This approach aims to integrate the previously underutilized physical property information into the spectral deconvolution, enabling accurate model selection and parameter estimation even in complex spectra containing noise and background.
The proposed method was applied to the spectral deconvolution of synthetic spectral data and IR spectra of polylactic acid, a representative biodegradable polymer, to verify its effectiveness.
\par

\section{Method}
\subsection{Problem formulation}
We address the problem of selecting spectral models, such as the number of peaks, by integrating the spectral data with physical prior knowledge. This knowledge includes the physical property value, as used by scientists.

\par
We define the dataset used in the proposed method.
We consider the spectral data $\{x_i, y_i\}_{i=1}^{N}$ consisting of $N$ observation points.
$x_i$ represents the horizontal axis, such as wavenumber or wavelength; $y_i$ represents the spectral intensity at that point.
We define a spectral dataset as $Y:=\{y_i\}_{i=1}^{N}$. We assume $N'$ spectral datasets $Y_{j}:=\{y_{i,{j}}\}_{i=1}^N$ ($j=1, \dots, N'$) sharing the observation points $\{x_i\}_{i=1}^{N}$ but representing distinct physical states, such as datasets containing material properties.
We define the paired dataset as \(\mathcal{D} = \{(Y_j^{\mathrm{obs}}, z_j)\}_{j=1}^{N'}\), where \(Y_j^{\mathrm{obs}}\) is the observed spectrum of sample \(j\) and \(z_j\) is the corresponding physical property.
This dataset $\mathcal{D}$ constitutes the dataset assumed by the proposed method.
\par
We formulate the problem of selecting a spectral model $M$ by combining the spectral data $\mathcal{Y}$ and physical prior knowledge $Z$ (e.g., material properties), as follows.
\begin{eqnarray}
\argmax_{M} p({M}|Z,\mathcal{Y})
\end{eqnarray}
Note that we first formulate the problem in terms of the full Bayesian model posterior.
As shown below, this posterior can be decomposed into a spectrum-only evidence term and a physical-property-consistency term.
The proposed method then adopts the latter term as the model-selection criterion, in accordance with the aim of this study.

\subsection{Model}
This section presents a Bayesian hierarchical model for spectral deconvolution that incorporates physical property values, mirroring scientific practices (Figure \ref{graph2}).
The proposed framework consists of two layers. The first is a spectral deconvolution model analyzing the spectrum. The second is a regression model that relates the physical property values to the spectral data. 
After describing the individual layers, we explain the spectral deconvolution model of the process followed by scientists.

\subsubsection{Spectral deconvolution model}
The spectral deconvolution model represents the spectral data as a linear combination of basis functions, such as Gaussian functions.
The regression of spectral data using the spectral deconvolution model yields information such as the number of peaks, peak positions, and peak variances \cite{nagata201282}.
Consider the spectral data consisting of $N$ observations $\{x_i, y_i\}_{i=1}^{N}$.
The spectral deconvolution model, utilizing Gaussian basis functions subject to independent and identically distributed (i.i.d.) normal noise, is expressed as follows:

\begin{figure}[h] 

%\vspace*{\fill}
\includegraphics[width=0.5\linewidth]{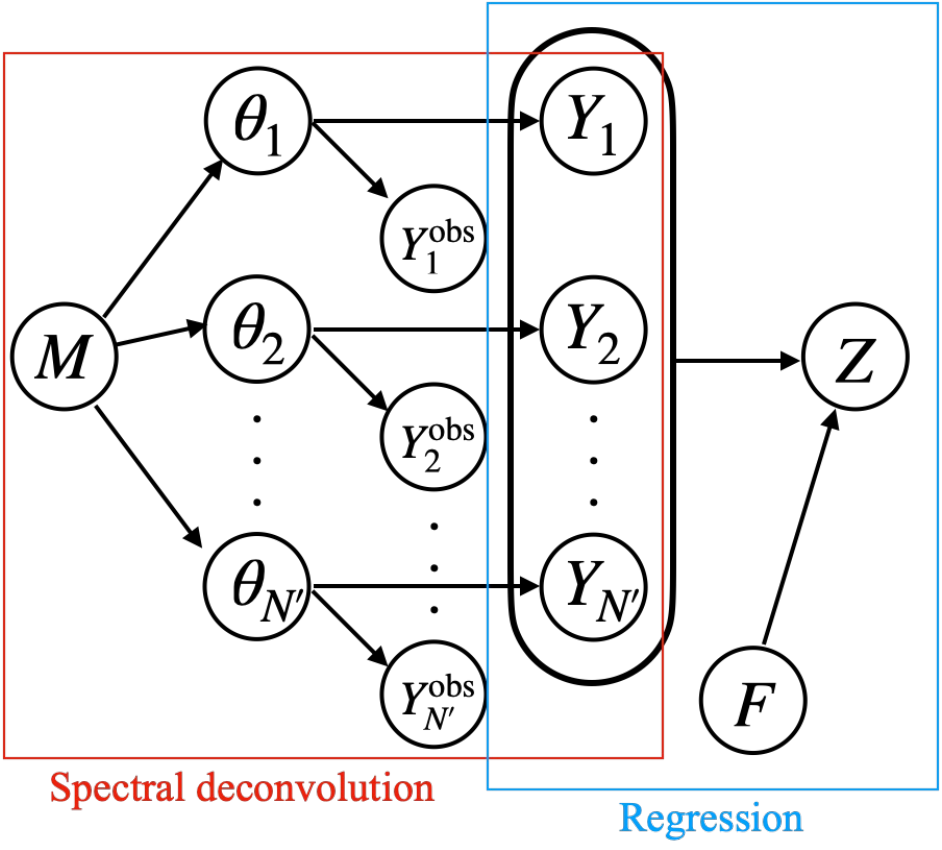}
%\raggedleft
\caption{
Graphical model of the proposed method. $M$ denotes the spectral model. For each sample \(j\), the spectral parameter \(\theta_j\) defines both the posterior predictive spectrum \(Y_j\) and the observed spectrum \(Y_j^{\mathrm{obs}}\) through the spectral likelihood model. The physical property $Z$ is modeled by Gaussian process regression using the latent function value $F$ conditioned on the spectral information. The red (blue) box indicates the components of spectral deconvolution (Gaussian process regression).}
\label{graph2}
\end{figure}

\begin{eqnarray}
\label{likelihood}
p(Y | \theta) &=& \prod_{i=1}^{N} \frac{1}{\sqrt{2\pi}\sigma} \exp\left(-\frac{(y_i - g (x_i; \theta))^2}{2\sigma^2}\right).
\end{eqnarray}
$g$ denotes the spectral fitting function, $\theta$ the regression parameters, and $\sigma$ the standard deviation of the noise. The representative examples of $g$ include

\begin{eqnarray}
\label{basis}
g (x; \theta) &=& \sum_{j=1}^{M} w_j \exp\left(-\frac{(x - \mu_j)^2}{2a_j^2}\right).
\end{eqnarray}
$M$ denotes the number of peaks in the fitting model, and $\theta = \{w_j, \mu_j, a_j\}_{j=1}^{M}$ represents the set of parameters.

\par

\subsubsection{Physical-property regression model}\label{GP}
The physical-property regression model connects the spectral data to the corresponding physical property values.
We employed Gaussian process regression as the Bayesian physical-property regression model in this study.
In the physical-property regression model, a single sample constitutes a pair of spectral data $Y:=\{y_i\}_{i=1}^{N}$ and a physical property value $z$.
This can be extended to multiple samples, as follows. Given a set of spectral and measurement data of sample size $N'$, $\{\mathcal{Y},Z\}:=\{Y_j,z_j\}_{j=1}^{N'}$, the Gaussian process regression model \cite{PRML} is formulated as

\begin{eqnarray}
{\it p}(Z|F,\mathcal{Y})=\left({\frac{\alpha}{2\pi}}\right)^{\frac{N'}{2}}\exp\left[-\frac{1}{2}\left(Z - F\right)^T \alpha I\left(Z - F\right)\right].
\end{eqnarray}
We define $F := F(\mathcal{Y}) = (f(Y_1), \dots, f(Y_{N'}))$. The prior distribution $p(F)$ is assumed to be an $N'$-dimensional normal distribution with zero mean and covariance matrix $K$. The hyperparameter $\alpha$ denotes the noise precision, which is the inverse of the noise variance in the likelihood model.
\begin{eqnarray}
    p(F)=\left(\frac{1}{(2\pi)^{N'} |K|}\right)^{1/2}\exp\left(-\frac{1}{2}F^T K^{-1} F\right)
\end{eqnarray}
Here, $k(Y_i, Y_j)$ denotes a kernel function representing the similarity between $Y_i$ and $Y_j$.

\subsubsection{Model of spectral deconvolution process followed by scientists}
We modeled the method followed by scientists for extracting high-precision information by analyzing the spectral data in combination with physical prior knowledge.
The simplest physical prior knowledge is the value of a property related to a spectrum.
Given the spectral data and corresponding physical properties, scientists analyze the spectral features, such as specific regions or peaks, that are correlated with variations in the physical properties.
The decision of scientists to focus on specific regions while down-weighting others corresponds to the operation of extracting only the dimensions of relevant variables.
Thus, the above represents a variable selection process in regression analysis, where the spectral intensities at each wavenumber serve as explanatory variables predicting the physical property values.
To reflect this requirement in the model, it is effective to model a predictive relationship in which spectral structures serve as explanatory variables for the physical property values. 
Incorporating this perspective into the model, we treat spectral features as predictors of the physical properties, so that spectral components relevant to the target property are preferentially weighted during model selection. Introducing a causal link from the spectra to the physical properties enables predictive performance to guide spectral-structure estimation, ensuring that physically meaningful components are prioritized.
\par
By mirroring this scientific spectral deconvolution process, the causal relationships can be modeled as depicted in the graphical model in Figure \ref{graph2}.
Conventional spectral deconvolution methods focus solely on estimating the number of peaks and their parameters from the spectral data, without accounting for their causal relationships with the physical properties.
In contrast, the spectral deconvolution method followed by scientists estimates the number of peaks and their parameters while maintaining a causal link between the spectral data and physical property values.
This represents a causal relationship in which specifying a spectral model (e.g., the number of peaks) and parameters (e.g., positions and variances) generates a spectrum from which the measured value $z$ is derived.
The following statistical model is derived from this graphical model.
\begin{eqnarray}
p(M,\vartheta,\mathcal{Y},\mathcal{Y}^{\text{obs}},F,Z) &=& {\it p}(Z|F,\mathcal{Y}) {\it p}(F){\it p}(\mathcal{Y}|\vartheta) \notag \\
&\quad& \times {\it p}(\mathcal{Y}^{\mathrm{obs}}|\vartheta) {\it p}(\vartheta|M) {\it p}(M). %\tag{14}
\end{eqnarray}
Here, let $\vartheta := \{\theta_j\}_{j=1}^{N'}$; then, ${\it p}(\mathcal{Y}|\vartheta), {\it p}(\mathcal{Y}^{\mathrm{obs}}|\vartheta), {\it p}(\vartheta|M)$ are defined as follows:
\begin{eqnarray}
\label{Y|theta}
{\it p}(\mathcal{Y}|\vartheta) := \prod_{j=1}^{N'} {\it p}(Y_j|\theta_j), 
\\
{\it p}(\mathcal{Y}^{\mathrm{obs}}|\vartheta) := \prod_{j=1}^{N'} {\it p}(Y^{\mathrm{obs}}_j|\theta_j),\label{Y_obs|theta}\\
{\it p}(\vartheta|M) := \prod_{j=1}^{N'} {\it p}(\theta_j|M).
\end{eqnarray}

\par
Given the spectral data $\mathcal{Y}^{\mathrm{obs}}$ and property values $Z$, the posterior $p(M|Z, \mathcal{Y}^{\mathrm{obs}})$ for the spectral model $M$ is obtained by marginalizing the joint probability.
\begin{flalign}
p(M|Z,\mathcal{Y}^{\mathrm{obs}})&\propto p(Z,\mathcal{Y}^{\mathrm{obs}}, M)\nonumber\\ 
&= \int  dF d\mathcal{Y} d\vartheta p(Z,F,\mathcal{Y}, \mathcal{Y}^{\mathrm{obs}}, \vartheta, M)
\end{flalign}
Rearranging the marginalization integral leads to the following interpretable form.
\begin{flalign}
p(M|Z,\mathcal{Y}^{\mathrm{obs}})&\propto \int  d\mathcal{Y} \,\left[ \left\{\int dF \:p(Z|F, \mathcal{Y}) p(F)\right\}
  \left\{{{\int d\vartheta \,p(\mathcal{Y}|\vartheta){\it p}(\mathcal{Y}^{\mathrm{obs}}|\vartheta) p(\vartheta|M) }}\right\}p(M)\right]\nonumber\\%\tag{15}
&=\int d\mathcal{Y}\: p(Z|\mathcal{Y})\left\{{{\int d\vartheta \,p(\mathcal{Y}|\vartheta){\it p}(\vartheta|\mathcal{Y}^{\mathrm{obs}},M)}}\right\}{\it p}(\mathcal{Y}^{\mathrm{obs}}|M)p(M)\label{eq_predict_post}\\
&={\it p}(\mathcal{Y}^{\mathrm{obs}}|M)p(M)\int d\mathcal{Y}\: p(Z|\mathcal{Y})p(\mathcal{Y}|\mathcal{Y}^{\mathrm{obs}},M)\\
&=p(\mathcal{Y}^{\mathrm{obs}}|M)p(M)
E_{p(\mathcal{Y}|\mathcal{Y}^{\mathrm{obs}},M)}
\left[p(Z|\mathcal{Y})\right],
\label{M|Z,Y}
\end{flalign}
where, the $\vartheta$ integral marginalizes the spectral deconvolution model, the $F$ integral marginalizes the Gaussian process regression model, and the $\mathcal{Y}$ integral marginalizes over the connection between layers in the hierarchical Bayesian spectral deconvolution model for scientists.
Each marginalization will be explained in the next section.\par

Assuming a uniform prior of $p(M)$ over candidate models, model comparison based on \(p(M|Z,\mathcal{Y}^{\mathrm{obs}})\) is equivalent to comparison based on the model evidence \(p(Z,\mathcal{Y}^{\mathrm{obs}}|M)\). 
Equation~(\ref{M|Z,Y}) shows that the full Bayesian model evidence can be decomposed into two conceptually distinct contributions:
\begin{equation}
p(Z,\mathcal{Y}^{\mathrm{obs}}|M)
=p(\mathcal{Y}^{\mathrm{obs}}|M)
E_{p(\mathcal{Y}|\mathcal{Y}^{\mathrm{obs}},M)}
\left[p(Z|\mathcal{Y})\right].
\end{equation}
The first contribution is the spectrum-only evidence \(p(\mathcal{Y}^{\mathrm{obs}}|M)\), which evaluates how well the candidate spectral model \(M\) explains the observed spectra themselves.
The second contribution is the physical-property consistency term,
\begin{equation}
E_{p(\mathcal{Y}|\mathcal{Y}^{\mathrm{obs}},M)}
\left[p(Z|\mathcal{Y})\right],
\end{equation}
which evaluates whether the latent spectral structures inferred under model \(M\), after conditioning on the observed spectra, can consistently explain the physical-property values.
This distinction becomes explicit by taking the negative logarithm of the model evidence.
The full Bayesian free energy is defined as
\begin{equation}
FE_{\mathrm{full}}(M)
=
-\log p(Z,\mathcal{Y}^{\mathrm{obs}}|M).
\end{equation}
Using the decomposition above, this free energy can be written as
\begin{equation}
FE_{\mathrm{full}}(M)
=
FE_{\mathrm{spec}}(M)
+
FE_{\mathrm{phys}}(M),
\end{equation}
where
\begin{equation}
FE_{\mathrm{spec}}(M)
=
-\log p(\mathcal{Y}^{\mathrm{obs}}|M)
\end{equation}
is the spectrum-only Bayesian free energy, and
\begin{equation}
FE_{\mathrm{phys}}(M)
=
-\log p(Z|\mathcal{Y}^{\mathrm{obs}},M)
=
-\log
E_{p(\mathcal{Y}|\mathcal{Y}^{\mathrm{obs}},M)}
\left[p(Z|\mathcal{Y})\right]
\end{equation}
is the physical-property-consistency free energy.
These two terms evaluate different aspects of the candidate model:
\(FE_{\mathrm{spec}}\) measures the fit to the observed spectra, whereas \(FE_{\mathrm{phys}}\) measures the consistency between the inferred spectral structures and the physical-property values.
The purpose of this study is not to select the spectral model that best fits the observed spectra alone, but to select the model whose inferred spectral structures are most informative for the physical-property values.
Therefore, in the proposed method, we use the physical-property-consistency term as the model-selection criterion.
We define the physical-property-informed model posterior as
\begin{equation}
\tilde{p}(M|Z,\mathcal{Y}^{\mathrm{obs}})
\propto
E_{p(\mathcal{Y}|\mathcal{Y}^{\mathrm{obs}},M)}
\left[p(Z|\mathcal{Y})\right].
\end{equation}
The corresponding physical-property-informed free energy is defined as
\begin{equation}
\widetilde{FE}(M)
=
-\log \tilde{p}(M|Z,\mathcal{Y}^{\mathrm{obs}}).
\end{equation}
In the proposed method, the candidate spectral model with the smallest \(\widetilde{FE}(M)\) is selected.

\subsection{Posterior inference for the model
}
Estimating the physical-property-informed model posterior $\tilde{p}(M|Z,\mathcal{Y}^{\mathrm{obs}})$ for the scientific spectral model, which encodes the physical properties, requires evaluating the marginalization integrals over $\theta$, $F$, and $\mathcal{Y}$, as detailed in Eq.~\eqref{M|Z,Y}.
These operations correspond to the marginalization within the spectral deconvolution model, Gaussian process regression model, and interface linking the two layers.
Marginalization proceeds sequentially over $\vartheta$, $F$, and $\mathcal{Y}$. 
This section presents the method for each step in this sequence.

\subsubsection{Marginalization of spectral deconvolution model}\label{spectrum_con}
The method for sampling from the posterior of the spectral deconvolution model using Replica Exchange Monte Carlo is described here.
The purpose of the marginalization calculation for the spectral deconvolution model, $\int d\theta \,p(\mathcal{Y}|\vartheta){\it p}(\mathcal{Y}^{\mathrm{obs}}|\vartheta)\, p(\vartheta|M)$, is to compute a sample set from the predictive distribution $p(\mathcal{Y}|\mathcal{Y}^{\mathrm{obs}})$.
Sampling from this predictive distribution $p(\mathcal{Y}|\mathcal{Y}^{\mathrm{obs}})$ can be achieved using the following procedure, as seen in Eq.~\eqref{eq_predict_post}.
The method proceeds by first drawing the parameters $\vartheta$ from the posterior distribution $p(\vartheta|\mathcal{Y}^{\mathrm{obs}},M)$. Then, sampling $\mathcal{Y}$ from the spectral deconvolution likelihood $p(\mathcal{Y}|\vartheta)$ (Eq.~\eqref{likelihood}) conditioned on these values yields the sample set from the predictive distribution $p(\mathcal{Y}|\mathcal{Y}^{\mathrm{obs}})$.
Because the likelihood function $p(\mathcal{Y}|\vartheta)$ is Gaussian, sampling from it is numerically straightforward. Therefore, the primary computational challenge lies in sampling from the posterior distribution $p(\vartheta|\mathcal{Y}^{\mathrm{obs}},M)$.

\par
Sampling from posterior distributions in high-dimensional statistical models is often performed using Markov chain Monte Carlo methods. Because posterior distributions in spectral deconvolution can be strongly multimodal, standard Metropolis--Hastings sampling~\cite{hastings1970monte} may require many iterations to move between modes and can converge slowly~\cite{2025mcmc, Marinari_1992}. 
This study employed the Replica Exchange Monte Carlo (REMC) method, an extended-sampling method~\cite {IBA_2001}, to address these sampling challenges. 
Constructing an extended artificial ensemble and exchanging states can efficiently enhance Markov chain mixing using extended-sampling methods~\cite{IBA_2001}.
We employed an established approach~\cite{nagata201282} that marginalizes the spectral model while avoiding this difficulty.
\par
Sampling from the posterior distribution $p(\vartheta|\mathcal{Y}^{\mathrm{obs}},M)$ for the spectral deconvolution model does not require the simultaneous use of all spectral data.
The likelihood function $p(\mathcal{Y}^{\mathrm{obs}}|\vartheta)$ factorizes into the individual spectral likelihoods $p({Y_j}^{\mathrm{obs}}|\theta)$, as shown in Eq.~\eqref{Y_obs|theta}.
By Bayes' theorem, the posterior distribution $p(\vartheta|\mathcal{Y}^{\mathrm{obs}},M)$ is proportional to the likelihood function ${\it p}(\mathcal{Y}^{\mathrm{obs}}|\vartheta)p(\vartheta|M)$. Therefore, the posterior distribution $p(\vartheta|\mathcal{Y}^{\mathrm{obs}}, M)$ also corresponds to the product of the posterior distributions ${\it p}(\theta|{Y_j}^{\mathrm{obs}})$ for individual spectra.
Sampling from the posterior distribution $p(\vartheta|\mathcal{Y}^{\mathrm{obs}},M)$ can be achieved by sampling from the posterior distribution of the individual spectra ${\it p}(\theta|{Y_j}^{\mathrm{obs}})$.
Hence, we describe a method for sampling from the posterior distribution ${\it p}(\theta|{Y_j}^{\mathrm{obs}})$ of a single spectral data point $Y^{\mathrm{obs}}:=Y_j^{\mathrm{obs}}$ using the replica exchange method.\par
Building on this, we define the probability distribution \( p(\theta|Y^{\mathrm{obs}},M) \) by introducing the inverse temperature parameter \( \beta\)  into the probability distribution $p_\beta (\theta|Y^{\mathrm{obs}},M)$ of the parameter set.
$p_\beta (\theta|Y^{\mathrm{obs}},M)$ is calculated as
\begin{align}
\label{eq:inverse_temparature}
&p_\beta (\theta|Y^{\mathrm{obs}},M) \propto \exp(-N \beta E(\theta,M))p(\theta), \\%\tag{16},
&E(\theta, M) := \frac{1}{2N\sigma^2} \sum_{i=1}^{N} (y_{i} - g (x_{i}; \theta, M))^2.& 
\end{align}
As can be seen from this equation, when \(\beta=1\), the sampling matches the target distribution \(p(\theta|Y^{\mathrm{obs}})\).
 However, reducing \(\beta\) corresponds to sampling from the prior distribution as the distribution approaches it. 
REMC is an efficient method for simultaneous sampling from different temperatures $\beta_l$, enabling the simultaneous distribution of the parameter values $\{\theta_l\}_{l=1}^L$ corresponding to $\beta_l$ to be obtained.
\begin{equation}
{\it  P}(\theta_1,\theta_2\cdots \theta_L|Y^{\mathrm{obs}},M)=\prod_{l=1}^L{\it  P}_{\beta_l}(\theta_l|Y^{\mathrm{obs}},M)
\label{yuudo_EX}
\end{equation}
is sampled by repeating the following procedure.
\begin{itemize}
\item[{\bf 1}]{\bf 
 Sampling from the individual distribution ${\it p}(\theta_l|Y^{\mathrm{obs}},\beta_l)$}\par
 By using sampling methods such as the Metropolis--Hastings algorithm\cite{hastings1970monte}, we sample the parameter $\theta_l$ from ${\it p}_{\beta_l}(\theta_l|Y^{\mathrm{obs}},M)$.
\item[{\bf 2}] {\bf 
Probabilistically swap sampling at each $\beta$
}\par
Swap $\theta_l$ and $\theta_{l+1}$ with the probability $\min(1,r)$.
\begin{equation}
\begin{split}
\;\;\;\;\; r &= \frac{{\it  p}(\theta_1,\cdots, \theta_{l+1}, \theta_l,\cdots,\theta_L|Y^{\mathrm{obs}},M)}{{\it  p}(\theta_1,\cdots, \theta_l, \theta_{l+1},\cdots,\theta_L|Y^{\mathrm{obs}},M)}\\
&=\frac{{\it  p}_{\beta_l}(\theta_{l+1}|Y^{\mathrm{obs}},M){\it  p}_{\beta_{l+1}}(\theta_l|Y^{\mathrm{obs}},M)}{{\it  p}_{\beta_l}(\theta_l|Y^{\mathrm{obs}},M){\it  p}_{\beta_{l+1}}(\theta_{l+1}|Y^{\mathrm{obs}},M)}\\
&=\exp\left\{N[\beta_{l+1} - \beta_{l}][E(\theta_{l+1},M) - E(\theta_l,M)]\right\}
\end{split}
\end{equation}

\end{itemize}\par
Exchanging the states across multiple inverse temperatures $\beta$ helps the sampling to converge faster for $p(\theta|\mathcal{Y}^{\mathrm{obs}})$ at $\beta=1$. 
Combining the samples from these spectral posteriors $p(\theta|\mathcal{Y}^{\mathrm{obs}},M)$ creates the global posterior set $p(\vartheta|\mathcal{Y}^{\mathrm{obs}},M)$. 
Finally, drawing $\mathcal{Y}$ from the Gaussian likelihood $p(\mathcal{Y}|\vartheta)$ using these parameters gives the final sample set $\{\mathcal{Y}_k\}_{k=1}^{N_{samp}}$. This set forms the spectral predictive distribution $p(\mathcal{Y}|\mathcal{Y}^{\mathrm{obs}},M)$ with $N_{samp}$ samples.

\subsubsection{Marginalization of Physical-Property Regression Models}
\label{sec_marginalize_regression_model}
This section describes the marginalization calculation for the physical-property regression model defined in Sec. \ref{GP}.
The spectral samples \(\{Y_k\}_{k=1}^{N_{\mathrm{samp}}}\) drawn from \(p(Y|\vartheta)\) are combined with the measurement data \(Z\) to form the dataset $\{\mathcal{Y}_k, Z\}_{k=1}^{N_{samp}}$.
$N_{samp}$ denotes the sample size drawn from the predictive distribution $p(\mathcal{Y}|\vartheta)$.
While performing this calculation, it is important to note that the physical property $Z$ does not change across the samples $k$.
For each sampled set $\left\{\mathcal{Y}_k,Z\right\}$, the marginal likelihood $p(Z|\mathcal{Y}_k)$ can be analytically computed, as follows.
Denoting the $N' \times N'$ identity matrix as $I_{N'}$ and defining $\Lambda = K + \alpha^{-1} I_{N'}$, Gaussian process regression yields the conditional probability $p(Z|{\mathcal{Y}}_k)$ after integrating out $F$, as shown in the following equation.

\begin{eqnarray}
&&p(Z|{\mathcal Y}_k) = \int dFp(Z|F,{\mathcal Y}_k)p(F)\nonumber  \\
&=&\left(\frac{1}{2\pi}\right)^{N'/2}|\Lambda^{-1}|^{1/2}\exp\left\{-\frac{1}{2}Z^T\Lambda^{-1}Z\right\}\end{eqnarray}
Thus, the set of $N_{samp}$ marginalized likelihoods $p(Z|\mathcal{Y}_k)$, denoted as $\{p(Z|\mathcal{Y}_k)\}_{k=1}^{N_{samp}}$, is obtained.

\subsubsection{Marginalization of models of spectral deconvolution processes used by scientists}
\label{sec_marginalize_scientist_model}
Marginalization over $\mathcal{Y}$ means taking the expectation (average value) of $p(Z|\mathcal{Y})$ with respect to the probability distribution $p(\mathcal{Y}|\mathcal{Y}^{\mathrm{obs}},M)$, where $\mathcal{Y}$ is the set of all possible values, $\mathcal{Y}^{\mathrm{obs}}$ represents the observed values, and $M$ denotes the model. This is shown in the following equation.

\begin{align}
\label{eq:expectated}
\tilde{p}(M|Z,\mathcal{Y}^{\mathrm{obs}})&\propto 
\int  d\mathcal{Y}\ p(Z|\mathcal{Y})p(\mathcal{Y}|\mathcal{Y}^{\mathrm{obs}},M)\nonumber\\
&=E_{p(\mathcal{Y}|\mathcal{Y}^{\mathrm{obs}},M)}\left[p(Z|\mathcal{Y})\right]
\end{align}
The expectation in Eq.~\eqref{eq:expectated} was evaluated by Monte Carlo averaging over the sampling set
\(\{p(Z|\mathcal{Y}_k)\}_{k=1}^{N_{\mathrm{samp}}}\).
Here, each \(\mathcal{Y}_k\) was sampled from the predictive distribution
\(p(\mathcal{Y}|\mathcal{Y}^{\mathrm{obs}},M)\),
which was generated using the sampling sequence of the parameter set \(\vartheta\) obtained from the spectral deconvolution model.
Thus, the physical-property-informed model posterior was approximated as
\begin{equation}
\label{eq:mc_approx}
\tilde{p}(M|Z,\mathcal{Y}^{\mathrm{obs}})\propto
E_{p(\mathcal{Y}|\mathcal{Y}^{\mathrm{obs}},M)}\left[p(Z|\mathcal{Y})\right]
\approx
\frac{1}{N_{\mathrm{samp}}}
\sum_{k=1}^{N_{\mathrm{samp}}}
p(Z|\mathcal{Y}_k).
\end{equation}
Building on this approximation, computing the physical-property-informed free energy ($\widetilde{\rm FE}$) as
\[\widetilde{FE}(M)=-\log E_{p(\mathcal{Y}|\mathcal{Y}^{\mathrm{obs}},M)}\left[p(Z|\mathcal{Y})\right] \approx
-\log\left[\frac{1}{N_{\mathrm{samp}}}
\sum_{k=1}^{N_{\mathrm{samp}}}
p(Z|\mathcal{Y}_k)\right] \]
enables spectral model selection.

\section{Demonstration}
This section evaluates the effectiveness of the proposed method.
As outlined in the Introduction, the proposed method estimates the number of peaks in complex spectra dominated by noise or where the background component model remains unknown.
Therefore, we generated artificial datasets characterized by either noise dominance or an unknown background to verify the effectiveness of the proposed method. 
The proposed method was applied to actual measurement data—the IR spectra of polylactic acid—to demonstrate its effectiveness in practical environments.
We compared the results with those of an existing Bayesian spectral deconvolution method, which relies solely on spectral data without considering physical properties.
The detailed results for each experimental case are presented below.

\subsection{Validation Using Noise-Dominant Synthetic Data}
To evaluate whether the proposed method can robustly estimate the number of peaks even when noise is as dominant as the signal, we verified the effectiveness of the proposed method using noise-dominated artificial spectral data.
This dataset, comprising paired spectra and physical property values, was generated as follows.
First, we generated the spectral data $\{x_i,y_i\}_{i=1}^{N}$ featuring two distinct peaks.
The position of the first peak $\mu_1$ was drawn from a uniform distribution over the range $3.5 \leqq x \leqq 4.75$, with its intensity $w_1$ comparable to the noise standard deviation ($\sigma = 0.2$). 
The position of the second peak, $\mu_2$, fell within $5.25 \leqq x \leqq 6.5$, whereas its intensity $w_2$ was uniformly sampled in the range $1.0 \leqq w_2 \leqq 5.0$ to ensure it was significantly larger than the noise.
The measurement domain consisted of 600 equally spaced points $x_i$ in the interval $3.0 \le x \le 7.0$, resulting in a total sample size of $N=600$.
Based on this configuration, six spectral sets were generated by randomly varying the peak positions within the specified interval and intensities within the specified ranges, according to the criteria described above.
We defined the physical property $\Delta$ of each spectrum as equaling ten times the peak separation: $\Delta = 10|\mu_1-\mu_2|$.
This procedure yielded the spectral dataset $\mathcal{Y} \in \mathbb{R}^{600 \times 6}$ containing $N'=6$ spectra, each with $N=600$ data points. We concurrently defined the physical property vector $Z = (z_1 = \Delta_1, z_2 = \Delta_2, \dots, z_6 = \Delta_6) \in \mathbb{R}^6$.
\par
The parameter settings and prior distributions for the proposed method are described below.
The spectral model parameters $\theta = (\mu, a, w)$ were given prior distributions for use in the marginalization of the spectral deconvolution model (Sec.~\ref{spectrum_con}) and in generating the sampling sequence from the posterior distribution (Eq.~\ref{eq:inverse_temparature}).
The prior $p(\mu)$ for the peak position $\mu$ was uniform over $3.25 \leqq \mu \leqq 6.5$; the prior $p(a)$ for the peak width $a$ was uniform over $0.005 \leqq a \leqq 0.05$; and the prior $p(w)$ for the peak intensity $w$ was uniform over $0.2 \leqq w \leqq 5$.
A burn-in period of 5000 steps was used for the REMC method, followed by 15,000 sampling steps. The temperatures were set at $L=40$ with $d=1.15$, and $\beta_l$ followed the function proposed by Nagata et al. \cite{exrate}.
\begin{equation}
\beta_l =
\begin{cases} 
0.0 & \quad \text{for } l = 1, \\
d^{{l-L}} & \quad \text{for } l = 2, 3, \dots, L
\end{cases}
\label{eq:temperature1}
\end{equation}
In the marginalization of the regression model for the physical property value (Section~\ref{sec_marginalize_regression_model}), the following kernel function was used.
\begin{eqnarray}
k(Y_i, Y_j) := C_0^2\left(1+\frac{\left|Y_i - Y_j\right|_2^2}{2C_1 C_2^2}\right)^{-C_1}
\end{eqnarray}
The kernel hyperparameters $\{C_0, C_1, C_2\}$ were estimated using the empirical Bayes method. These parameters maximize the likelihood function $p(Z|\mathcal{Y}^{\rm sample}_{i})$.
The sampling sequence for the spectral predictive distribution $p(\mathcal{Y}|\mathcal{Y}^{\mathrm{obs}},M)$ was calculated during the marginalization of the model encapsulating the analytical approach used by scientists (Sec.~\ref{sec_marginalize_scientist_model}). This calculation used the sampling sequence of the vector $\vartheta$, which is composed of the parameters $\theta$.
The sequence of $\vartheta$ was extracted at five-step intervals to minimize the correlation. This procedure ensured that the samples were i.i.d.
The expected value of $p(Z|\mathcal{Y})$ was computed based on $p(\mathcal{Y}|\mathcal{Y}^{\mathrm{obs}},M)$ (Eq.~\eqref{eq:expectated}).
The number of sampling sequences for $p(\mathcal{Y}|\mathcal{Y}^{\mathrm{obs}},M)$ was set to 100. This setting balances the computational efficiency with rapid convergence in the expectation calculation.
\par
\begin{figure}[t]

\centering
\includegraphics[width=0.95\columnwidth]{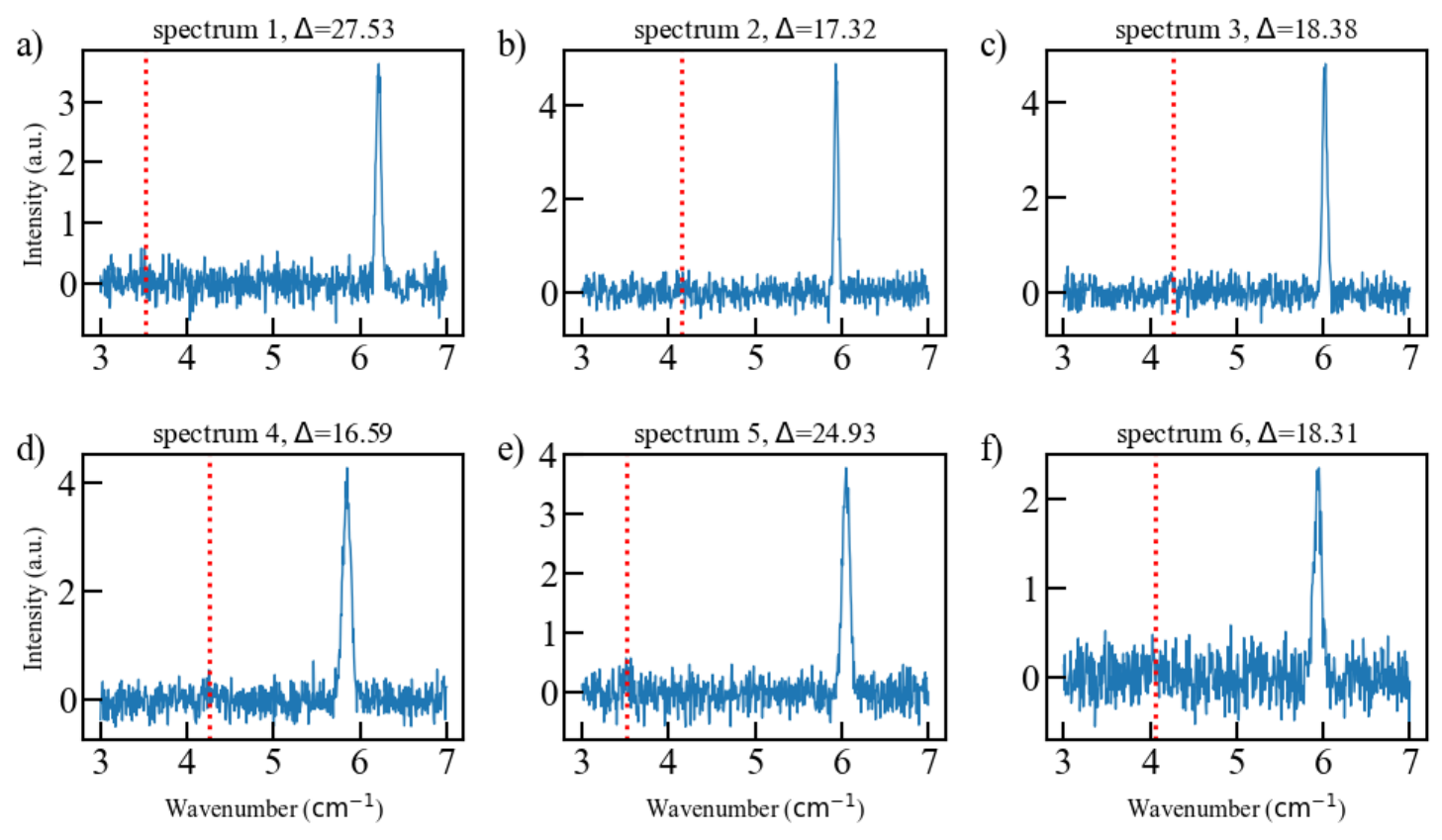}
\caption{
Noise-dominant synthetic spectral dataset used to validate the proposed method. Panels (a)--(f) show six samples generated under different conditions. In each panel, the blue solid line represents the observed spectrum $Y$, which was generated by adding high-intensity Gaussian noise to two Gaussian peaks. The left peak is completely buried in the noise, reproducing a situation in which peak estimation is difficult when using conventional methods. The red dashed lines indicate the true peak positions $(\mu_1, \mu_2)$, and the physical property value $\Delta$ shown above each panel is defined as $\Delta = 10|\mu_2 - \mu_1|$.}
\label{data_arti}
\end{figure}

\begin{table}[t]
\centering
\caption{
Comparison of model-selection results for the noise-dominant synthetic spectral dataset. For each candidate model, the table shows the Bayesian free energy \(FE\) calculated by conventional Bayesian spectral deconvolution, the physical-property-informed free energy \(\widetilde{FE}\) calculated by the proposed method, and the corresponding model probabilities obtained by normalizing \(\exp[-FE(M)]\) or \(\exp[-\widetilde{FE}(M)]\) over candidate models.}
\label{tab:result_IR}
\begin{tabular}{lcccc}
\toprule
& \multicolumn{2}{c}{\textbf{Bayesian spectral deconvolution}} & \multicolumn{2}{c}{\textbf{Proposed method}} \\
\cmidrule(lr){2-3} \cmidrule(lr){4-5}
Model & FE & Probability & $\mathbf{\widetilde{FE}}$ & Probability \\
\midrule
1-peak & $4.255\times 10^3$ & 1.0 & $3.977\times 10^3$ & 0 \\
2-peak & $4.347\times 10^3$ & 0 & $3.962\times 10^3$ & 1.0 \\
\bottomrule
\end{tabular}
\end{table}

\begin{figure}[t]
    \centering
    \includegraphics[width=0.8\columnwidth]{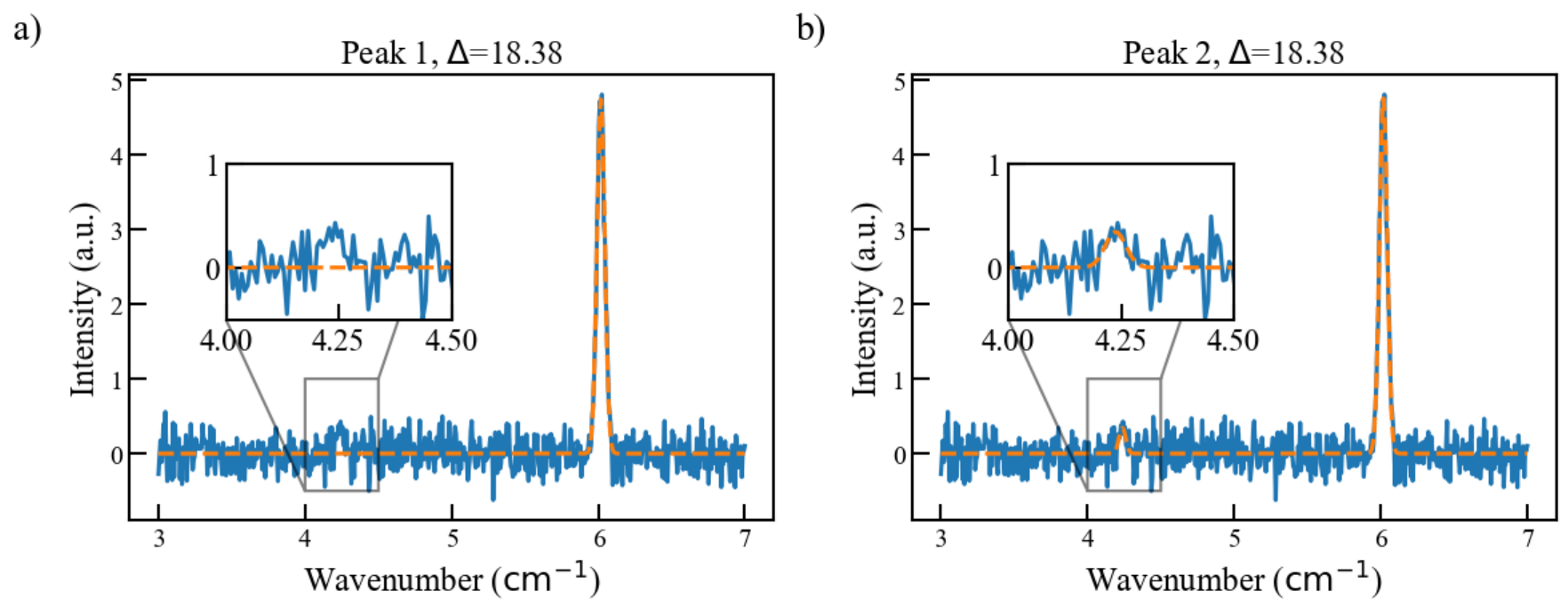}
\caption{
Example of fitting curves at the maximum a posteriori (MAP) solution for a synthetic spectrum ($\Delta = 18.38$). Panels (a) and (b) show the fitting results for the 1-peak and 2-peak models. The blue line represents the observed spectrum, and the orange dashed line represents the fitting curve at the MAP solution for each candidate model.}

    \label{arti_fit}
\end{figure}
The conventional Bayesian spectral deconvolution method and the proposed method were applied to the generated dataset $\{\mathcal{Y},Z\}$. 
The conventional method selected a single-peak spectral model, whereas the proposed method correctly selected a two-peak model (Table~\ref{tab:result_IR}).
Fig.~\ref{arti_fit} presents the fitting results for the MAP solutions of each model. 
The MAP solution is the estimate of the parameter $\theta$ maximizing the posterior distribution $p(\theta|Y)$ given the observed data $Y$. 
The spectral deconvolution model assumes uniform prior distributions. The calculation of the MAP solution $\theta_{\rm MAP}$ reduces to the maximization of the log-likelihood function.
Taking the logarithm of Eq.~(\ref{likelihood}) yields the following expression.
\begin{eqnarray}
\log p(Y | \theta) &=& \sum_{i=1}^{N} \left( -\frac{(y_i - g(x_i; \theta))^2}{2\sigma^2} - \log(\sqrt{2\pi}\sigma) \right)
\end{eqnarray}
The second term on the right-hand side is a constant independent of $\theta$. Maximizing this expression is equivalent to minimizing the squared error term $\sum_{i=1}^{N}(y_i - g(x_i; \theta))^2$. 
The MAP solution minimizes the discrepancy between the spectral data and fitted curve.
The proposed method correctly estimates the true peak positions and intensities (Fig.~\ref{arti_fit}). 
 The conventional method struggles with noise-dominant spectral data, where the peak and noise intensities are comparable. However, the proposed method effectively estimates the number of peaks in these challenging scenarios. These results confirm the validity of the approach for peak estimation.
\par

\subsection{Validation Using Synthetic Data with Unknown Backgrounds}
\label{sec:arti_unknown_bg}
The effectiveness of the proposed method was verified using artificial spectral data, which simulated spectra containing unknown background components.
The artificial dataset contained pairs of spectra and physical property values. The generation process is described below.
The spectral data $\{x_i,y_i\}_{i=1}^{N}$ were constructed with two distinct peaks.
The first peak position $\mu_1$ follows a uniform distribution over $3.5 \leqq x \leqq 4.0$. Its intensity $w_1$ was randomly sampled from a uniform distribution over $0.1 \leqq w_1 \leqq 0.2$. The second peak position $\mu_2$ is located within $5.5 \leqq x \leqq 6.0$. Its intensity $w_2$ was also drawn from a uniform distribution over $0.1 \leqq w_2 \leqq 0.2$.
Six hundred measurement points $x_i$ were distributed at equal intervals between $3.0 \leqq x \leqq 7.0$.
The total number of sample points, $N$, in the spectral data was 600.
Each spectrum was generated by first adding a background component based on a sigmoid function.
This background, $B(x)$, was defined using the parameters of amplitude $A$, slope $s$, and center position $c$, as follows:
$$B(x) = \frac{A}{1 + \exp(-s(x - c))}$$.
The parameters $A$, $s$, and $c$ were randomly sampled from the following uniform distributions: $0.15 \leqq A \leqq 0.20$, $4 \leqq s \leqq 6$, and $4.8 \leqq c \leqq 5.2$. This background component was then added to the peak signals and noise.
The spectral deconvolution assumed that the background structure was unknown to verify the effectiveness of the method. 
Six spectral samples were generated by randomly varying the peak positions, intensities, and background shapes.
The physical property $\Delta$ was defined as ten times the peak separation: $\Delta = 10|\mu_1 - \mu_2|$.
The resulting dataset $\mathcal{Y} \in \mathbb{R}^{600 \times 6}$ consisted of $N' = 6$ spectra with $N = 600$ data points each. The corresponding physical property values are $Z = (z_1 = \Delta_1, z_2 = \Delta_2, \dots, z_6 = \Delta_{6}) \in \mathbb{R}^{6}$.
\par
The spectral deconvolution process assumed an unknown background model. The asymmetric least squares (AsLS) method performed background removal without any specific model assumptions.
The AsLS method applies asymmetric weighting to the residuals between the observed data and estimated values \cite{Eilers2005}. This approach eliminates the influence of the peak components. The procedure estimates a smooth background without requiring a prior model.
This technique optimizes the tradeoff between the data fit and background smoothness.
The method estimates the background $B_{est}=\{b_i\}_{i=1}^{N}$ minimizing the following objective function $S$:
$$S = \sum_{i=1}^{N} w_i (y_i - b_i)^2 + \lambda \sum_{i=1}^{N} (\Delta^2 b_i)^2$$,
where \(\Delta^2\) denotes the second-order difference operator, and $w_i$ is an asymmetric weight determined by the sign of the residual.
For data points above the background ($y_i > b_i$), the weight is set to $w_i = p$, and $w_i = 1-p$ for all other cases.
$\lambda$ and $p$ denote the smoothing and asymmetry parameters, respectively.
In this verification, the parameters for the AsLS method were set to $\lambda = 1000$ and $p = 0.0001$.
Although the AsLS method is a powerful technique for subtracting unknown backgrounds, it has certain limitations. It may fail to completely remove the background or mistakenly eliminate actual spectral peak structures if the smoothing parameter is set inappropriately, if spectral smoothness varies significantly across the regions, or if the background and peak structures share similar smoothness characteristics.\par
The proposed method was applied to the preprocessed dataset.
Here, we describe the parameter settings and prior distributions used in its application.
In the marginalization of the spectral deconvolution model (Sec.~\ref{spectrum_con}), uniform prior distributions were assigned to obtain the sampling sequence from the posterior distribution (Eq.~\ref{eq:inverse_temparature}) of the spectral model parameters $\theta = (\mu, a, w)$, as follows.
Specifically, the prior distributions for the peak position $p(\mu)$, peak intensity $p(w)$, and peak width $p(a)$ were set as uniform distributions over the ranges $3.5 \leq \mu \leq 6.5$, $0.001 \leq w \leq 0.5$, and $0.005 \leq a \leq 0.18$, respectively.
Furthermore, in the REMC method used for sampling, the burn-in period was set to 5000 steps, followed by 15,000 sampling steps.
The number of temperatures was set to $L=40$ and $d=1.26$. Following the research by Nagata et al.\cite{exrate}, $\beta_l$ was set according to Eq.~(\ref{eq:temperature1}).

\begin{figure}[hb]

    \includegraphics[width=\columnwidth]{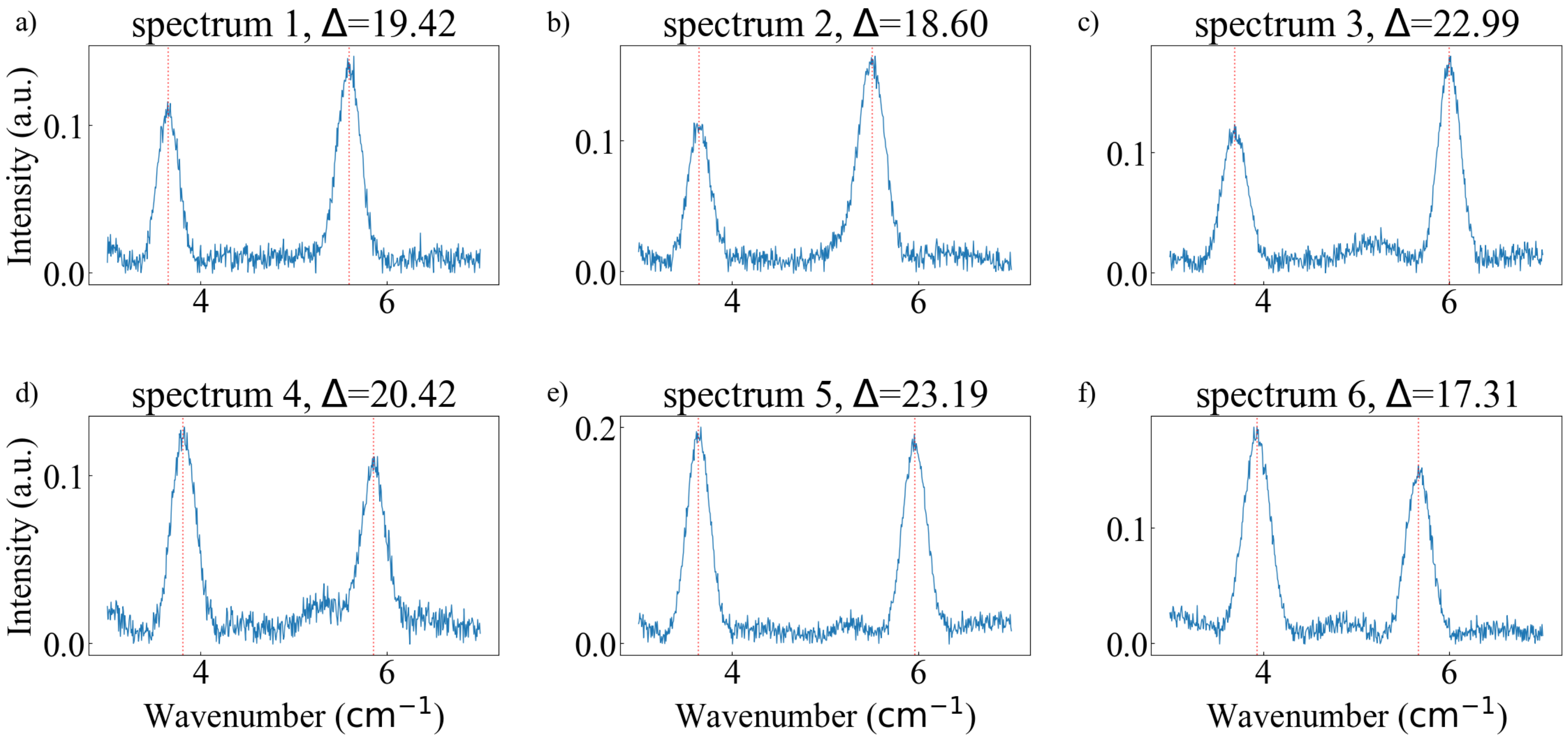}
\caption{
Synthetic spectral dataset containing unknown background components. Panels (a)--(f) show six different spectra generated by adding a sigmoidal background and Gaussian noise to two Gaussian peaks. The red dotted lines indicate the true peak positions $\mu_1$ and $\mu_2$, and $\Delta$ shown in the heading is defined as $10|\mu_2 - \mu_1|$.}

    \label{ARD_bg}
\end{figure}

\begin{figure}[htbp]
    \centering
    \includegraphics[width=\columnwidth]{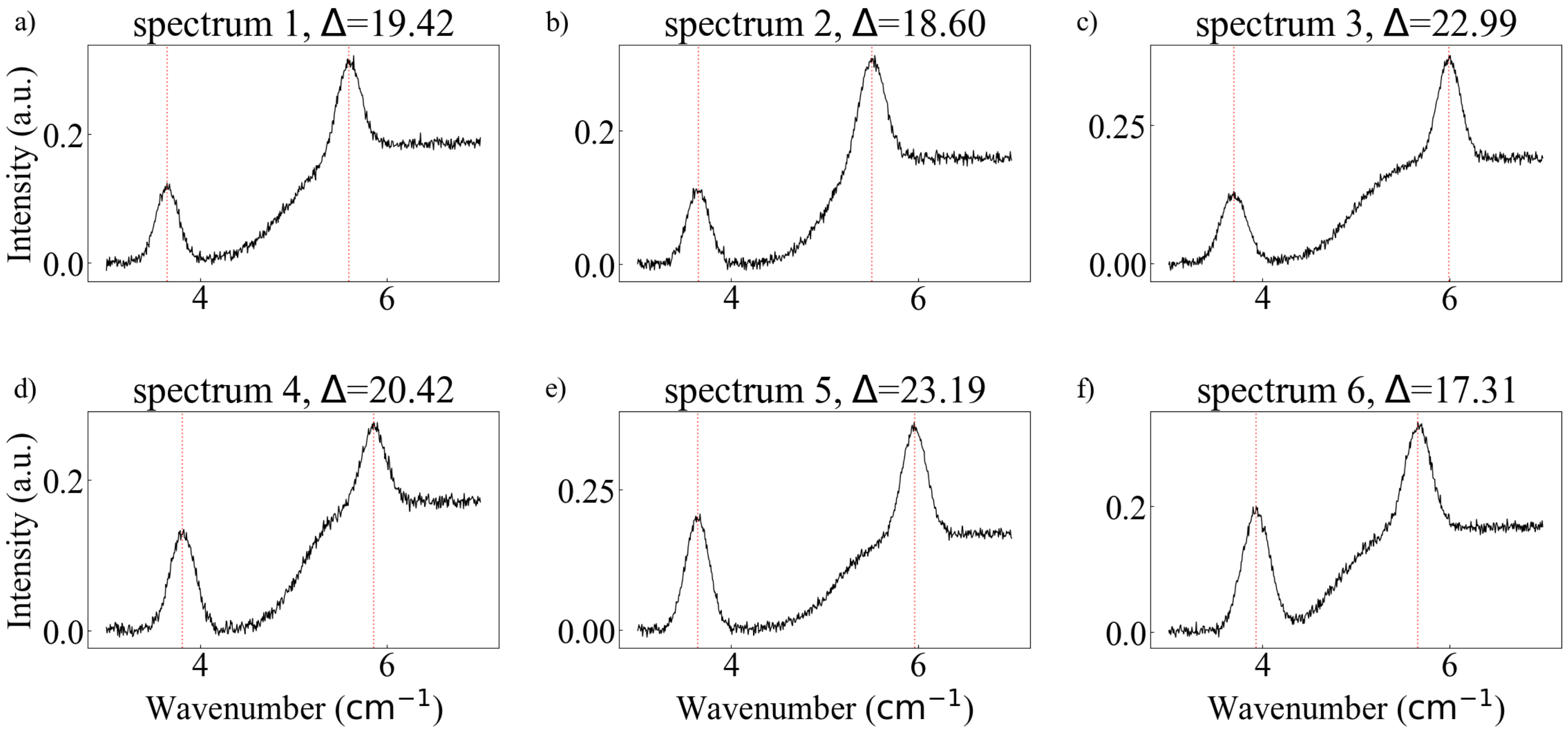} 
\caption{
Synthetic spectral dataset prior to background subtraction. This figure displays the raw, unprocessed spectra containing unknown sigmoidal background components, generated to evaluate the robustness of the analytical method under baseline fluctuations. Panels (a)--(f) simulate varying observational conditions by combining two ideal Gaussian peaks with a nonlinear sigmoidal background and observational Gaussian noise. The red dotted lines indicate the true peak center positions, $\mu_1$ and $\mu_2$. The value \(\Delta\) shown in each title is defined as ten times the peak separation, \(\Delta=10|\mu_2-\mu_1|\). These pre-correction data serve as a reference to visually demonstrate how the degree of peak overlap (varying $\Delta$) and the gradient of the background affect the overall spectral profile and complicate subsequent peak detection processes.
}
    \label{fig:raw_bg_dataset}
\end{figure}

\begin{table}[t]

\caption{
Comparison of peak-number model selection for synthetic spectra containing unknown background components. For each candidate model, the table shows the Bayesian free energy \(FE\) calculated by conventional Bayesian spectral deconvolution, the physical-property-informed free energy \(\widetilde{FE}\) calculated by the proposed method, and the corresponding model probabilities obtained by normalizing \(\exp[-FE(M)]\) or \(\exp[-\widetilde{FE}(M)]\) over candidate models.}
\label{tab:result_backs}
\begin{tabular}{lcccc}
\toprule
& \multicolumn{2}{c}{\textbf{Bayesian spectral deconvolution}} & \multicolumn{2}{c}{\textbf{Proposed method}} \\
\cmidrule(lr){2-3} \cmidrule(lr){4-5}
Model & FE & Probability & $\widetilde{\rm FE}$ & Probability \\
\midrule
2-peak & $9.503\times 10^3$ & 0.0 & $1.320\times 10^3$ & 0.73 \\
3-peak & $6.432\times 10^3$ & 1.0 & $1.321\times 10^3$ & 0.27 \\
\bottomrule
\end{tabular}
\end{table}
The marginalization of the physical-property regression model (Sec.~\ref{sec_marginalize_regression_model}) utilized the following kernel function:
\begin{eqnarray}
k(Y_i, Y_j) := C_0^2 \exp\left(-\frac{\left|Y_i - Y_j\right|_2^2}{2C_1^2}\right).
\end{eqnarray}
The kernel hyperparameters $\{C_0, C_1\}$ were estimated using the empirical Bayes method. These parameters maximize the likelihood function $p(Z|\mathcal{Y}^{\rm sample}_{i})$.
The marginalization of the model representing the approach used by scientists calculated the sampling sequence for the spectral predictive distribution $p(\mathcal{Y}|\mathcal{Y}^{\mathrm{obs}},M)$ (Sec.~\ref{sec_marginalize_scientist_model}).
This calculation utilized the sampling sequence of the vector $\vartheta$, which concatenates the parameters $\theta$ obtained by marginalizing the spectral deconvolution model for each spectrum.
The sequence $\vartheta$ was extracted at 25-step intervals to minimize the correlations. This procedure ensured that the sequence is i.i.d.
The expected value of $p(Z|\mathcal{Y})$ was computed based on $p(\mathcal{Y}|\mathcal{Y}^{\mathrm{obs}},M)$ (Eq.~\eqref{eq:expectated}).
The rapid convergence of the expectation calculation and requirement for computational efficiency determine the sample size. The number of sampling sequences for $p(\mathcal{Y}|\mathcal{Y}^{\mathrm{obs}},M)$ was set to 100.\par
\begin{figure}[t]
    \centering
    \includegraphics[width=\columnwidth]{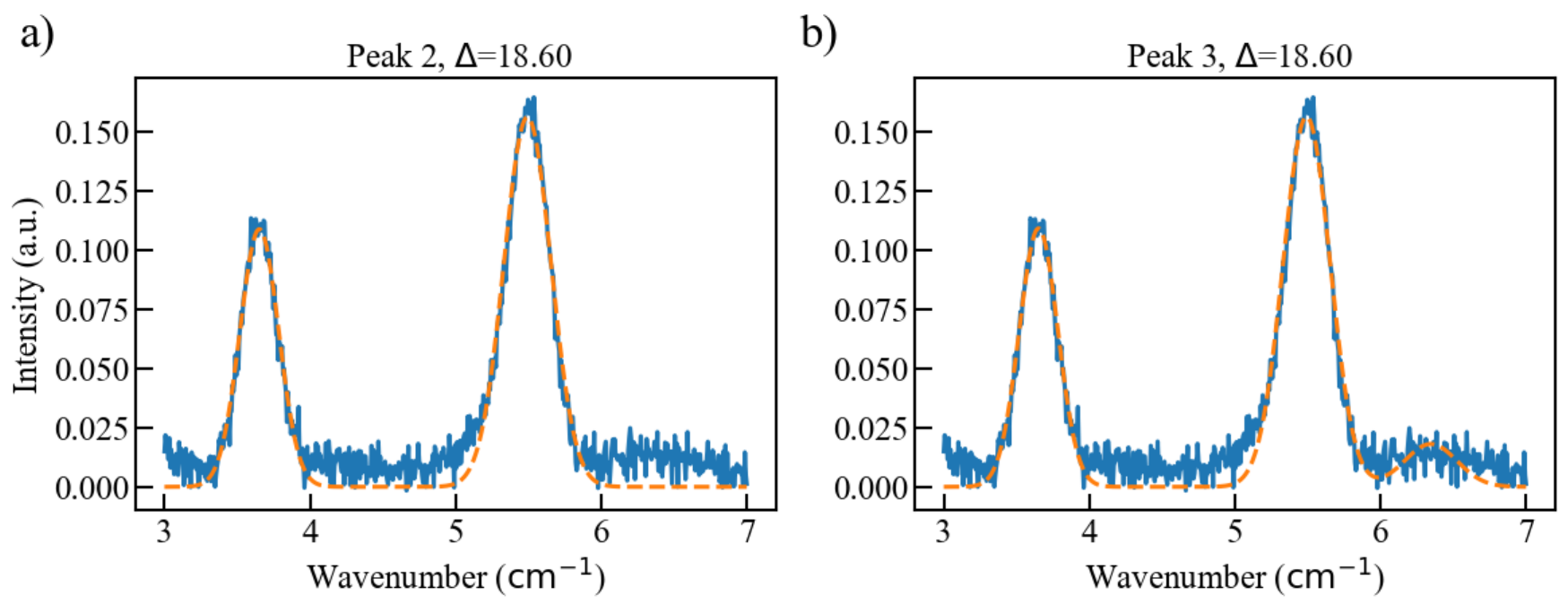}
\caption{
Example of fitting curves at the MAP solution for a synthetic spectrum ($\Delta = 18.60$). Panels (a) and (b) show the fitting results for the 2-peak and 3-peak models. The blue line represents the observed spectrum, and the orange dashed line represents the fitting curve at the MAP solution for each candidate model.}
    \label{backs_fit}
\end{figure}

We compared the conventional Bayesian spectral deconvolution with the proposed method using the artificial dataset $\{\mathcal{Y},Z\}$. 
Background removal preprocessing generated pseudo-peak structures. The conventional method selected a three-peak spectral model containing these artifacts. The proposed method correctly selected a two-peak model (Table~\ref{tab:result_backs}).
The spectral deconvolution results obtained by the proposed method accurately estimated the true peak positions and intensities (Fig.~\ref{backs_fit}).
Conventional methods struggle with spectral data containing unknown backgrounds. The proposed approach effectively determines the number of peaks in these challenging scenarios. These results confirm the validity of the method for noise-dominant data with complex backgrounds.

\subsection{Application to IR spectral data}
\begin{figure}[t]
\centering
\includegraphics[width=\linewidth]{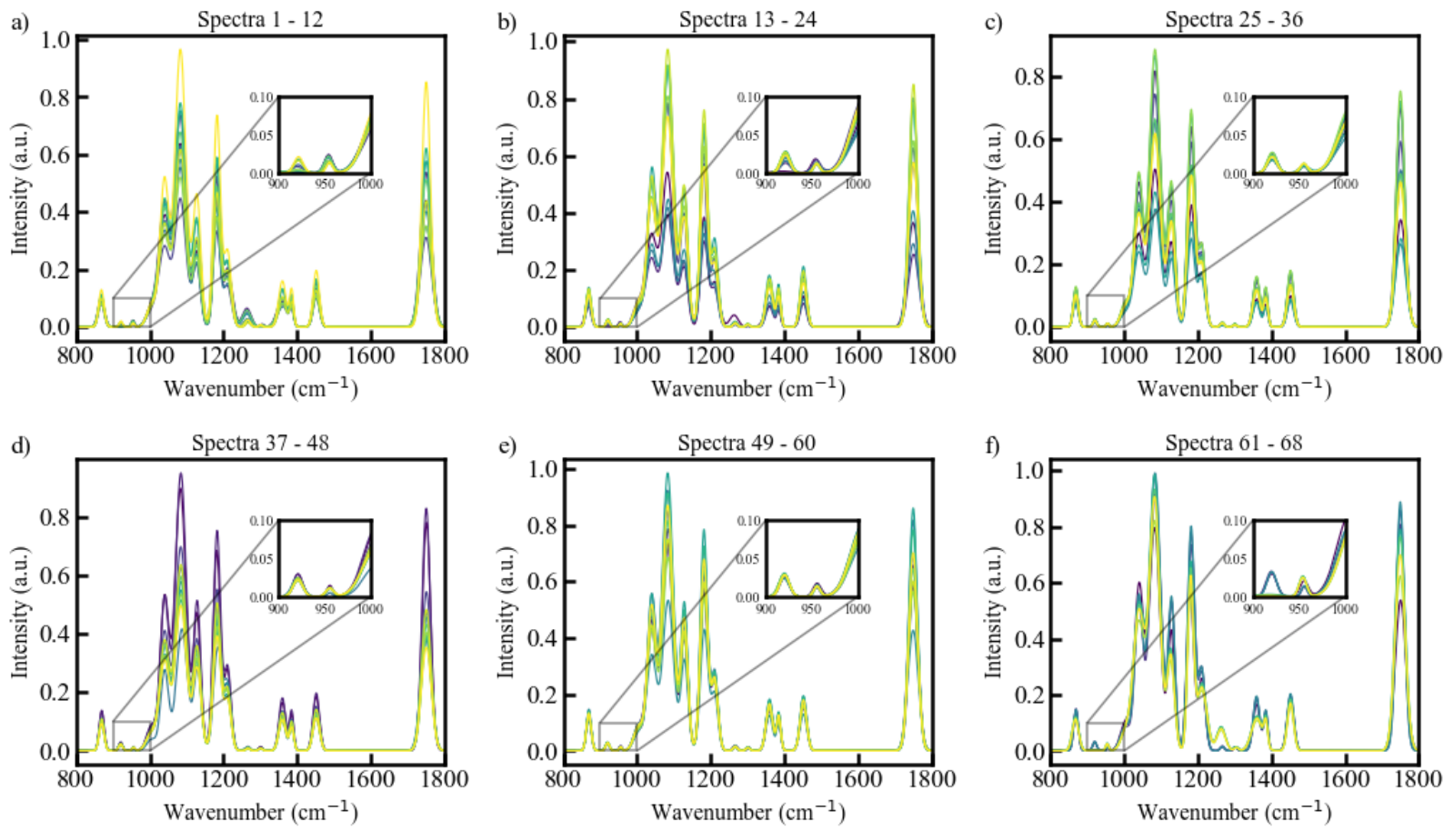} 
\caption{
IR spectral dataset of polylactic acid used as the real dataset. Each curve represents one measured spectrum plotted over the wavenumber range $800$--$1800~\mathrm{cm}^{-1}$. For visibility, the spectra are displayed in panels grouped according to the spectrum number: (a) 1--12, (b) 13--24, \ldots, (f) 61--68. These data constitute $Y = \{Y_j\}_{j=1}^{N'}$ in the hierarchical model.}

\label{IRdata}
\end{figure}

This study verified the effectiveness of the proposed method using actual IR spectra data of polylactic acid.
Crystallinity of a material signifies the ratio of its crystalline to amorphous phases. IR spectroscopy evaluates the degree of crystallinity \cite{Malone2024}, which governs the biodegradability of polylactic acid.
Degradation proceeds preferentially in amorphous regions, which are easily penetrated by water molecules and enzymes, whereas tightly packed molecular chains in crystalline regions inhibit degradation.
Understanding the microscopic crystalline state remains essential for predicting the biodegradability potential.
The wavenumber region from $900~\mathrm{cm}^{-1}$ to $1000~\mathrm{cm}^{-1}$ contains a group of low-intensity peaks \cite{Meaurio2006}, which respond sensitively to the polymer backbone conformation.
The peak at $956~\mathrm{cm}^{-1}$ indicates amorphous structures, whereas the $920~\mathrm{cm}^{-1}$ band marks regular helical $\alpha$ crystals.
 The balance between these regions indicates biodegradability. Accurate quantification relies on extracting small peaks in this region.
This study utilized 68 poly(lactic acid) samples synthesized under various crystallization temperatures and concentrations \cite{Takahashi2021}. The dataset consisted of IR spectra from $800~\mathrm{cm}^{-1}$ to $1800~\mathrm{cm}^{-1}$. It also included the degradation rates for each sample. 
Previous studies report 14 peak structures in this specific wavenumber region \cite{IR1,IR2}.
 The background of the poly(lactic acid) IR spectrum is generally unknown. Empirical preprocessing typically employs the AsLS method \cite{ALS}, which is a non-parametric technique for estimating the unknown background structure \cite{Meyns2023_AnalMethods_606_617}.
In this experiment, the existing spectral deconvolution methods and the proposed approach were applied to AsLS-preprocessed data. The verification focused on the extraction of peak structures between $900\text{ cm}^{-1}$ and $1000\text{ cm}^{-1}$. These peaks correlate strongly with the biodegradability of the material and are characterized by extremely small intensities. 
The analysis focused on the region between $900~\mathrm{cm}^{-1}$ and $1000~\mathrm{cm}^{-1}$. The evaluation determined whether to estimate one peak or two peaks in this range.
The estimation yielded a total peak count of either 14 or 13. The spectral data $Y$ consisted of 2074 evenly spaced points: $Y = (y_1, y_2, \dots, y_{2074})$.
As described above, the spectral data were preprocessed in advance using the AsLS method \cite{ALS}.
Figure~\ref{IRdata} shows the preprocessed spectral data.
The degradation rate $Dg$, defined as the percentage change in weight after each polymer was immersed in an enzymatic buffer at $37^\circ\mathrm{C}$ for 2 days, was used as the physical property value $Z$.
The resulting set of spectra, $\mathcal{Y} \in \mathbb{R}^{2074 \times 68}$, and degradation rates $Z = (z_1 = Dg_1, z_2 = Dg_2, \dots, z_{N'} = Dg_{N'}) \in \mathbb{R}^{68}$ were used as the datasets.

We describe the prior distributions and sampling parameters used in the demonstration of the proposed method.
The prior distributions were set as shown in Table~\ref{tab:model_settings2}.
For the prior distribution of the peak positions, we referred to the peak occurrence ranges identified in previous studies \cite{IR1,IR2}.
The number of burn-in steps was set to 5000, followed by 195{,}000 sampling steps.
The temperature parameters were set to $L = 70$ and $d = 1.35$, and $\beta_l$ was determined according to Eq.(~\ref{eq:temperature1}).
Because the real IR spectra contain unknown noise, the spectral noise variance was estimated based on the maximum likelihood by using the sampling results of Bayesian spectral deconvolution, following the process of Tokuda et al. \cite{Tokuda_2017}.
In the marginalization of the physical-property regression model (Sec.~\ref{sec_marginalize_regression_model}), we used the following kernel function:
\begin{eqnarray}
k(Y_i, Y_j) := C_0^2 \exp\left(-\frac{\left|Y_i - Y_j\right|_2^2}{2C_1^2}\right).
\end{eqnarray}
The hyperparameters of the kernel function, $\{C_0, C_1\}$, were estimated using an empirical Bayes method as the values that maximize the likelihood function $p(Z|\mathcal{Y}^{\rm sample}_{i})$.
In the marginalization of the spectral deconvolution process model representing the approach used by scientists, we computed a sample sequence from the predictive distribution of the spectra, $p(\mathcal{Y}|\mathcal{Y}^{\rm obs}, M)$. This was accomplished using the sample sequence of the vector $\vartheta$, which consists of the parameters $\theta$ obtained by marginalizing the spectral deconvolution model for each spectrum (Sec.~\ref{sec_marginalize_scientist_model}).
At this stage, to ensure that the sequence of $\vartheta$ obtained by marginalizing the spectral deconvolution model was approximately i.i.d., we computed the autocorrelation of the samples and extracted every 25th sample, as this interval reduced the correlation to a sufficiently low level.
Based on $p(\mathcal{Y}|\mathcal{Y}^{\rm obs}, M)$, we computed the expectation of $p(Z|\mathcal{Y})$ (Eq.~\eqref{eq:expectated}).
Considering the computational efficiency and relatively fast convergence of this expectation calculation, the number of samples drawn from $p(\mathcal{Y}|\mathcal{Y}^{\rm obs}, M)$ was set to 100.\par

\begin{table}[t]
\centering
\caption{
Prior distributions assigned to the parameters of each peak in the IR spectra of polylactic acid. The prior distribution of the peak position $\mu_j$ is given as a uniform distribution $U(\cdot,\cdot)$, and the corresponding wavenumber range (in units of $\mathrm{cm}^{-1}$) is specified for each $j$. In addition, uniform distributions are assigned to the peak width $a$ and peak intensity $w$ for all peaks, and these are used as the prior distributions of the spectral deconvolution model.}

\label{tab:model_settings2}
\begin{tabular}{lccc}
\toprule
\multicolumn{4}{c}{\textbf{Prior distributions of $\mu$}} \\
\cmidrule(r){1-2} \cmidrule(l){3-4}
Peak Number ($j$) & $p(\mu_j)$ & Peak Number ($j$) & $p(\mu_j)$ \\
\midrule
1  & $U(1740,\ 1780)$ & 8  & $U(1195,\ 1220)$ \\
2  & $U(1432,\ 1472)$ & 9  & $U(1110,\ 1150)$ \\
3  & $U(1348,\ 1388)$ & 10 & $U(1070,\ 1100)$ \\
4  & $U(1345,\ 1368)$ & 11 & $U(1025,\ 1065)$ \\
5  & $U(1300,\ 1315)$ & 12 & $U(915,\ 935)$   \\
6  & $U(1250,\ 1290)$ & 13 & $U(945,\ 960)$   \\
7  & $U(1160,\ 1195)$ & 14 & $U(860,\ 880)$   \\
\midrule
\multicolumn{4}{c}{\textbf{Prior distributions of $a$ and $w$}} \\
\cmidrule(r){1-4}
\multicolumn{2}{c}{Parameter} & \multicolumn{2}{c}{Setting} \\
\cmidrule(r){1-2} \cmidrule(l){3-4}
\multicolumn{2}{c}{$p(a)$} & \multicolumn{2}{c}{$U(0,\ 13)$} \\
\multicolumn{2}{c}{$p(w)$} & \multicolumn{2}{c}{$U(0,\ 1.2)$} \\
\bottomrule
\end{tabular}
\end{table}

\begin{figure}[H]

\centering
    \centering
        \includegraphics[width=\columnwidth]{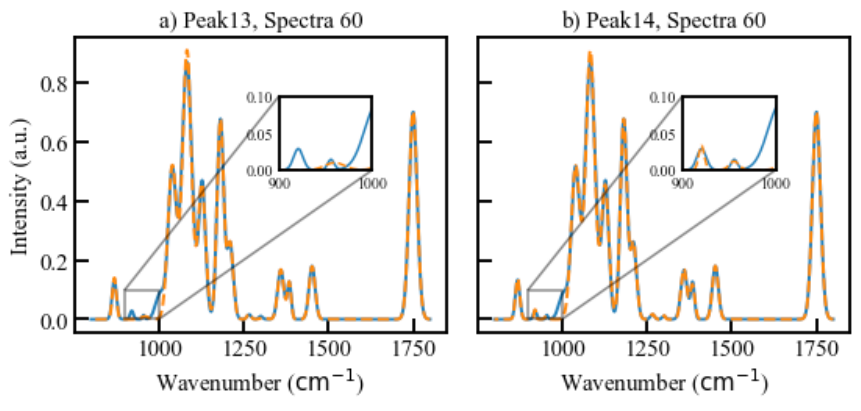}
\caption{
Results of MAP fitting for the IR spectrum of polylactic acid under the two candidate peak-number models. The blue line represents the observed spectrum, and the orange dashed line represents the MAP fit. Panels (a) and (b) show the results for 13 and 14 peaks, respectively. This figure illustrates how the assumed number of peaks affects the reconstruction of the target region.}

    \label{IR_fit}
\end{figure}

\begin{table}[t]
\centering
\caption{
Comparison of model-selection results for peak-number models applied to the IR spectral dataset. For each candidate model, the table shows the Bayesian free energy \(FE\) calculated by conventional Bayesian spectral deconvolution, the physical-property-informed free energy \(\widetilde{FE}\) calculated by the proposed method, and the corresponding model probabilities obtained by normalizing \(\exp[-FE(M)]\) or \(\exp[-\widetilde{FE}(M)]\) over candidate models. This table serves as an indicator for evaluating whether the proposed method can support the appropriate peak-number model even in the presence of pseudo-peaks arising from the background.}

\label{tab:result_comparison}
\begin{tabular}{lcccc}
\toprule
& \multicolumn{2}{c}{\textbf{Bayesian spectral deconvolution}} & \multicolumn{2}{c}{\textbf{Proposed method}} \\
\cmidrule(lr){2-3} \cmidrule(lr){4-5}
\textbf{Model} & \textbf{FE} & \textbf{Probability} & $\mathbf{ \widetilde{FE} }$ & \textbf{Probability} \\
\midrule
13-peak & $1.1277 \times 10^{-2}$ & 0.501 & $-1.0015\times 10^4$ & 0.0\\
14-peak & $1.6612 \times 10^{-2}$ & 0.499   & $-1.0042\times 10^4$ & 1.0 \\
\bottomrule
\end{tabular}
\end{table}
We applied the conventional Bayesian spectral deconvolution method and the proposed method to the generated dataset $\{\mathcal{Y}, Z\}$.
The conventional method assigned nearly equal posterior probabilities to the 13- and 14-peak models, whereas the proposed method strongly favored the 14-peak model, which is consistent with peak assignments reported in previous IR studies (Table~\ref{tab:result_comparison}).
More precisely, the conventional method assigned comparable posterior probabilities to the 13- and 14-peak models, whereas the proposed method strongly favored the 14-peak model, consistent with previous IR peak assignments. 
Figure~\ref{IR_fit} shows the fitting curve obtained from the MAP solution for a spectrum $Y_{60}$.
This fitting result suggests that the proposed method extracted two weak peaks in the region associated with degradation behavior. 
Appendix~\ref{all_IR} shows all the fitting curves.
These results confirm that the proposed method can effectively estimate the number of peaks even for real spectral data with an unknown background, a task that is difficult for the conventional method.\par

\subsection{Discussion}
\begin{figure}[htb]
  \centering
  \includegraphics[width=0.9\linewidth]{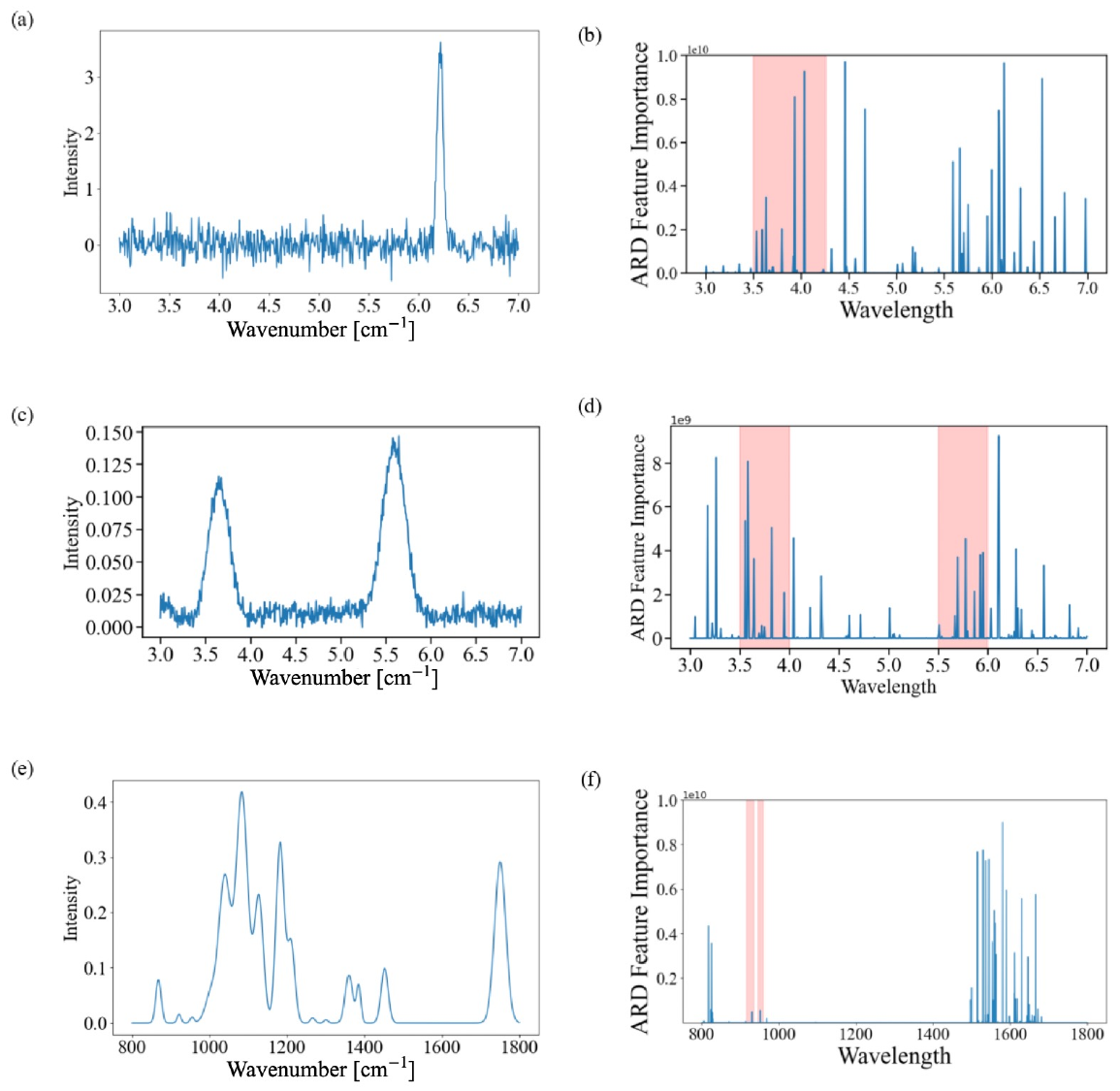}
  \caption{
  Correspondence between ARD-based feature importance and spectra. Panels (a), (c), and (e) show the observed spectra, and panels (b), (d), and (f) show the ARD-estimated feature importance for each coordinate, where larger values indicate greater contributions to predicting the physical property $Z$. The pairs (a,b), (c,d), and (e,f) correspond to the noise-dominant synthetic data, synthetic data with an unknown background, and polylactic acid IR data, respectively. Red-shaded regions indicate the possible true peak ranges in the synthetic data and the prior ranges of the two target peak positions in the real data.}
  \label{fig:ARD_summary}
\end{figure}
Using the proposed method, which mimics the spectral deconvolution process followed by scientists by integrating the physical-property information with spectral data, we confirmed that property-informed peak-number estimation can be achieved even for spectra containing high-intensity noise and background.
Depending on the information obtained from the physical properties, the method may also misestimate the number of peaks.
Such behavior was observed when a Hamiltonian model was estimated for an X-ray photoelectron spectroscopy (XPS) spectrum (see Appendix~\ref{XPS_result} for details).
This spectrum contains three peak structures, whereas the physical property is uniquely determined by the relative intensities of only two of them.
The peak structure extracted by the proposed method, which existing methods could not recover, is, by construction, related to the physical property.
Such an extracted peak structure is unlikely to be erroneous, and the proposed method should therefore be regarded not as a standalone method for estimating the correct peak structure, but as a method for recovering peak structures overlooked by existing Bayesian spectral deconvolution methods.
We do not claim that the proposed method can replace existing methods; rather, we argue that a more refined understanding of the spectral structure can be obtained by using it as a complementary tool alongside existing methods.
\par
We examined whether the proposed method behaves in a manner consistent with the analytical process followed by scientists, namely, focusing on important spectral regions based on the physical-property information.
To quantify which spectral regions were emphasized by the physical-property regression model, we applied Gaussian process regression with an ARD kernel, using the spectral data $\mathcal{Y}$ as the explanatory variable and the physical property $Z$ as the response variable.
The ARD kernel is a Gaussian process framework that introduces an independent length scale for each input dimension and identifies, in an empirical Bayes manner, the dimensions that contribute to the prediction.
When ARD is introduced into an RBF kernel, it takes the form
\begin{equation}
k(\mathbf{x},\mathbf{x}')=\sigma_f^2\exp\!\left(-\frac{1}{2}\sum_{i=1}^{d}\frac{(x_i-x_i')^2}{\ell_i^2}\right).
\end{equation}
A length scale $\ell_i$ is assigned to each dimension $i$.
A smaller $\ell_i$ allows the function to vary more sharply along that dimension. Therefore, the dimension can be interpreted as being more important for predicting the output $Z$.
In this study, the intensity at each wavenumber was treated as an input dimension, and the estimated importance
\begin{equation}
[{\rm importance\:\:of\:\:wave\:\:number}] \propto \ell_i^{-2}
\end{equation}
was visualized as a function of the wavenumber to quantify the spectral regions emphasized by the physical-property regression model.
The kernel used for physical-property regression in the proposed method does not include the length scale $\ell_i$ and therefore differs from the ARD kernel; strictly speaking, this analysis is not a direct evaluation of the proposed method itself, although it can still reveal the overall tendency of the emphasized spectral regions.
The distribution of important spectral structures estimated by ARD focused on specific regions around the assumed peak structures.
This suggests that the effectiveness of the proposed method is consistent with our original expectation.
The overall distribution of important spectral structures estimated by ARD was spread over a wide region with a complex shape, rather than being confined to specific regions around the assumed peak structures (Fig.~\ref{fig:ARD_summary}).
Such a complex distribution of importance would be difficult for humans to use effectively.
The ability to exploit information from such complex, important spectral regions may be regarded as an advantage of the proposed method, which hierarchically integrates the two-stage scientific process of modeling the relationship between the physical properties and spectra, as well as the spectra themselves.

\section{Conclusion}
In this study, we explicitly formulated, as a probabilistic model, the mapping between physical properties (or chemical structure) and spectral shape that scientists implicitly use in spectral deconvolution.
Thus, we developed a framework that enables peak-number estimation under high-intensity noise and unknown background, conditions under which conventional Bayesian spectral deconvolution based only on spectral data often fails.
The proposed method is formulated as a hierarchical Bayesian model that (i) estimates the spectral structure while preserving the uncertainties in the peak number, position, and intensity through Bayesian spectral deconvolution,
and (ii) describes the relationship between the estimated spectrum and physical properties through Gaussian process regression, thereby feeding physical-property information back into the spectral model selection, that is, peak-number estimation.
For strongly multimodal posterior distributions, REMC is used to perform stable marginalization, and the number of peaks is selected based on posterior model probabilities.
\par
In synthetic experiments where the noise intensity was comparable to the peak intensity, the conventional method selected one peak under conditions in which noise was as dominant as the peaks,
whereas the proposed method correctly selected two peaks by exploiting consistency with the physical properties.
In synthetic experiments with an unknown background, the conventional method could select an incorrect model because of pseudo-peaks introduced by preprocessing,
whereas the proposed method correctly identified the true number of peaks.
When applied to real IR spectra of polylactic acid, the method 
extracted weak peaks related to biodegradability in a manner consistent with previous IR studies,
while retaining valid peak-number estimation under an unknown background.
\par
The significance of this framework lies not only in fitting spectral shapes but also in enabling model selection from the viewpoint of spectral structures that can explain the physical property values.
As a result, it can extract and present candidate peaks buried in noise or background in a manner consistent with the goal of exploring the mechanisms underlying the physical property.
However, the results suggest that peaks unrelated to the target physical property may be overlooked when that property is determined uniquely by using only a subset of the peak information.
Future research will extend this framework by integrating multiple physical properties and other measurements, including structural information, to establish a more general foundation for automated spectral analysis and knowledge extraction in materials science.

\subsection*{Acknowledgments}
This work was supported by JSPS KAKENHI grant numbers 22K13979, 23K28150, 24K22309, 24H00247, 25K00986, and 25H01470; JST; and PRESTO grant numbers JPMJPR212A, JST CREST JPMJCR2431, and NEDO JPNP22100843-0.

\section{References}

\bibliography{main}

\clearpage

\appendix
\section{Fitting results of this study}
This section presents the detailed fitting curves for all samples of the artificial data and polylactic acid IR spectral data discussed in Section 3.
Specifically, we compare the spectral model based on the MAP solution, which minimizes the error estimated by the proposed method, with the observed spectral data.
In each figure shown below, the solid blue line represents the observed spectral data, and the dashed orange line represents the fitting curve reconstructed by the proposed method.

\subsection{Fitting results for artificial data}\label{all_arti}
This subsection presents the fitting results for all six artificial spectra generated for different peak-to-peak distances $\Delta$.
The figure shows that regardless of the value of $\Delta$, that is, the degree of peak overlap, the proposed method (orange dashed line) accurately captures the true two-peak structure underlying the observed data (blue solid line).
Even when the noise intensity matches the peak intensity, the method provides appropriate peak-number and peak-shape estimates without overfitting false peaks.
\clearpage
\begin{figure}[]
\centering
\foreach \i in {0,1,2}{
  \begin{subfigure}{0.8\linewidth}
    \centering
    \includegraphics[width=\linewidth]{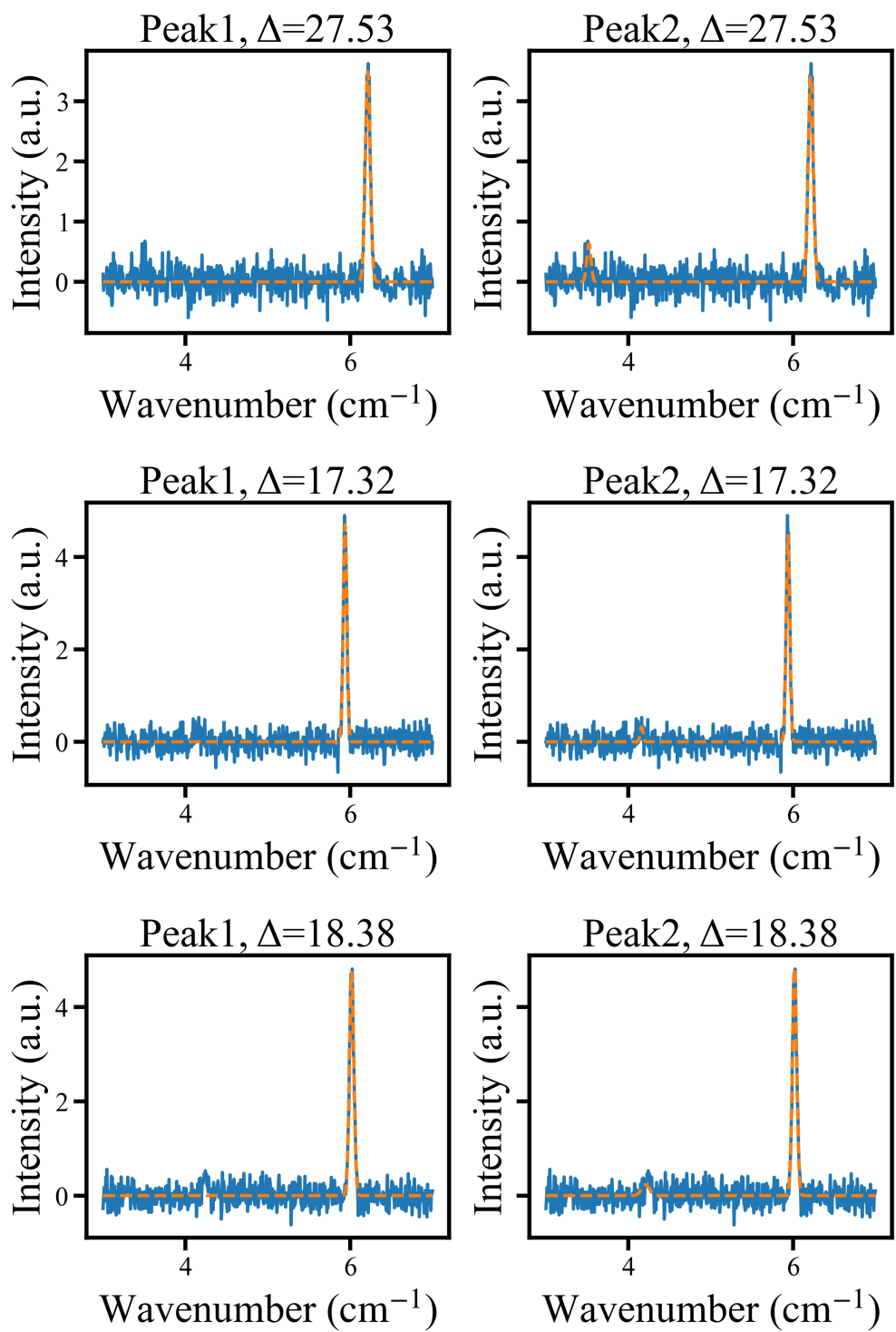}
  \end{subfigure}\par\medskip}
  
\caption{
Fitting results for artificial data 1 to 3. The solid blue line indicates the observed spectral data. The dashed orange line shows the reconstructed fitting curve obtained using the proposed method.}
\end{figure}

\begin{figure}[tbp]
\centering
\foreach \i in {3,4,5}{
  \begin{subfigure}{0.8\linewidth}
    \centering
    \includegraphics[width=\linewidth]{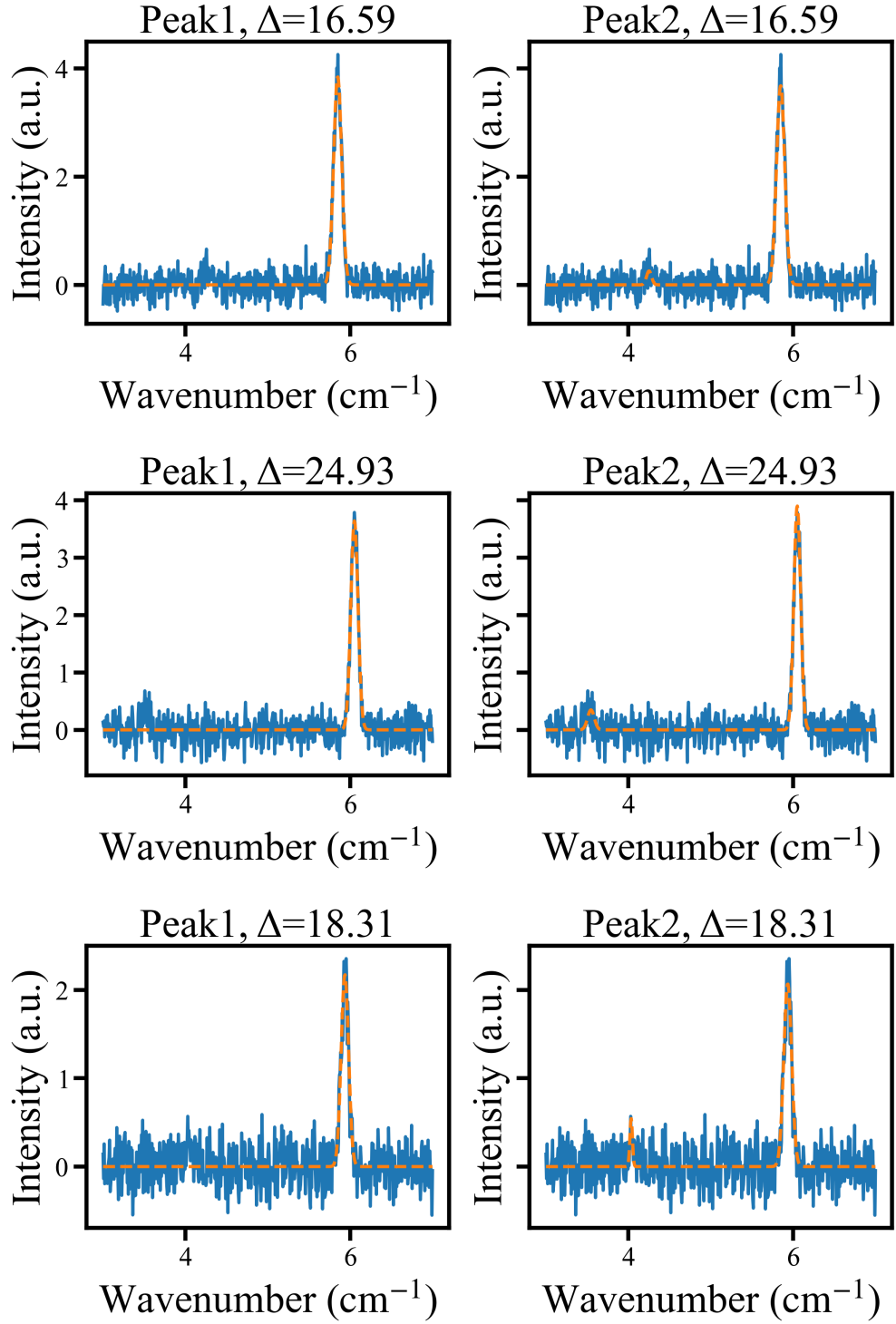}
  \end{subfigure}\par\medskip
}
\caption{
Fitting results for artificial data 4 to 6. The solid blue line indicates the observed spectral data. The dashed orange line shows the reconstructed fitting curve obtained using the proposed method.}
\end{figure}

\clearpage
\subsection{IR Spectrum Fitting Results}\label{all_IR}
\begin{figure}[H]
    \centering
    \begin{subfigure}[b]{0.45\linewidth}
        \centering
        \includegraphics[width=\linewidth]{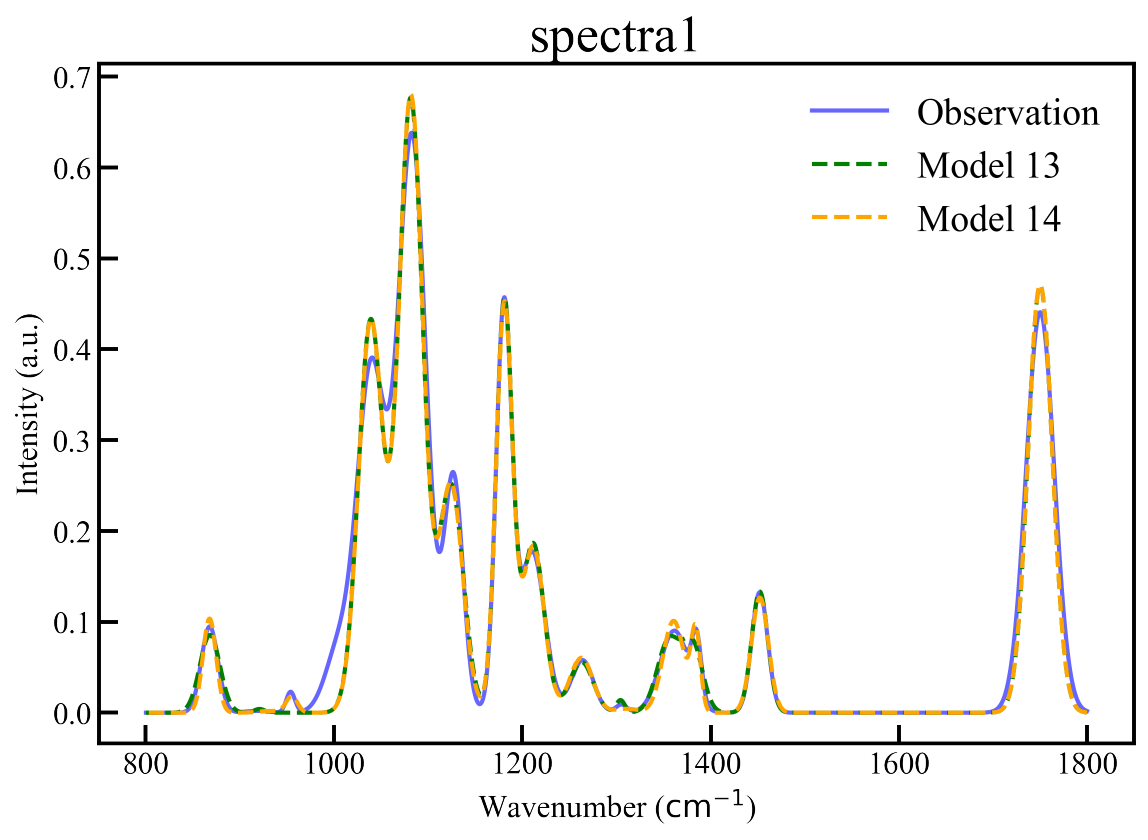}
        \caption{Spectrum 1}
    \end{subfigure}\hfill
    \begin{subfigure}[b]{0.45\linewidth}
        \centering
        \includegraphics[width=\linewidth]{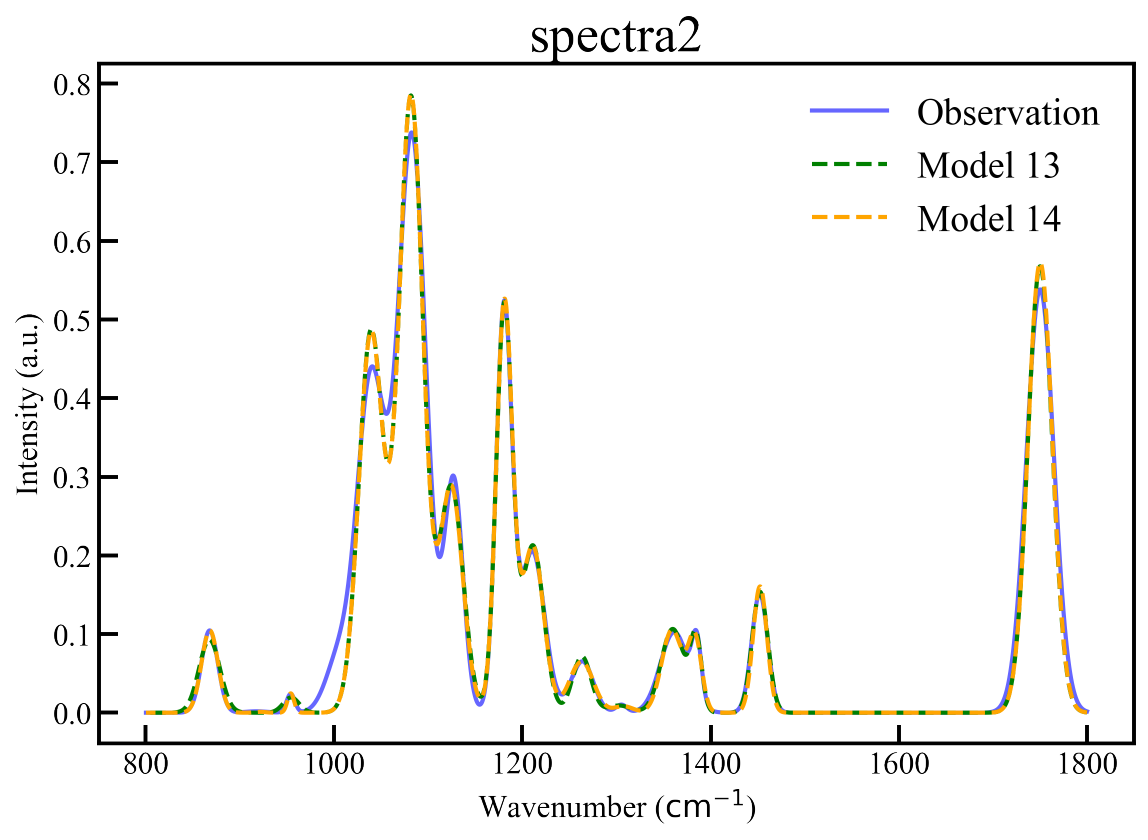}
        \caption{Spectrum 2}
    \end{subfigure}

    \vspace{2em}
    \begin{subfigure}[b]{0.45\linewidth}
        \centering
        \includegraphics[width=\linewidth]{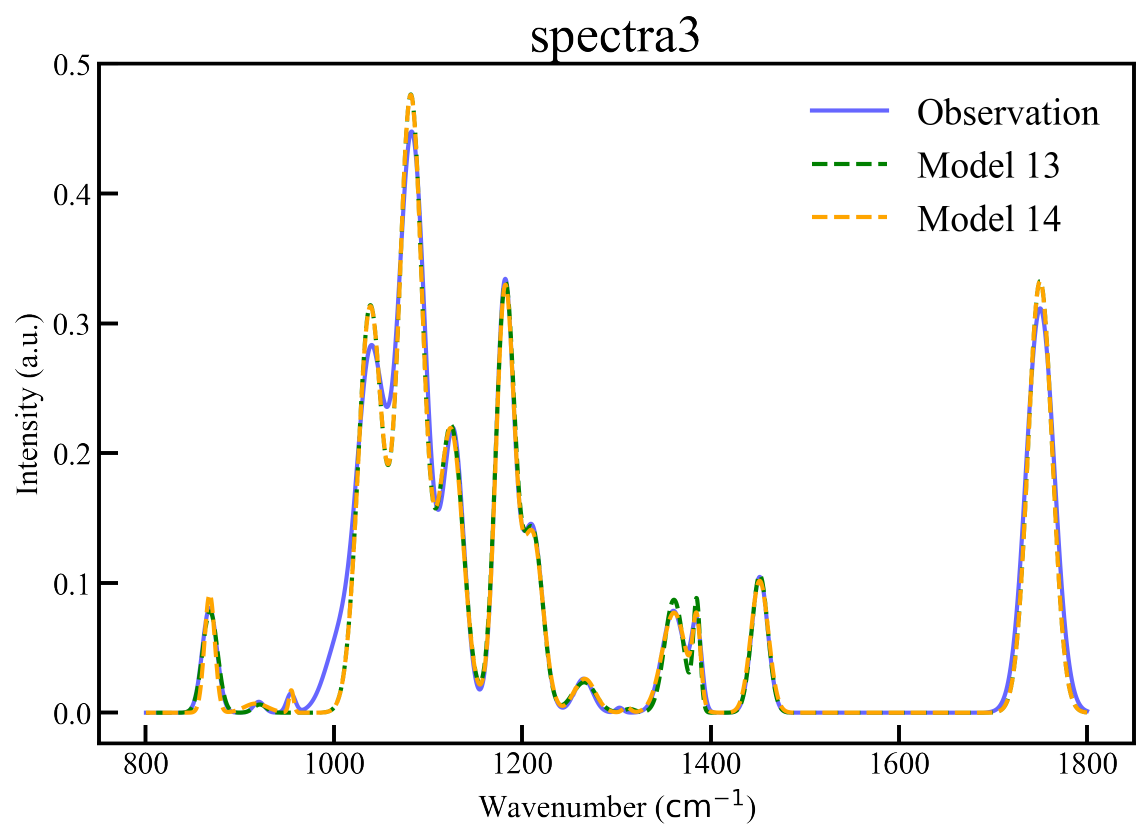}
        \caption{Spectrum 3}
    \end{subfigure}\hfill
    \begin{subfigure}[b]{0.45\linewidth}
        \centering
        \includegraphics[width=\linewidth]{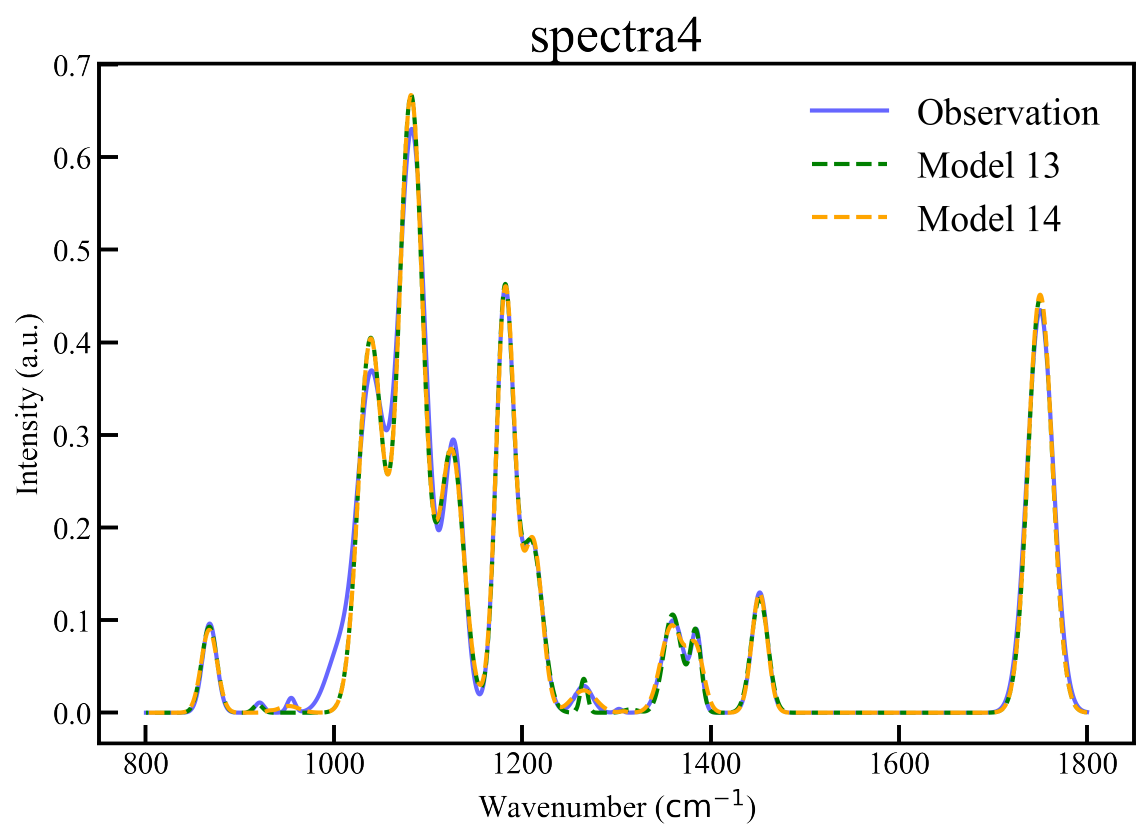}
        \caption{Spectrum 4}
    \end{subfigure}

    \vspace{2em}
    \begin{subfigure}[b]{0.45\linewidth}
        \centering
        \includegraphics[width=\linewidth]{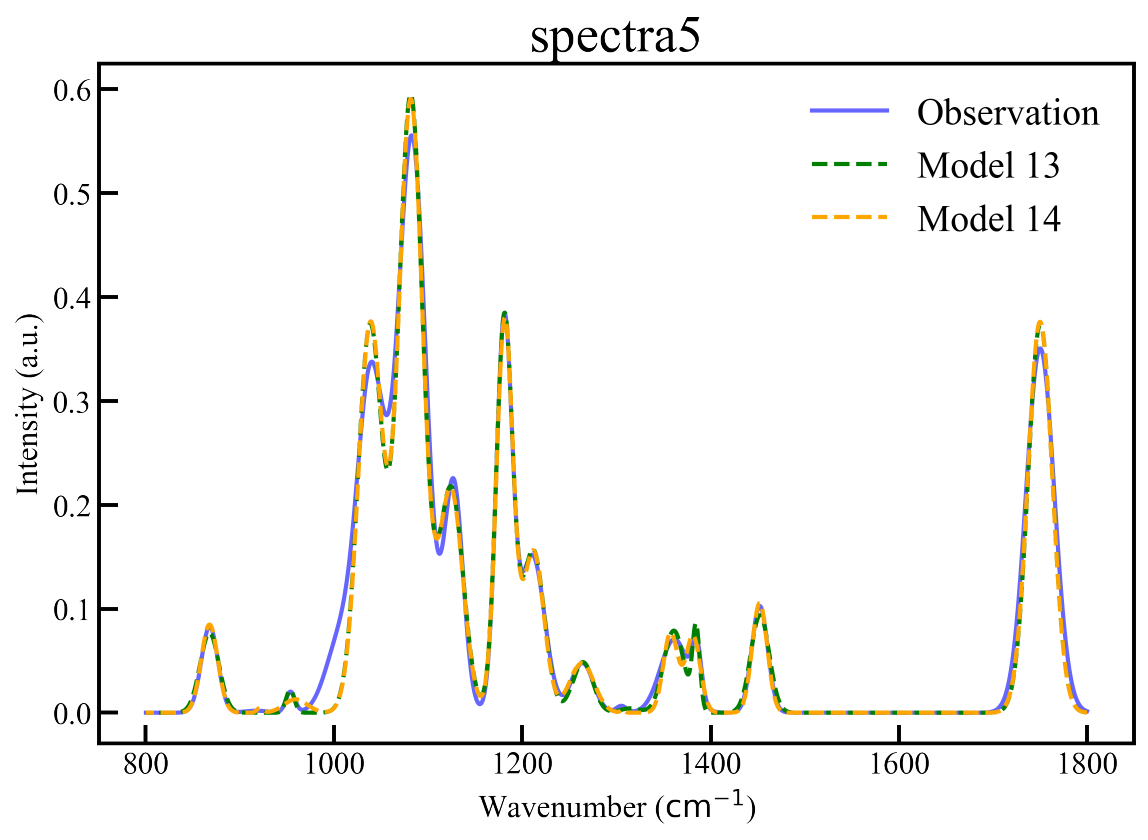}
        \caption{Spectrum 5}
    \end{subfigure}\hfill
    \begin{subfigure}[b]{0.45\linewidth}
        \centering
        \includegraphics[width=\linewidth]{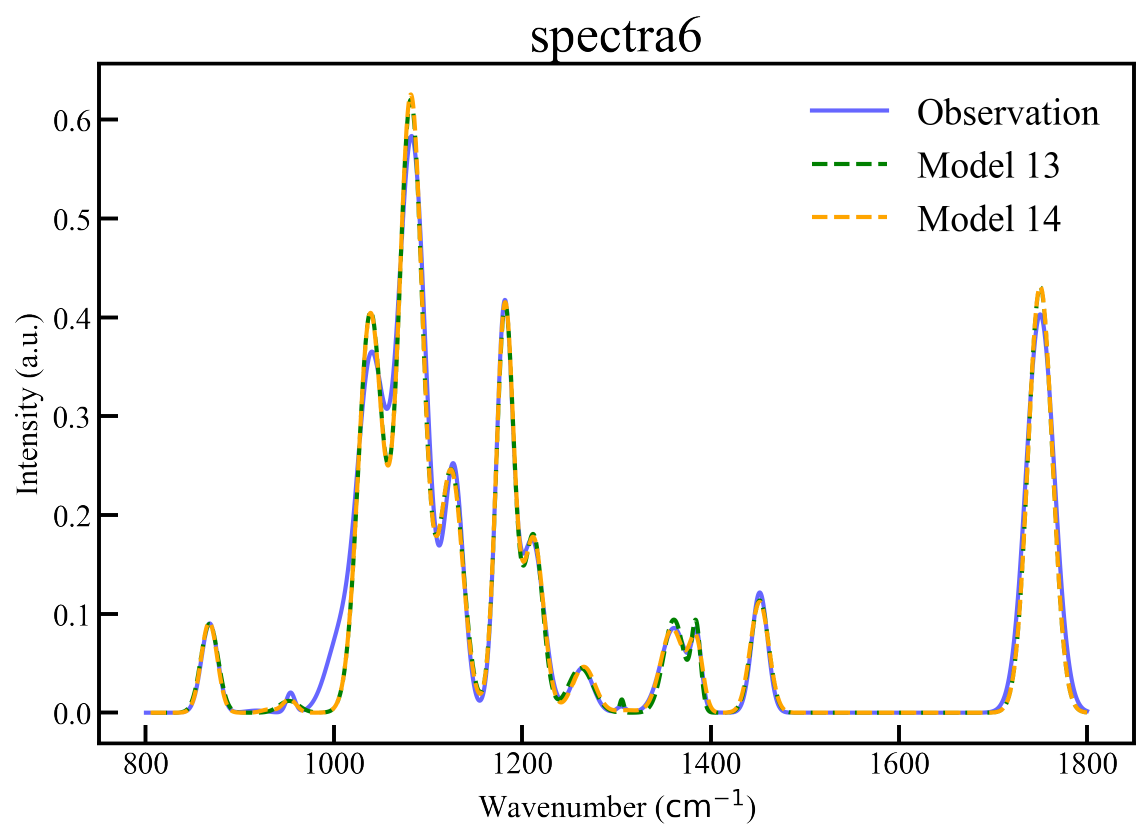}
        \caption{Spectrum 6}
    \end{subfigure}

    \caption{IR fitting results (Spectra 1--6). The blue solid line represents the observed spectral data, and the orange (green) dashed line represents the fitting curve for the 14-peak (13-peak) model.}
    \label{IR_first}
\end{figure}

\begin{figure}[p]
    \centering
    \begin{subfigure}[b]{0.45\linewidth} \centering \includegraphics[width=\linewidth]{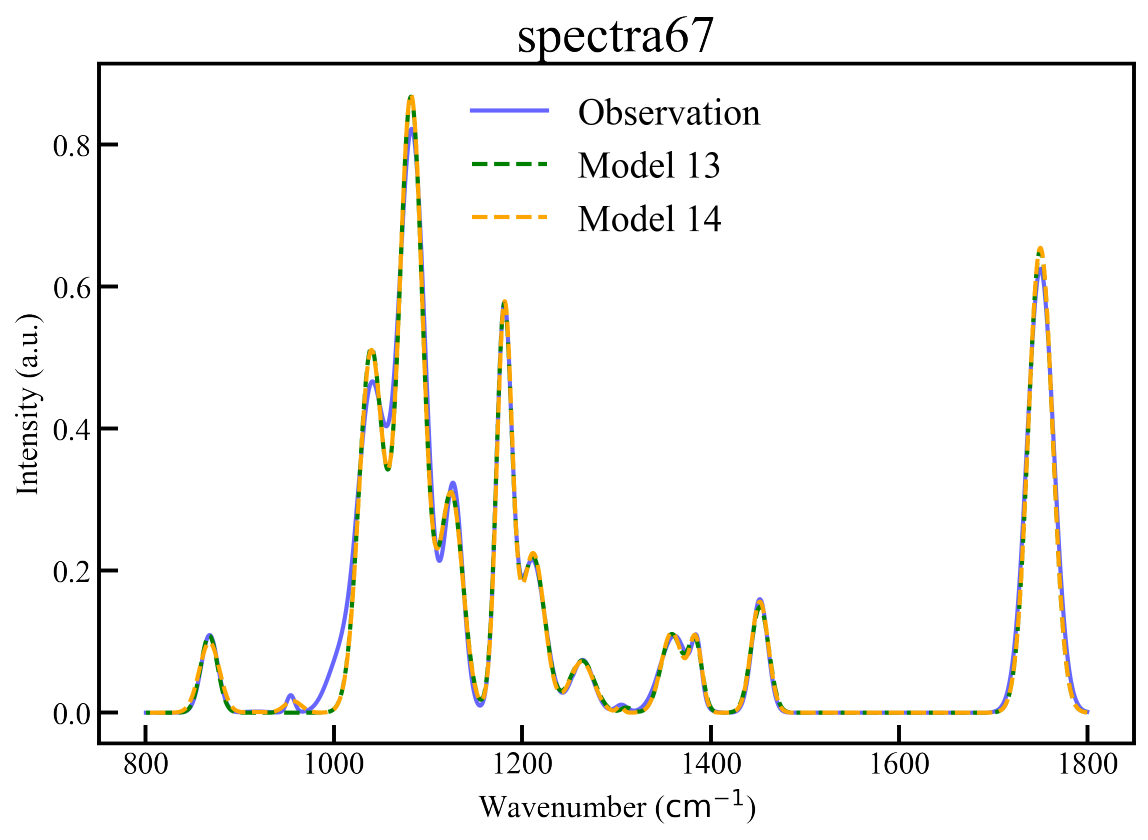} \caption{Spectrum 67} \end{subfigure}\hfill
    \begin{subfigure}[b]{0.45\linewidth} \centering \includegraphics[width=\linewidth]{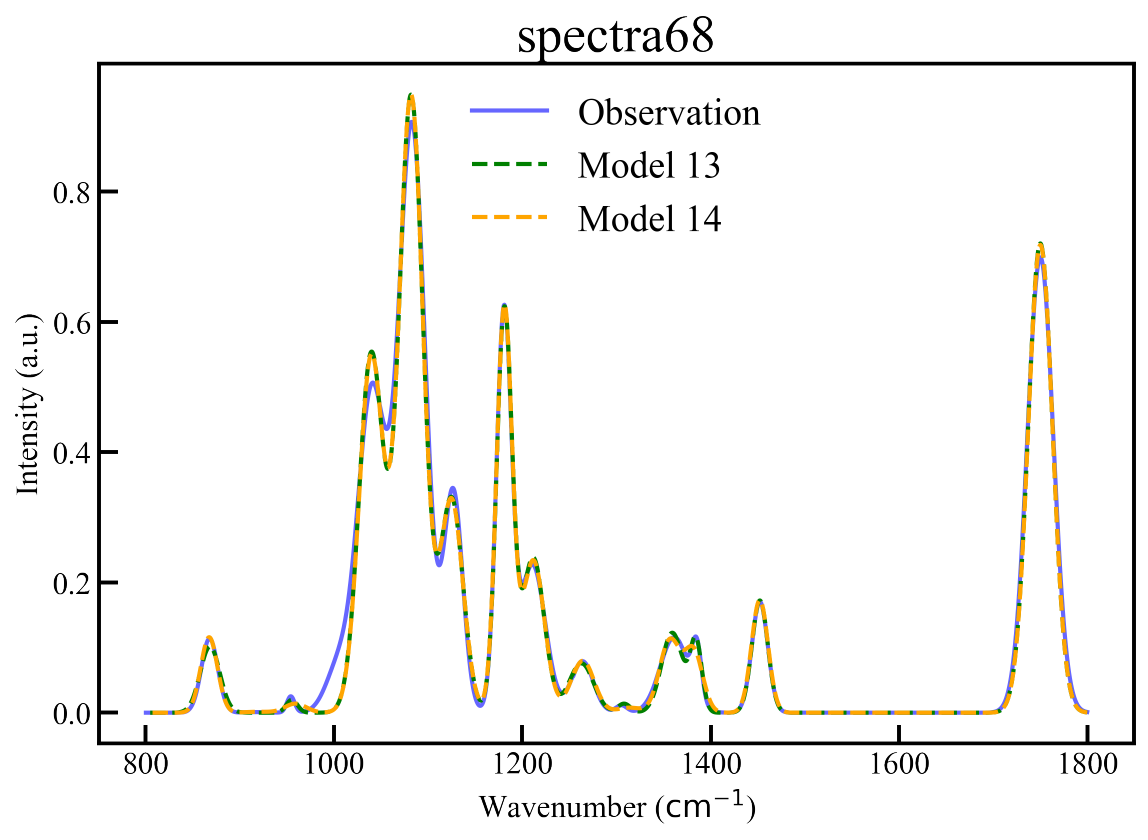} \caption{Spectrum 68} \end{subfigure}
    \caption{IR fitting results (Spectra 67--68). The blue solid line represents the observed spectral data, and the orange (green) dashed line represents the fitting curve for the 14-peak (13-peak) model.}
    \label{IR_last}
\end{figure}

This subsection lists the fitting results for each model applied to the IR spectra of polylactic acid obtained under 68 conditions with different crystallization temperatures and concentrations (Figs.~\ref{IR_first}--\ref{IR_last}).
These results indicate that the proposed method stably extracts two weak peaks in the biodegradability-related region.
\clearpage

\clearpage
\begin{figure}[p]
    \centering
    \begin{subfigure}[b]{0.45\linewidth}
        \centering
        \includegraphics[width=\linewidth]{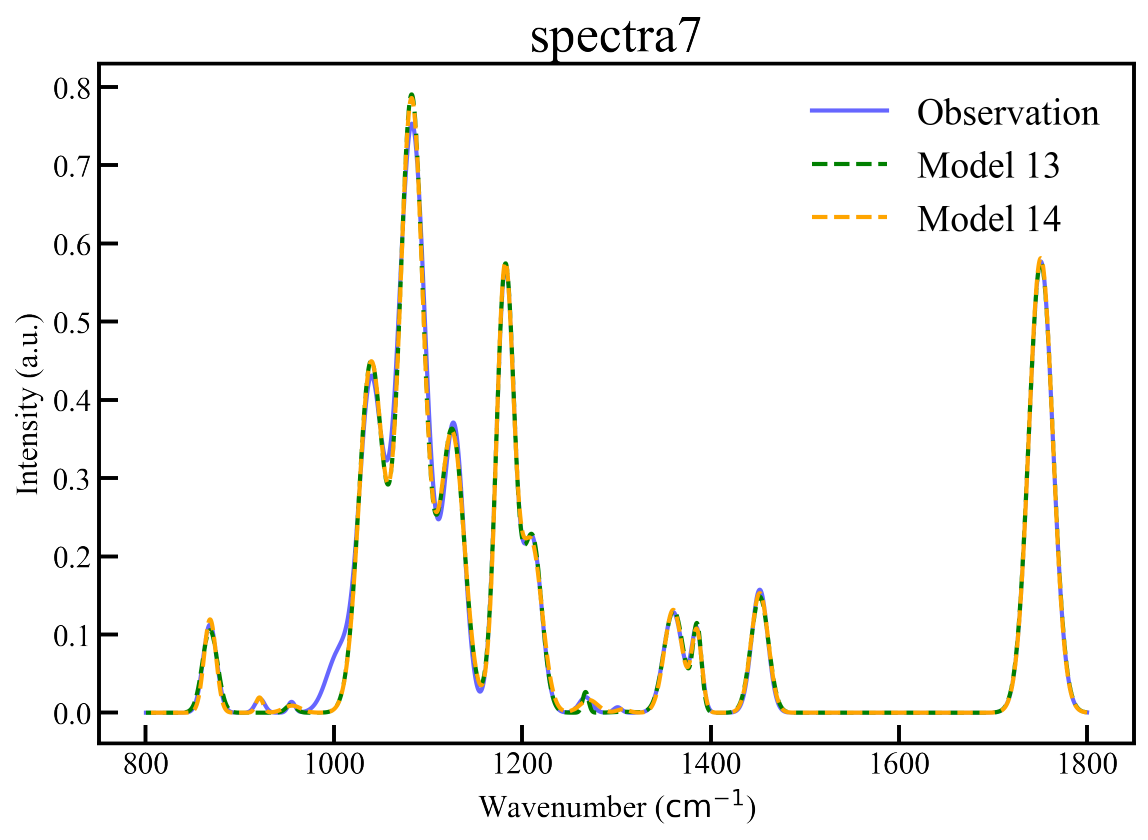}
        \caption{Spectrum 7}
    \end{subfigure}\hfill
    \begin{subfigure}[b]{0.45\linewidth}
        \centering
        \includegraphics[width=\linewidth]{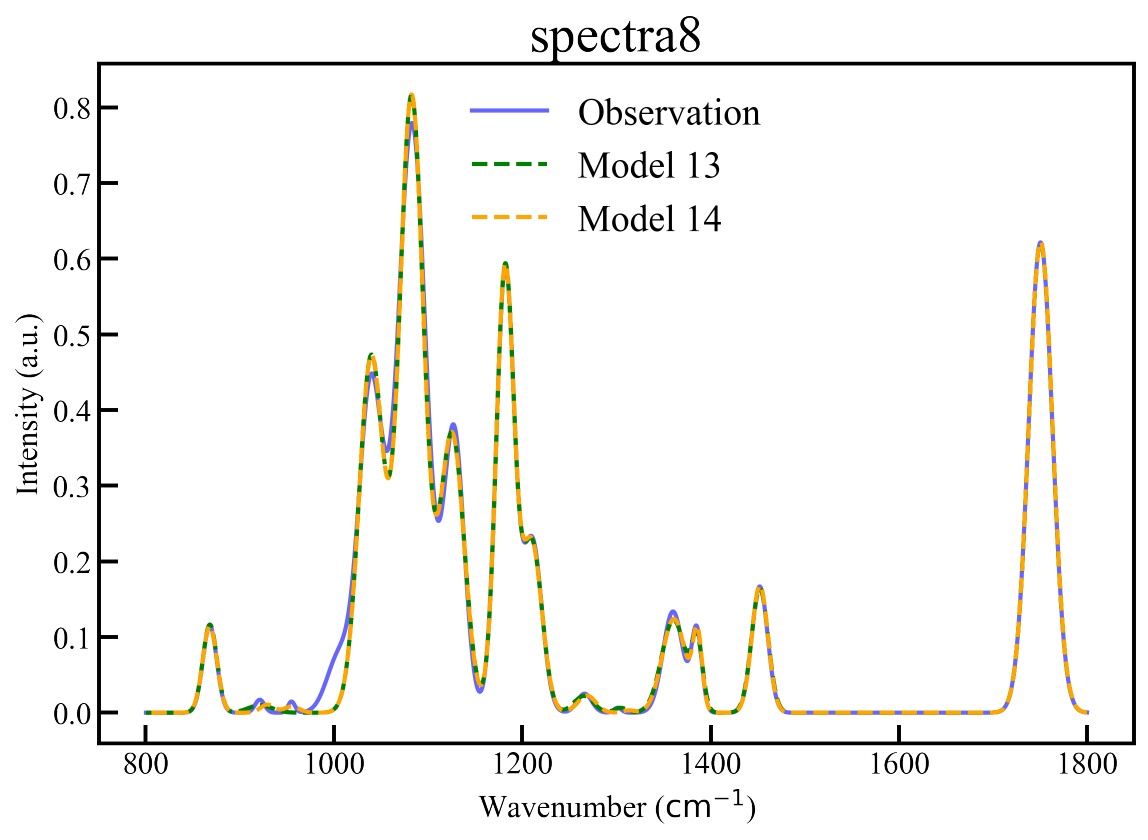}
        \caption{Spectrum 8}
    \end{subfigure}
    
    \vspace{2em}
    \begin{subfigure}[b]{0.45\linewidth}
        \centering
        \includegraphics[width=\linewidth]{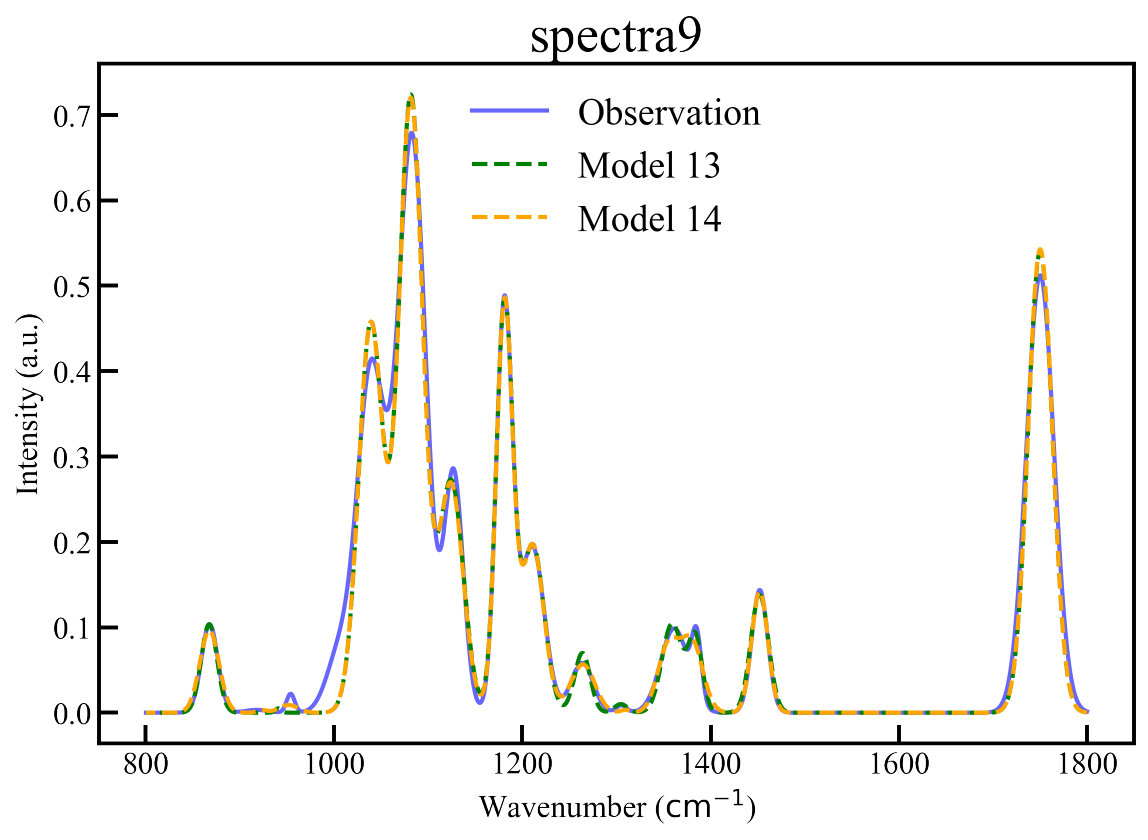}
        \caption{Spectrum 9}
    \end{subfigure}\hfill
    \begin{subfigure}[b]{0.45\linewidth}
        \centering
        \includegraphics[width=\linewidth]{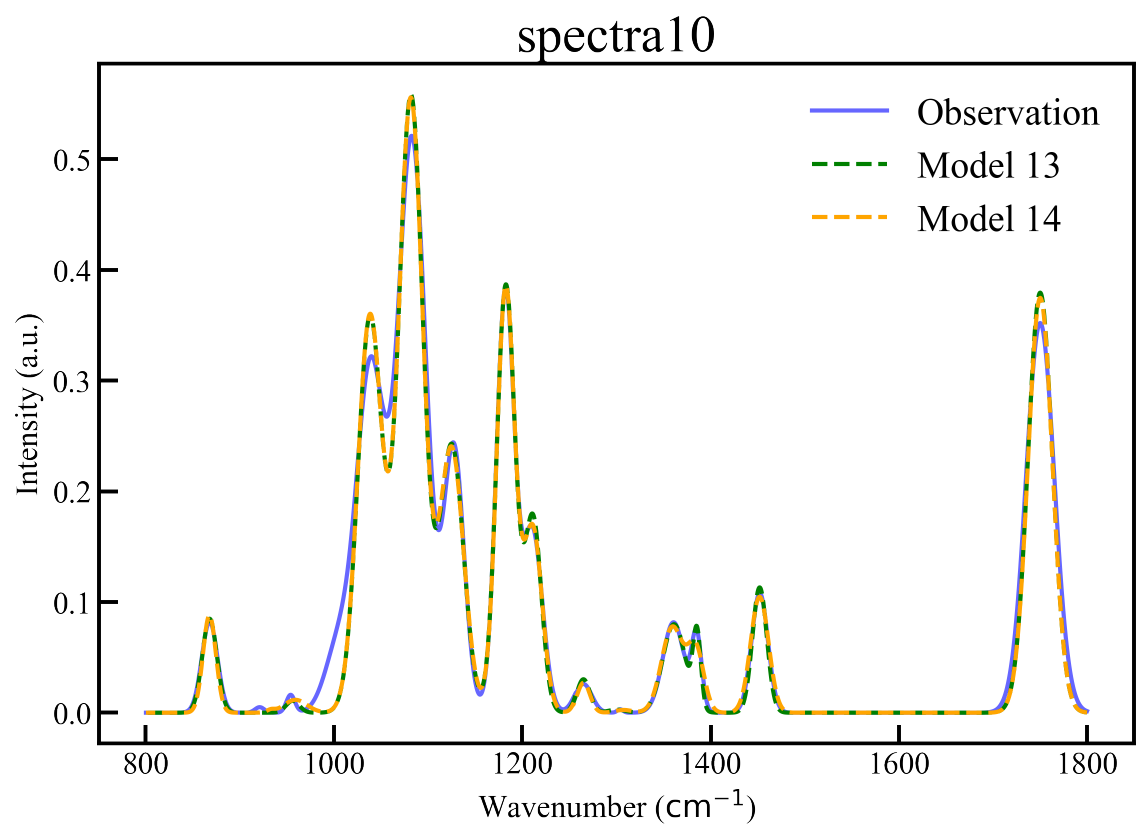}
        \caption{Spectrum 10}
    \end{subfigure}

    \vspace{2em}
    \begin{subfigure}[b]{0.45\linewidth}
        \centering
        \includegraphics[width=\linewidth]{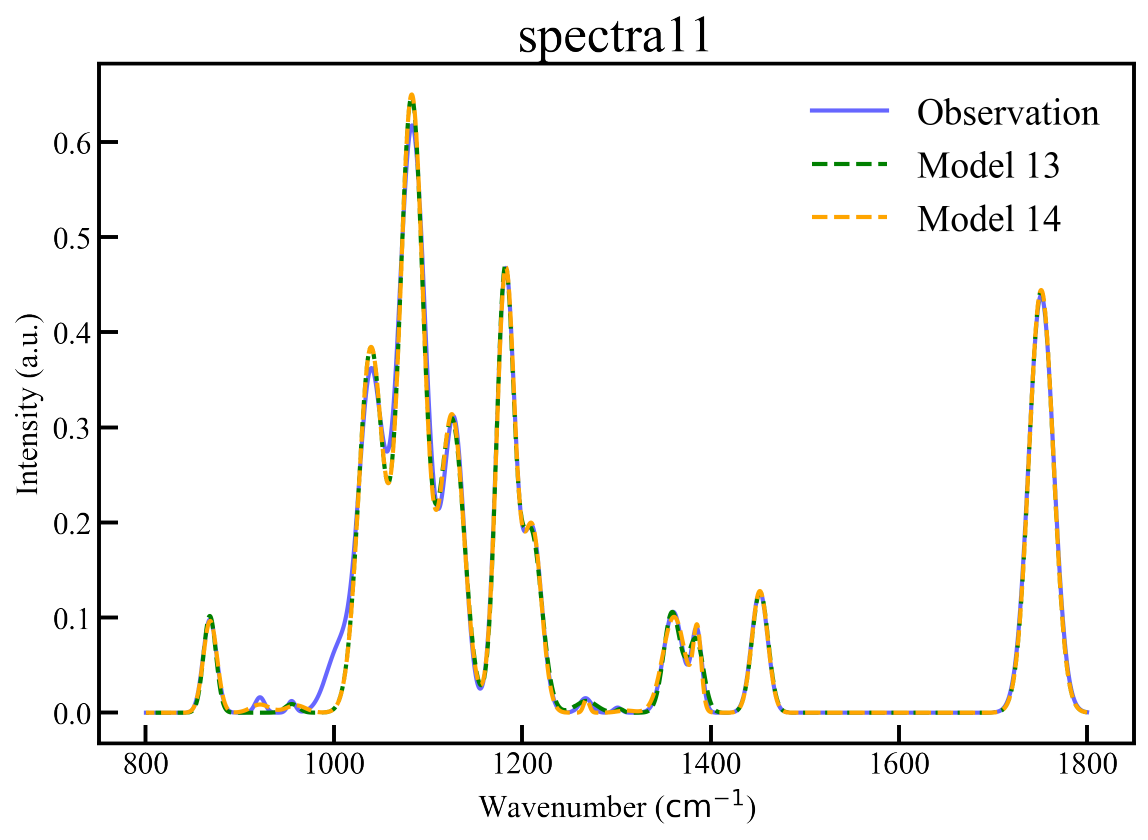}
        \caption{Spectrum 11}
    \end{subfigure}\hfill
    \begin{subfigure}[b]{0.45\linewidth}
        \centering
        \includegraphics[width=\linewidth]{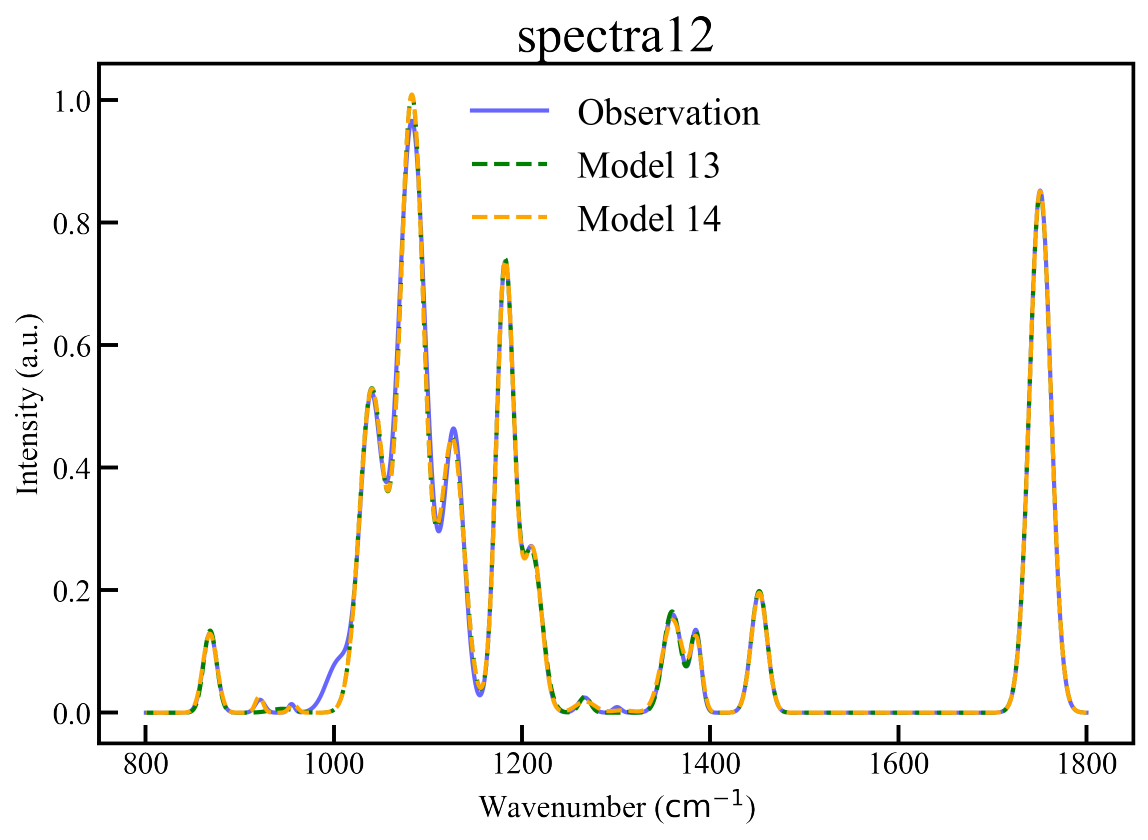}
        \caption{Spectrum 12}
    \end{subfigure}

    \caption{IR fitting results (Spectra 7--12). The blue solid line represents the observed spectral data, and the orange (green) dashed line represents the fitting curve for the 14-peak (13-peak) model.}
\end{figure}

\clearpage
\begin{figure}[p]
    \centering
    \begin{subfigure}[b]{0.45\linewidth} \centering \includegraphics[width=\linewidth]{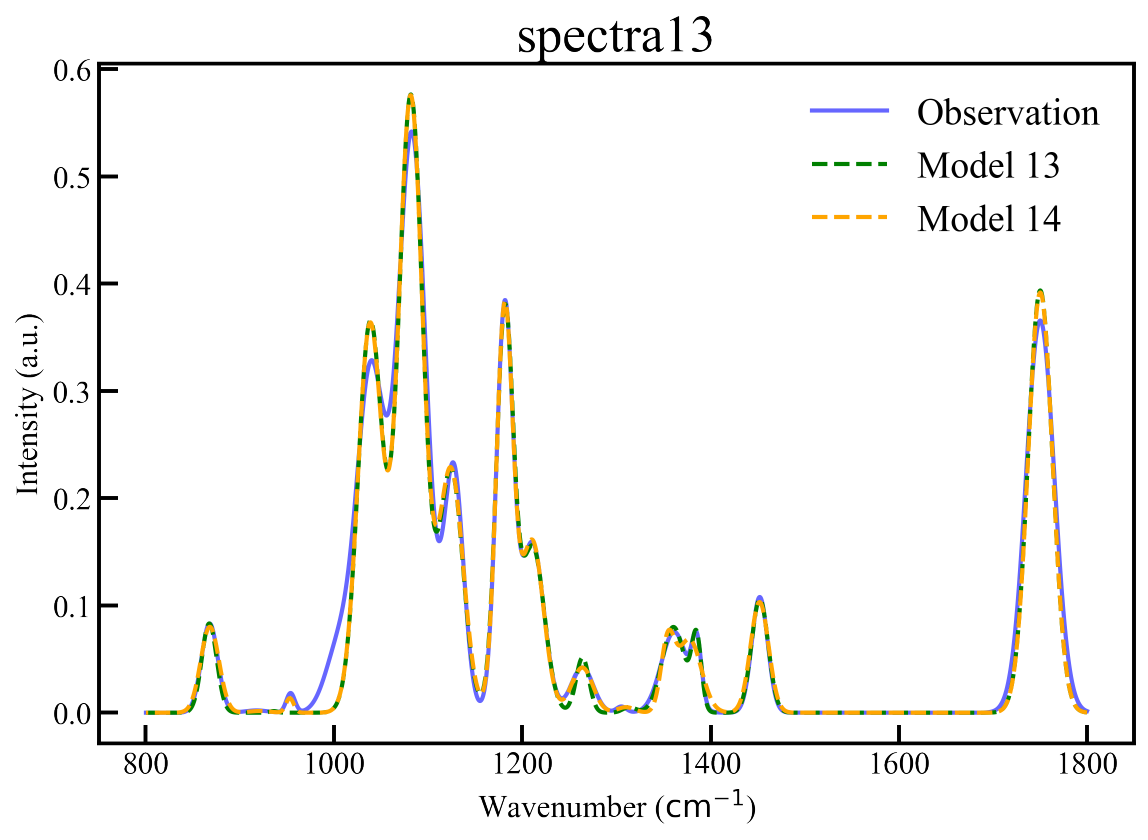} \caption{Spectrum 13} \end{subfigure}\hfill
    \begin{subfigure}[b]{0.45\linewidth} \centering \includegraphics[width=\linewidth]{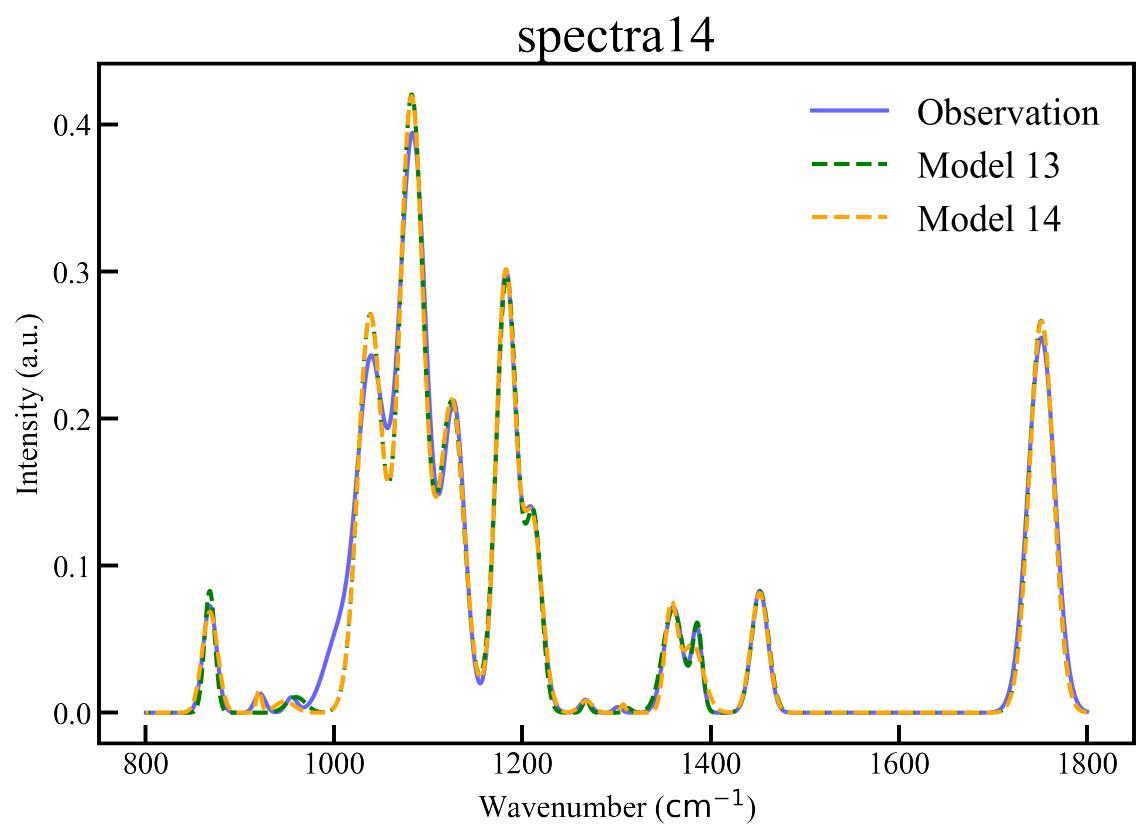} \caption{Spectrum 14} \end{subfigure}
    \vspace{2em}
    \begin{subfigure}[b]{0.45\linewidth} \centering \includegraphics[width=\linewidth]{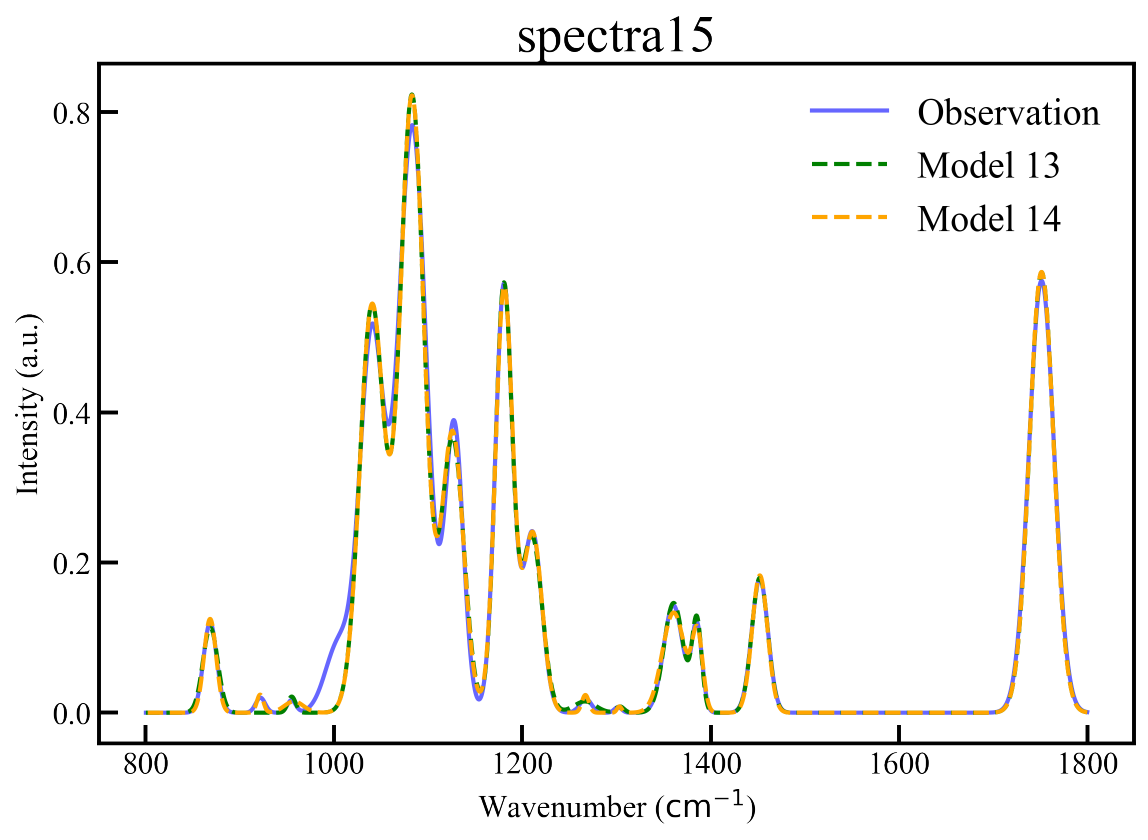} \caption{Spectrum 15} \end{subfigure}\hfill
    \begin{subfigure}[b]{0.45\linewidth} \centering \includegraphics[width=\linewidth]{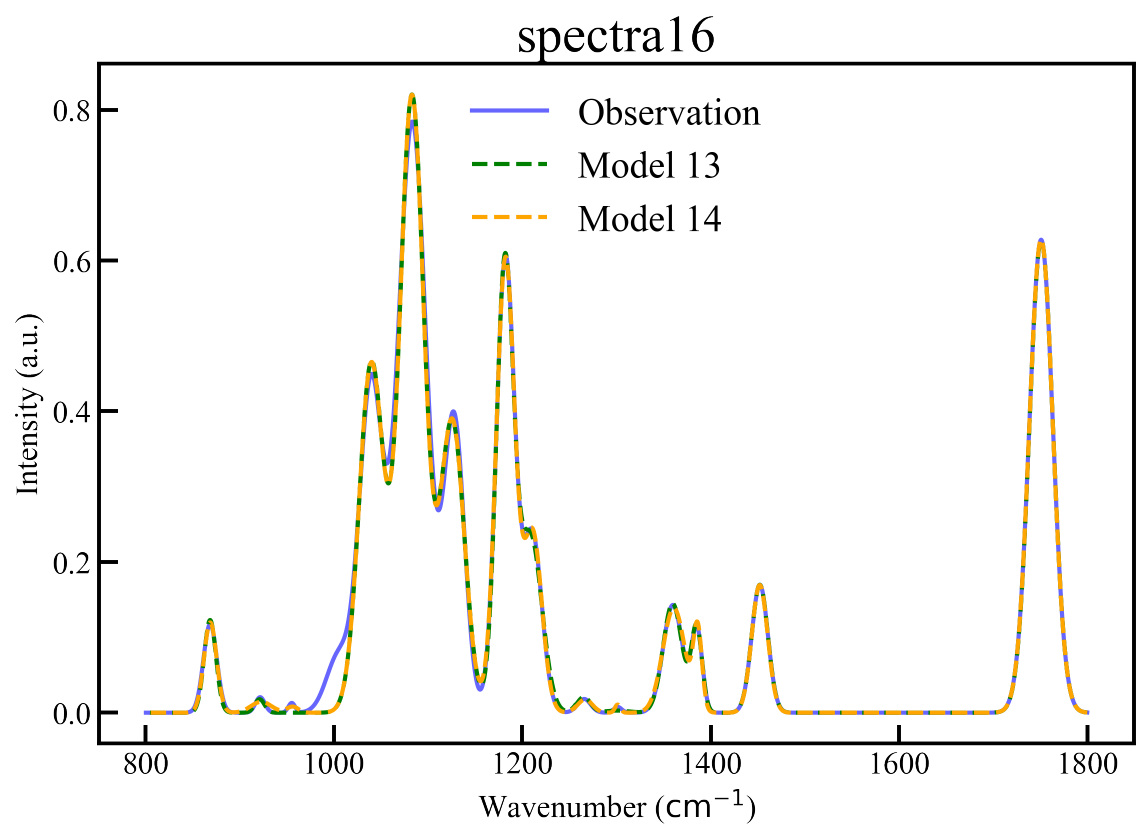} \caption{Spectrum 16} \end{subfigure}
    \vspace{2em}
    \begin{subfigure}[b]{0.45\linewidth} \centering \includegraphics[width=\linewidth]{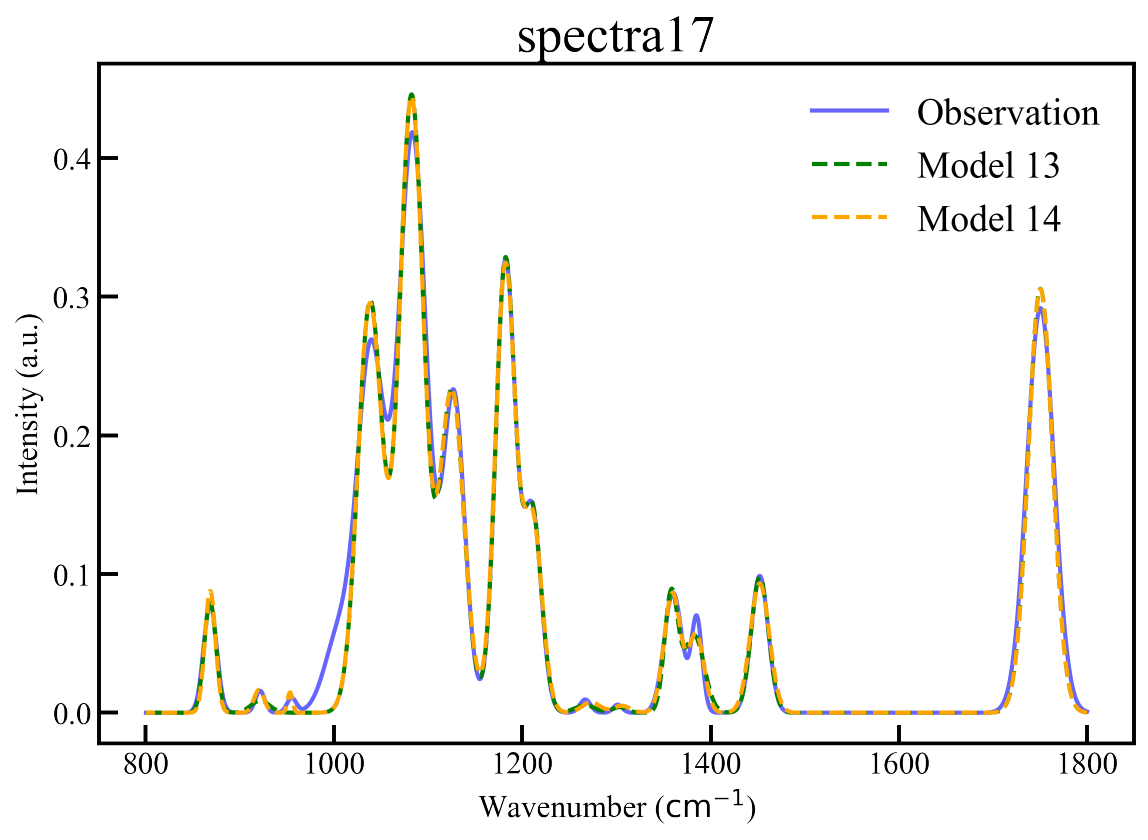} \caption{Spectrum 17} \end{subfigure}\hfill
    \begin{subfigure}[b]{0.45\linewidth} \centering \includegraphics[width=\linewidth]{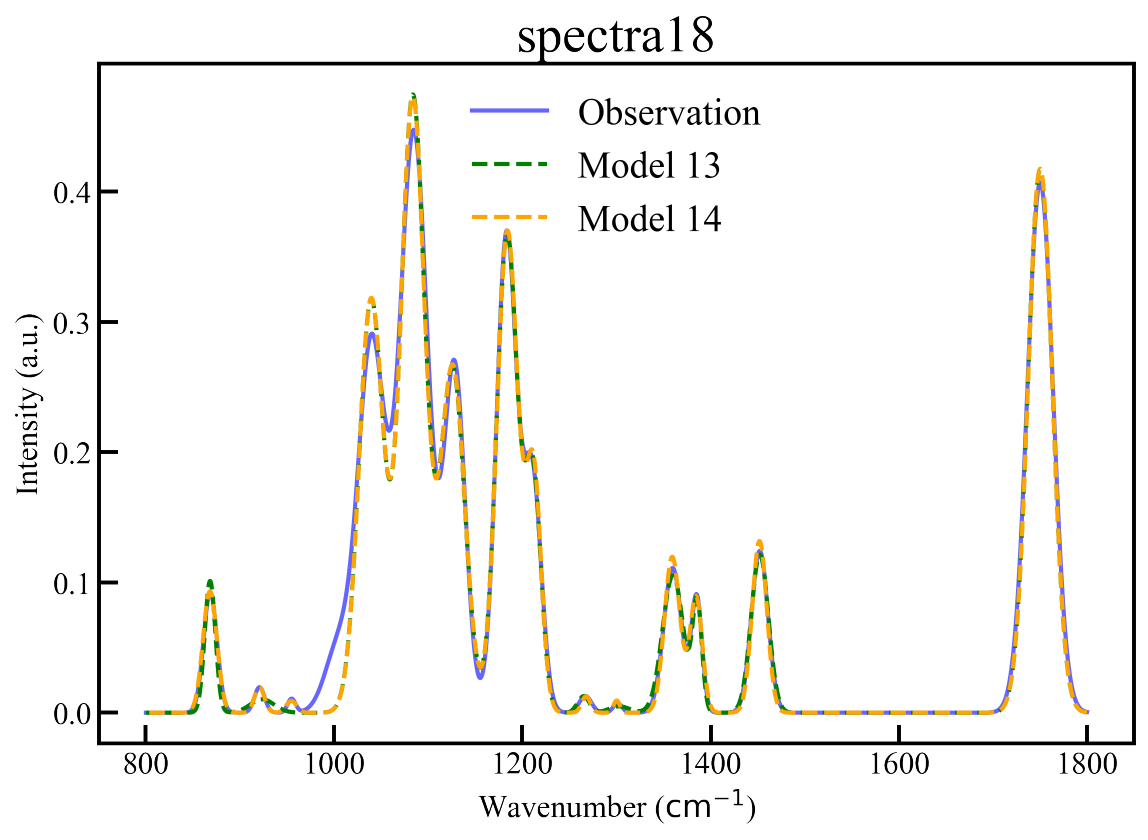} \caption{Spectrum 18} \end{subfigure}
    \caption{IR fitting results (Spectra 13--18). The blue solid line represents the observed spectral data, and the orange (green) dashed line represents the fitting curve for the 14-peak (13-peak) model.}
\end{figure}

\clearpage
\begin{figure}[p]
    \centering
    \begin{subfigure}[b]{0.45\linewidth} \centering \includegraphics[width=\linewidth]{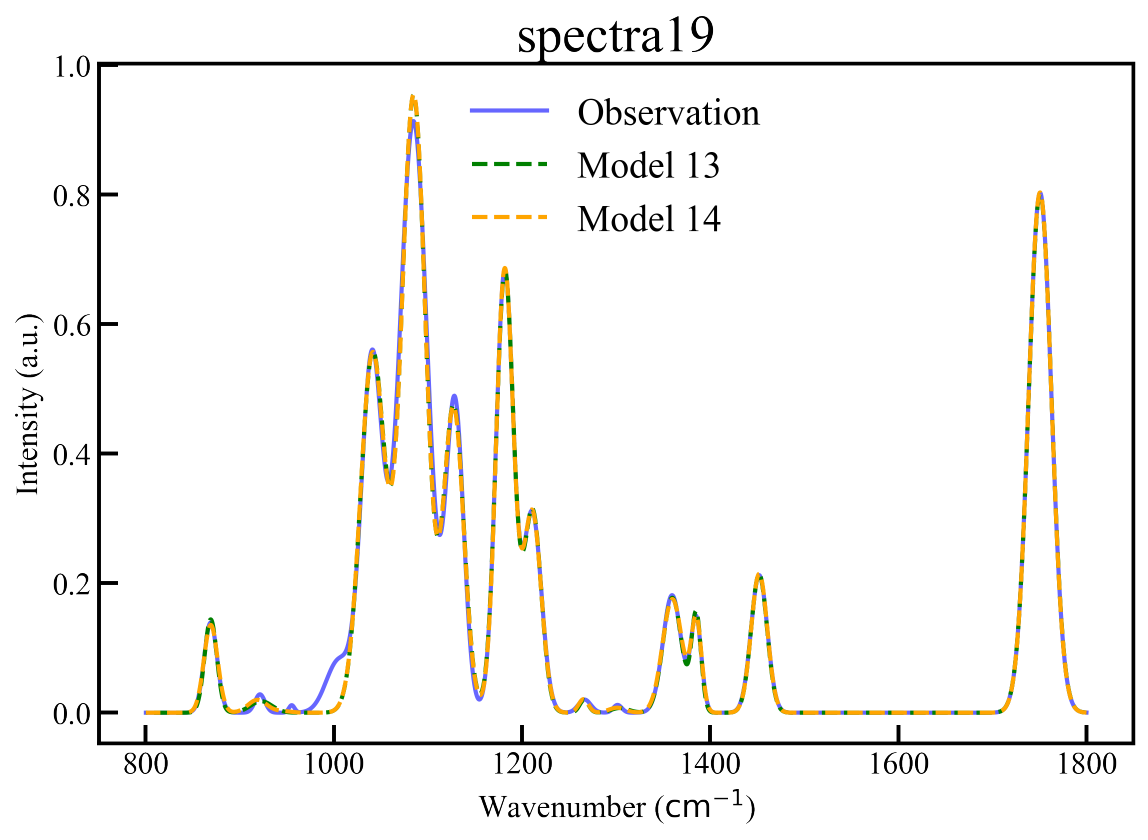} \caption{Spectrum 19} \end{subfigure}\hfill
    \begin{subfigure}[b]{0.45\linewidth} \centering \includegraphics[width=\linewidth]{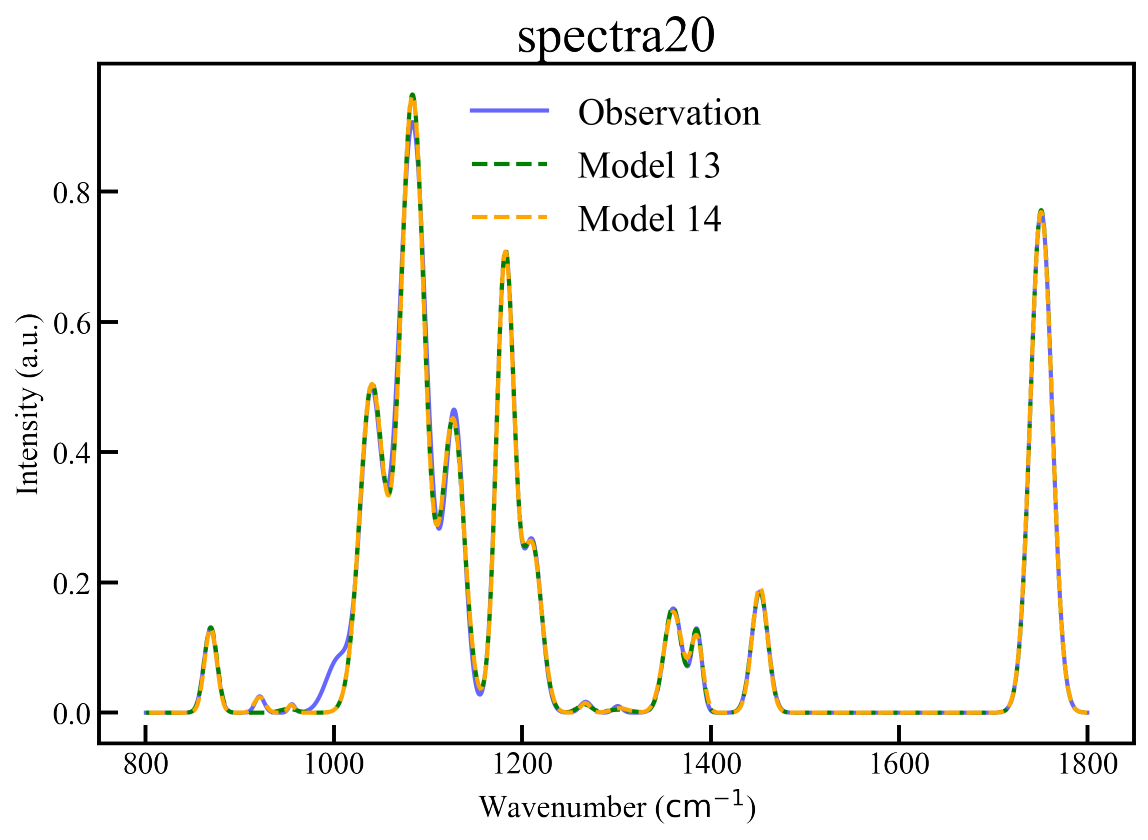} \caption{Spectrum 20} \end{subfigure}
    \vspace{2em}
    \begin{subfigure}[b]{0.45\linewidth} \centering \includegraphics[width=\linewidth]{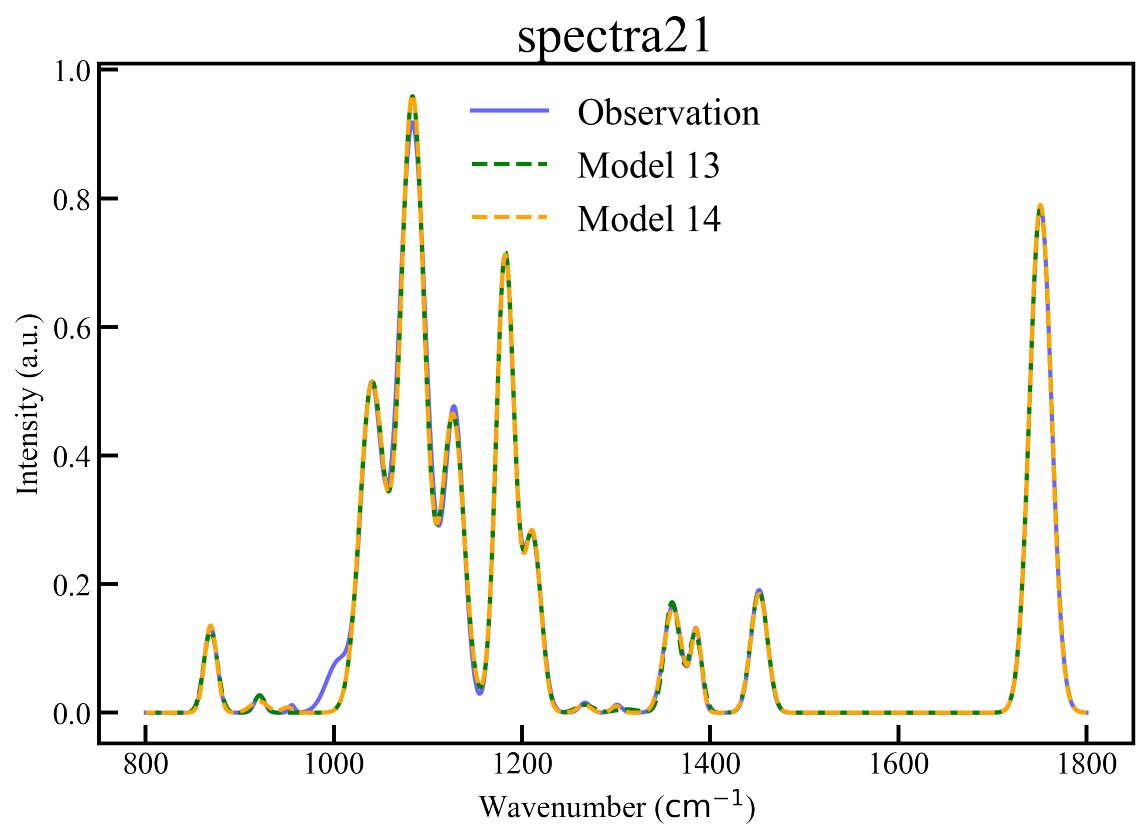} \caption{Spectrum 21} \end{subfigure}\hfill
    \begin{subfigure}[b]{0.45\linewidth} \centering \includegraphics[width=\linewidth]{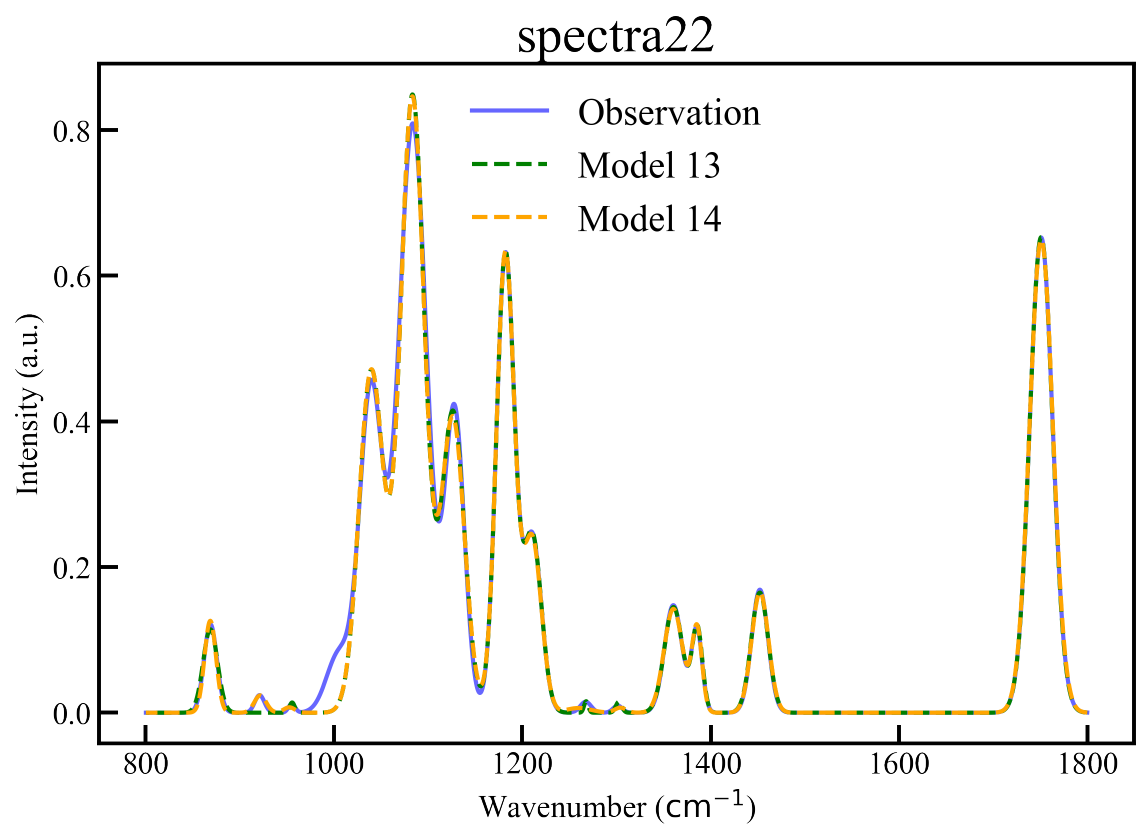} \caption{Spectrum 22} \end{subfigure}
    \vspace{2em}
    \begin{subfigure}[b]{0.45\linewidth} \centering \includegraphics[width=\linewidth]{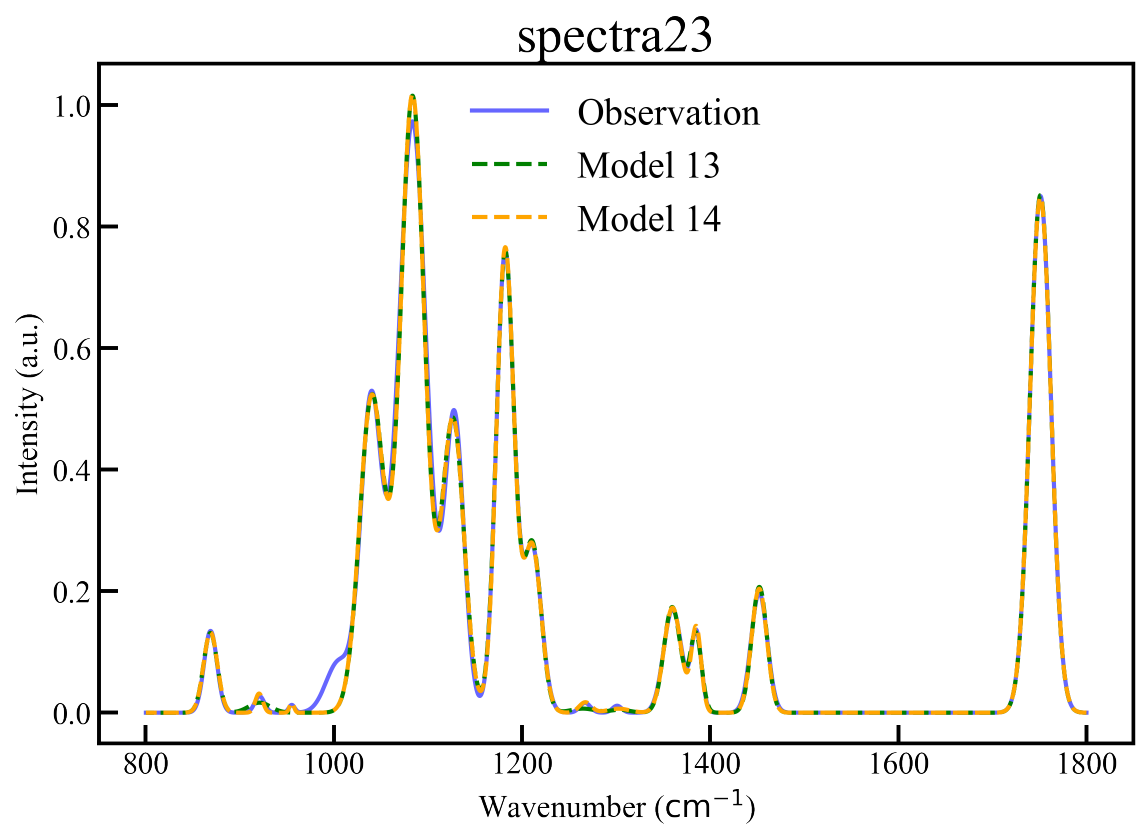} \caption{Spectrum 23} \end{subfigure}\hfill
    \begin{subfigure}[b]{0.45\linewidth} \centering \includegraphics[width=\linewidth]{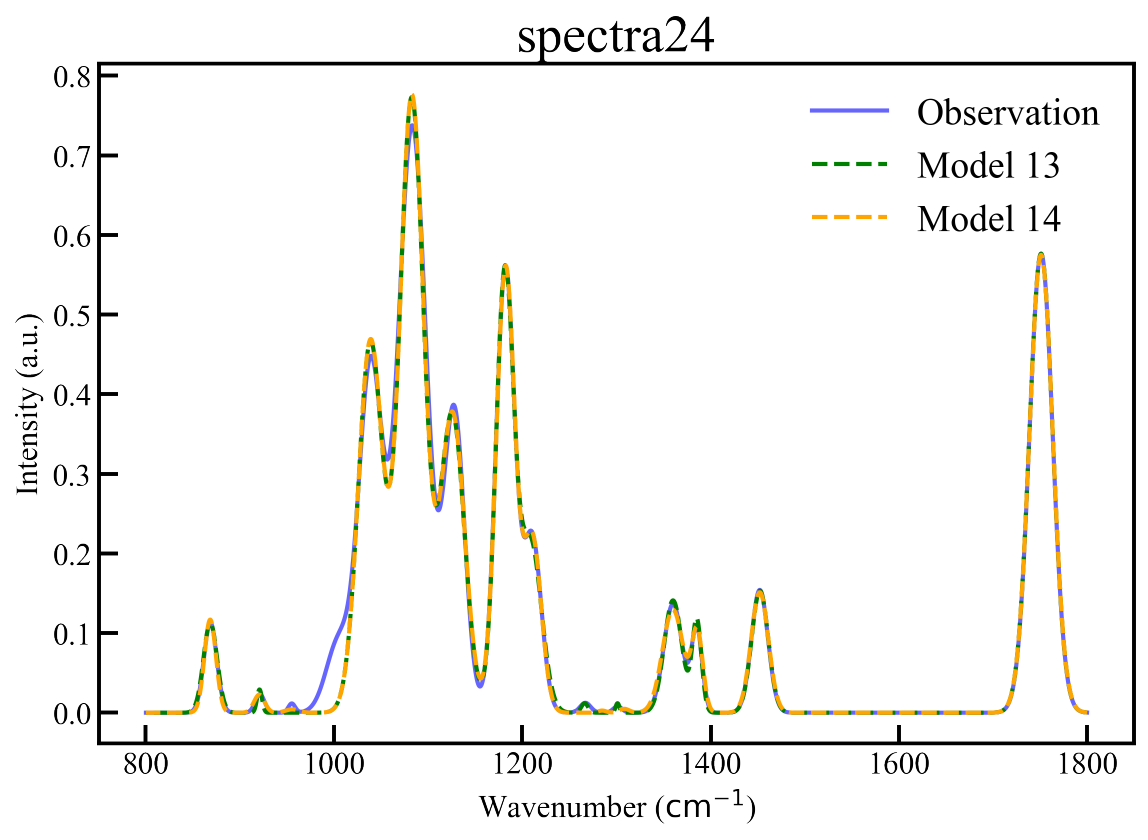} \caption{Spectrum 24} \end{subfigure}
    \caption{IR fitting results (Spectra 19--24). The blue solid line represents the observed spectral data, and the orange (green) dashed line represents the fitting curve for the 14-peak (13-peak) model.}
\end{figure}

\clearpage
\begin{figure}[p]
    \centering
    \begin{subfigure}[b]{0.45\linewidth} \centering \includegraphics[width=\linewidth]{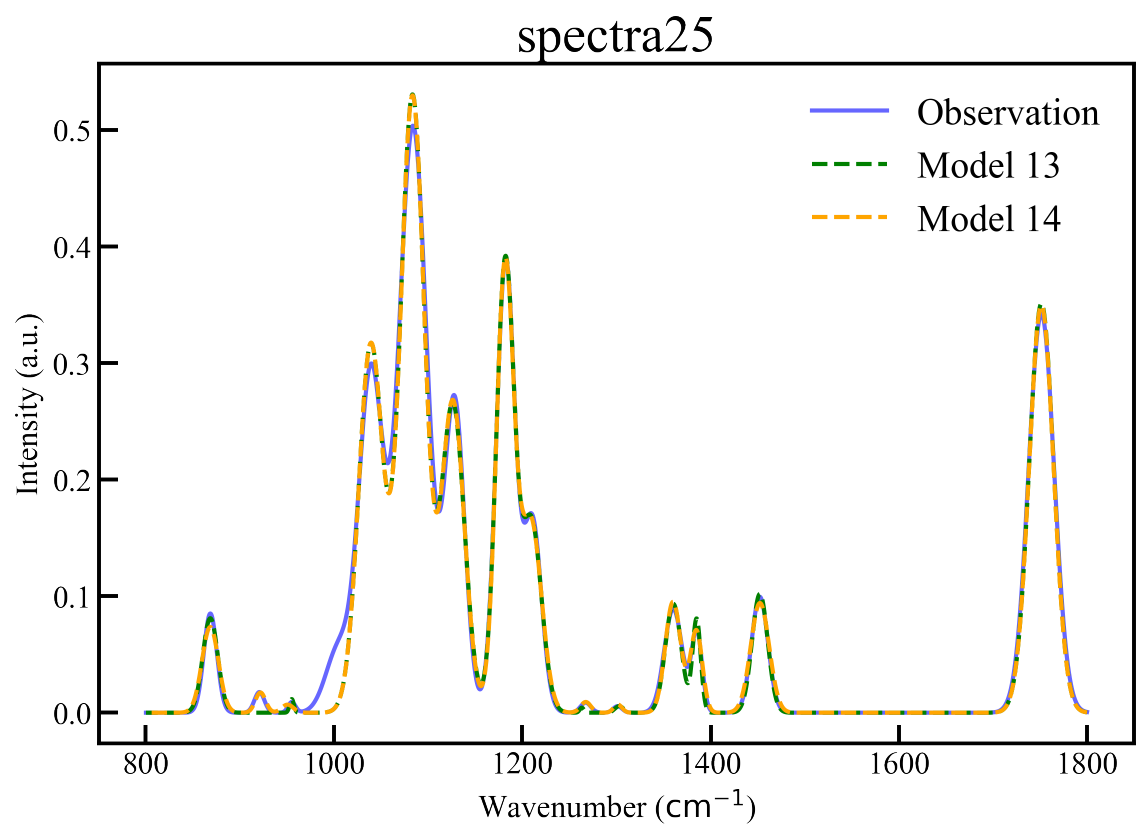} \caption{Spectrum 25} \end{subfigure}\hfill
    \begin{subfigure}[b]{0.45\linewidth} \centering \includegraphics[width=\linewidth]{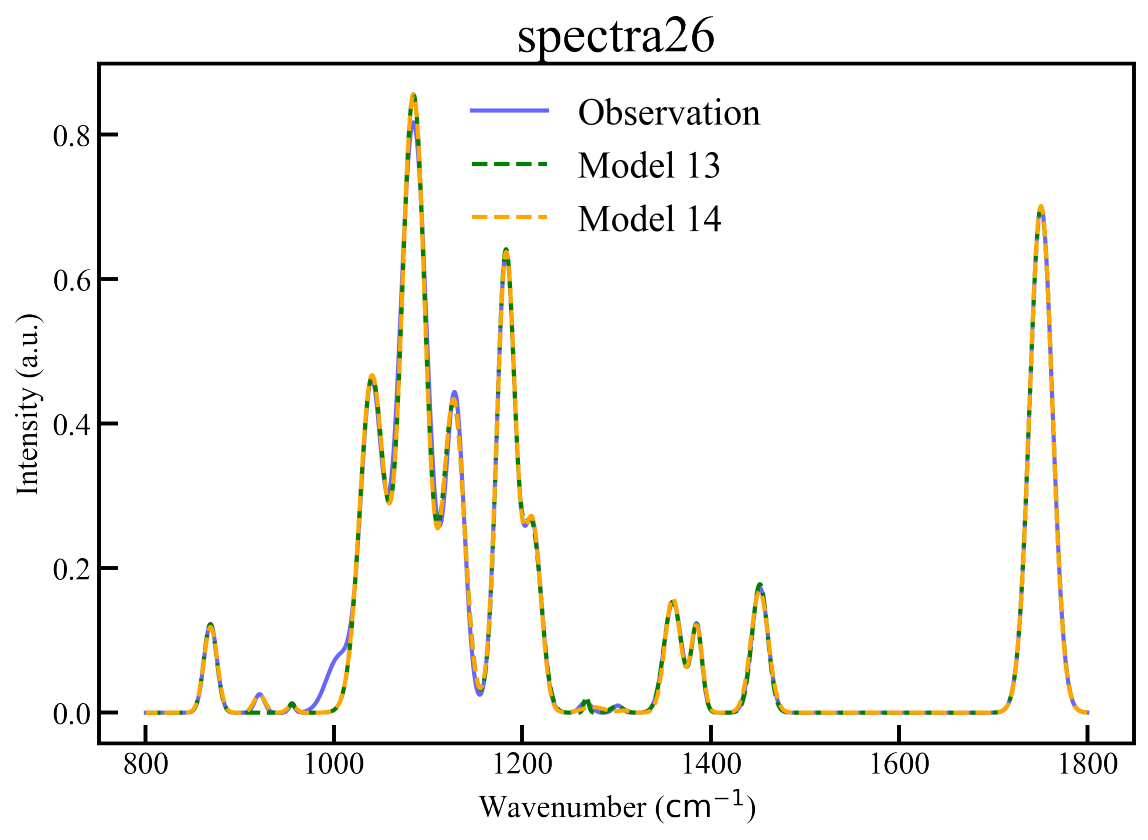} \caption{Spectrum 26} \end{subfigure}
    \vspace{2em}
    \begin{subfigure}[b]{0.45\linewidth} \centering \includegraphics[width=\linewidth]{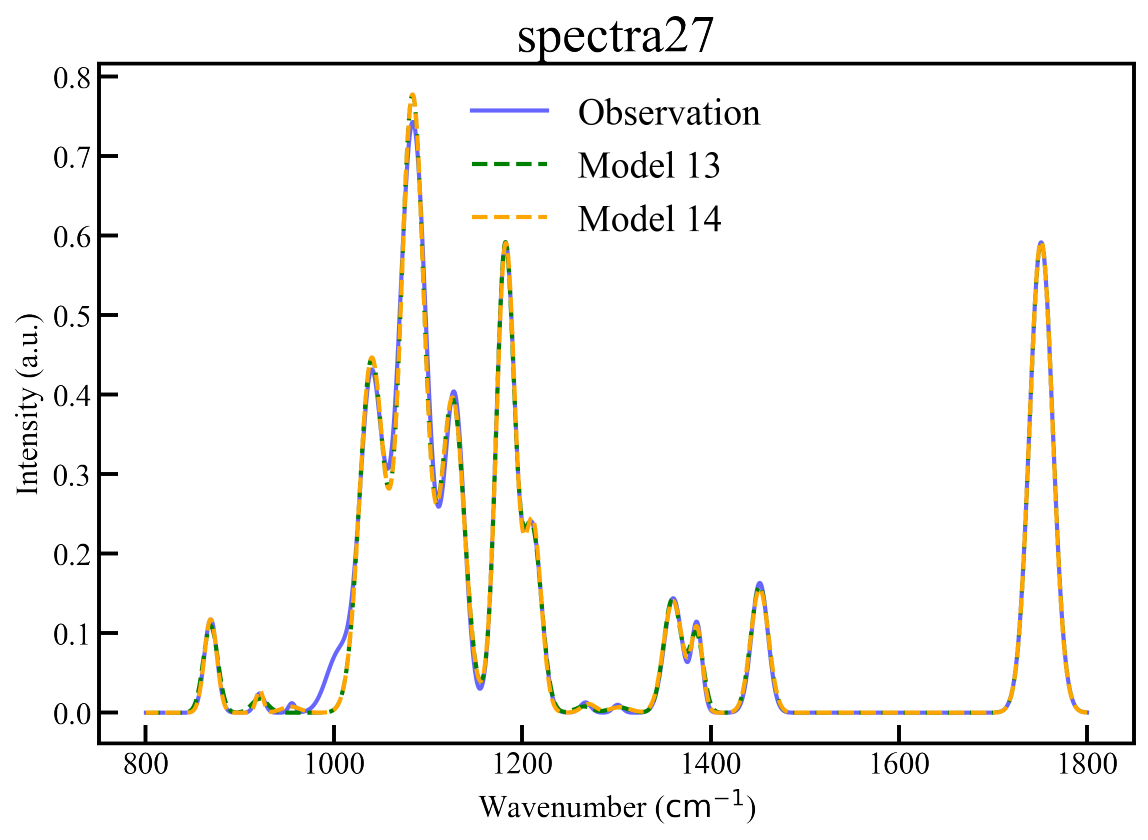} \caption{Spectrum 27} \end{subfigure}\hfill
    \begin{subfigure}[b]{0.45\linewidth} \centering \includegraphics[width=\linewidth]{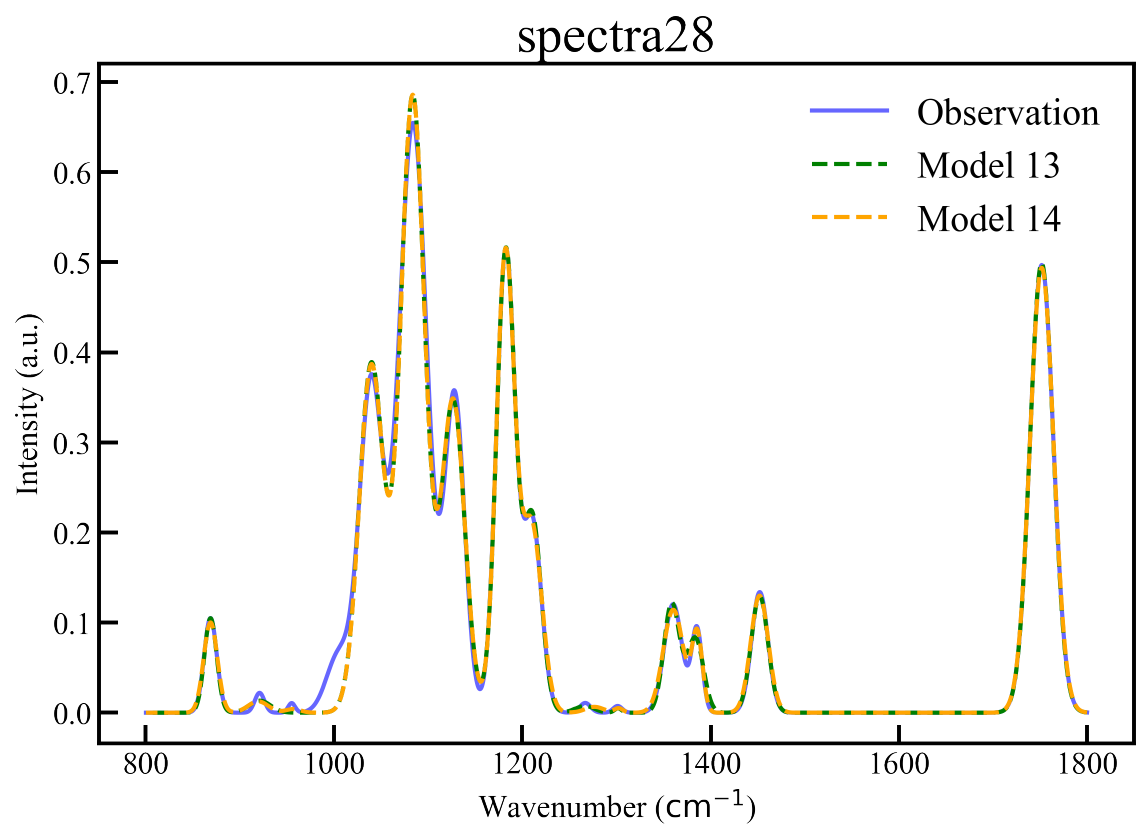} \caption{Spectrum 28} \end{subfigure}
    \vspace{2em}
    \begin{subfigure}[b]{0.45\linewidth} \centering \includegraphics[width=\linewidth]{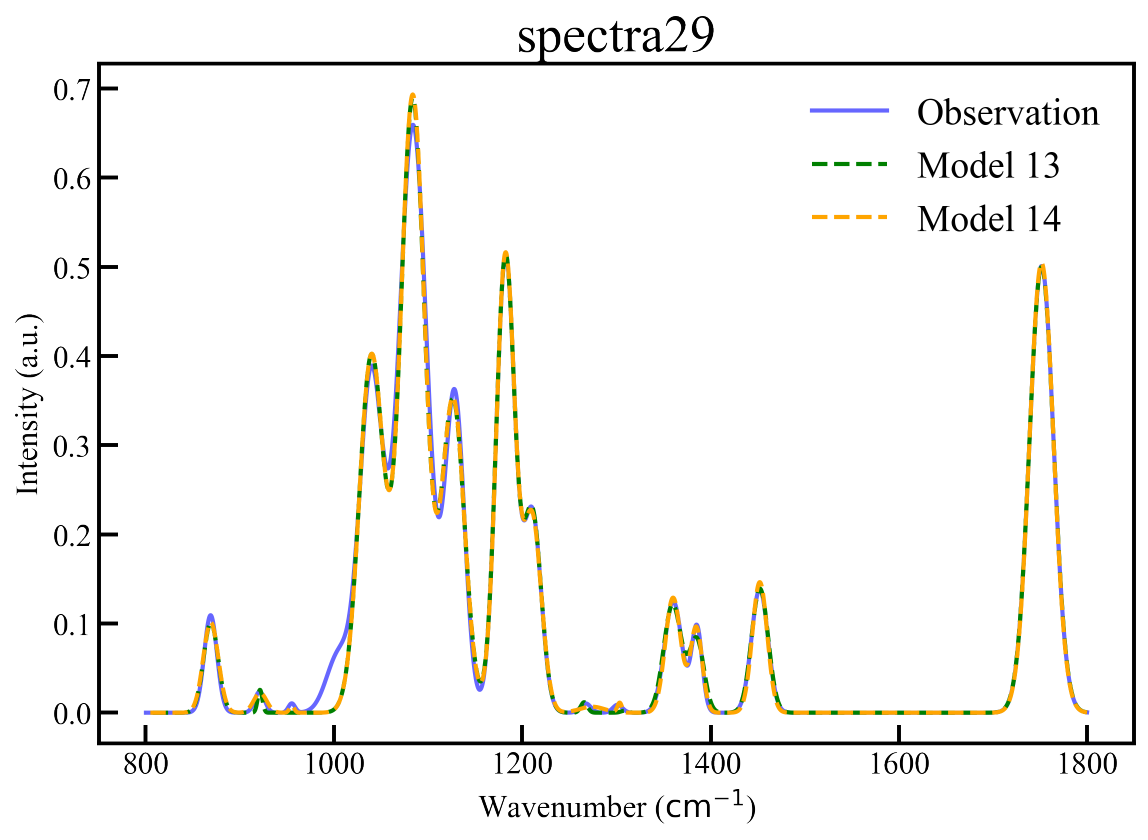} \caption{Spectrum 29} \end{subfigure}\hfill
    \begin{subfigure}[b]{0.45\linewidth} \centering \includegraphics[width=\linewidth]{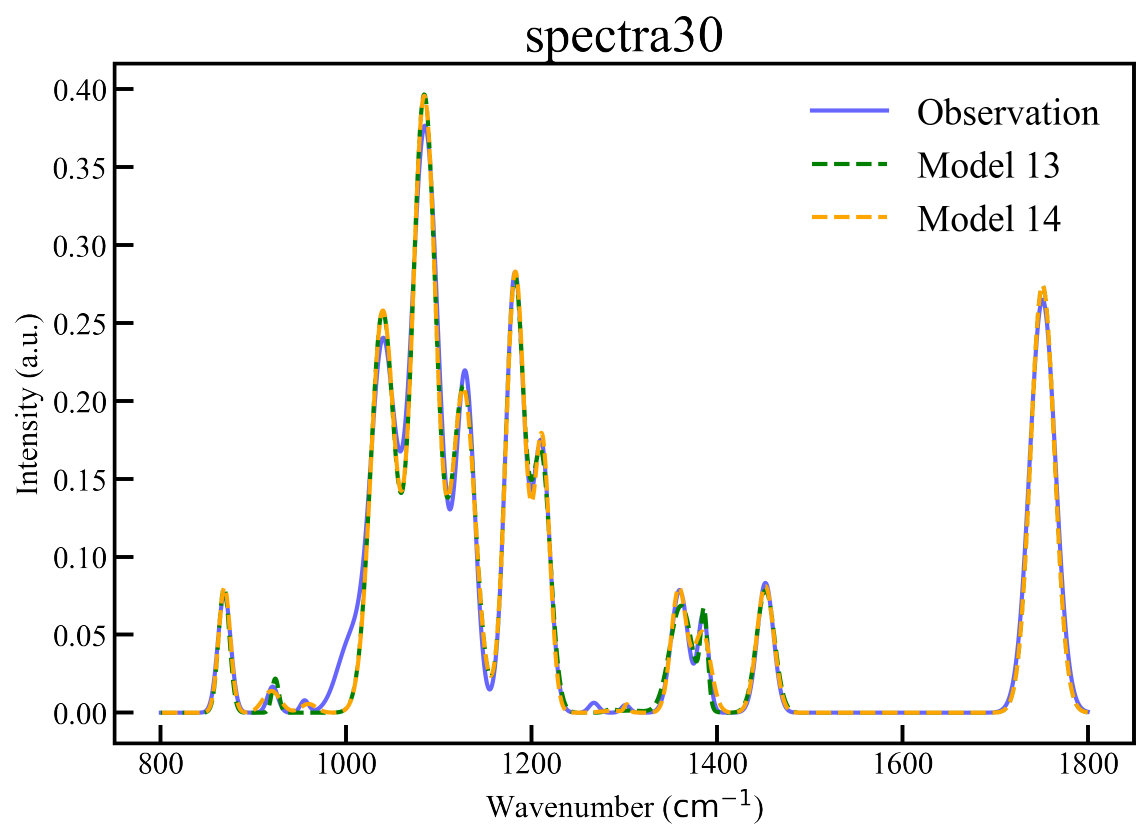} \caption{Spectrum 30} \end{subfigure}
    \caption{IR fitting results (Spectra 25--30). The blue solid line represents the observed spectral data, and the orange (green) dashed line represents the fitting curve for the 14-peak (13-peak) model.}
\end{figure}

\clearpage
\begin{figure}[p]
    \centering
    \begin{subfigure}[b]{0.45\linewidth} \centering \includegraphics[width=\linewidth]{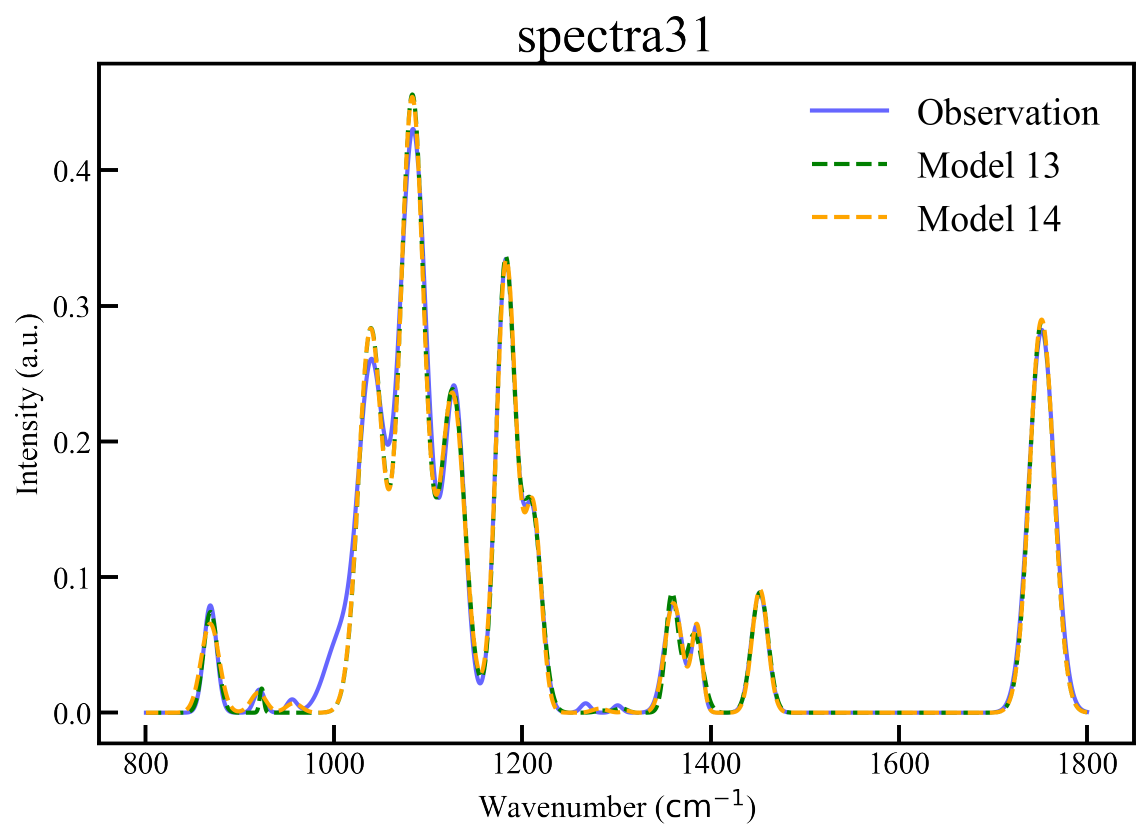} \caption{Spectrum 31} \end{subfigure}\hfill
    \begin{subfigure}[b]{0.45\linewidth} \centering \includegraphics[width=\linewidth]{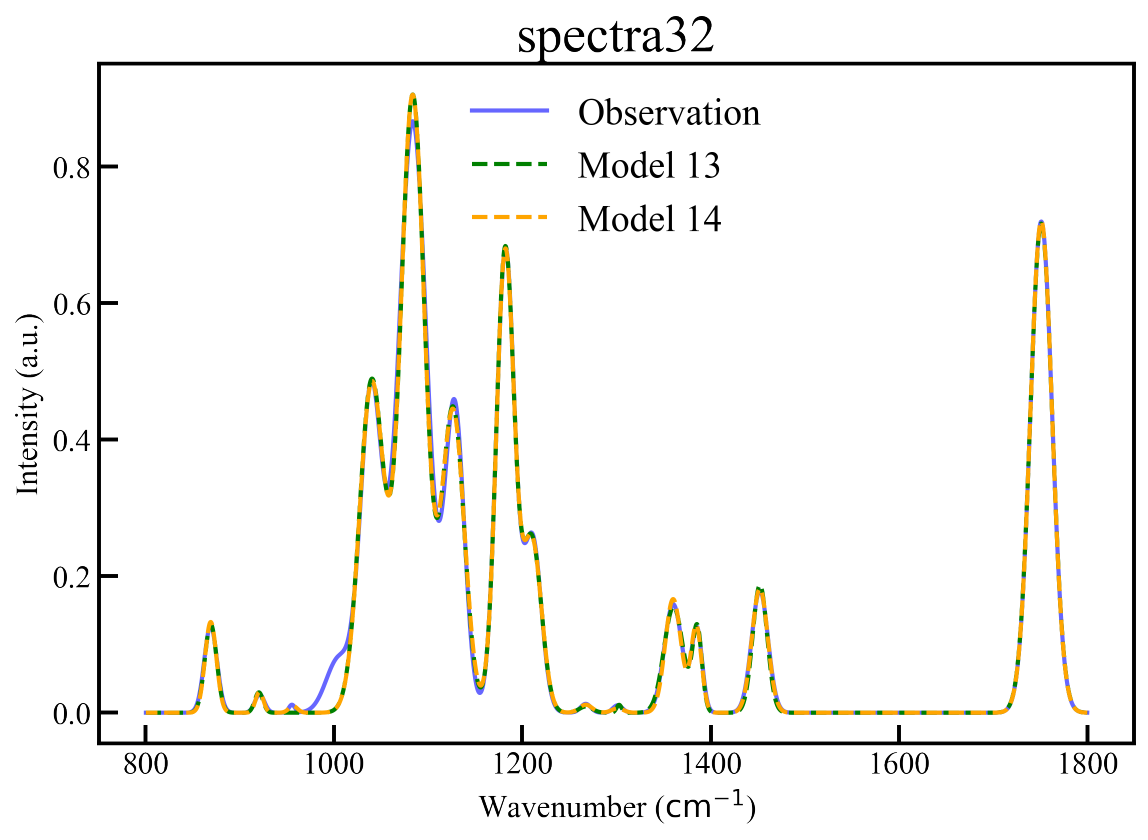} \caption{Spectrum 32} \end{subfigure}
    \vspace{2em}
    \begin{subfigure}[b]{0.45\linewidth} \centering \includegraphics[width=\linewidth]{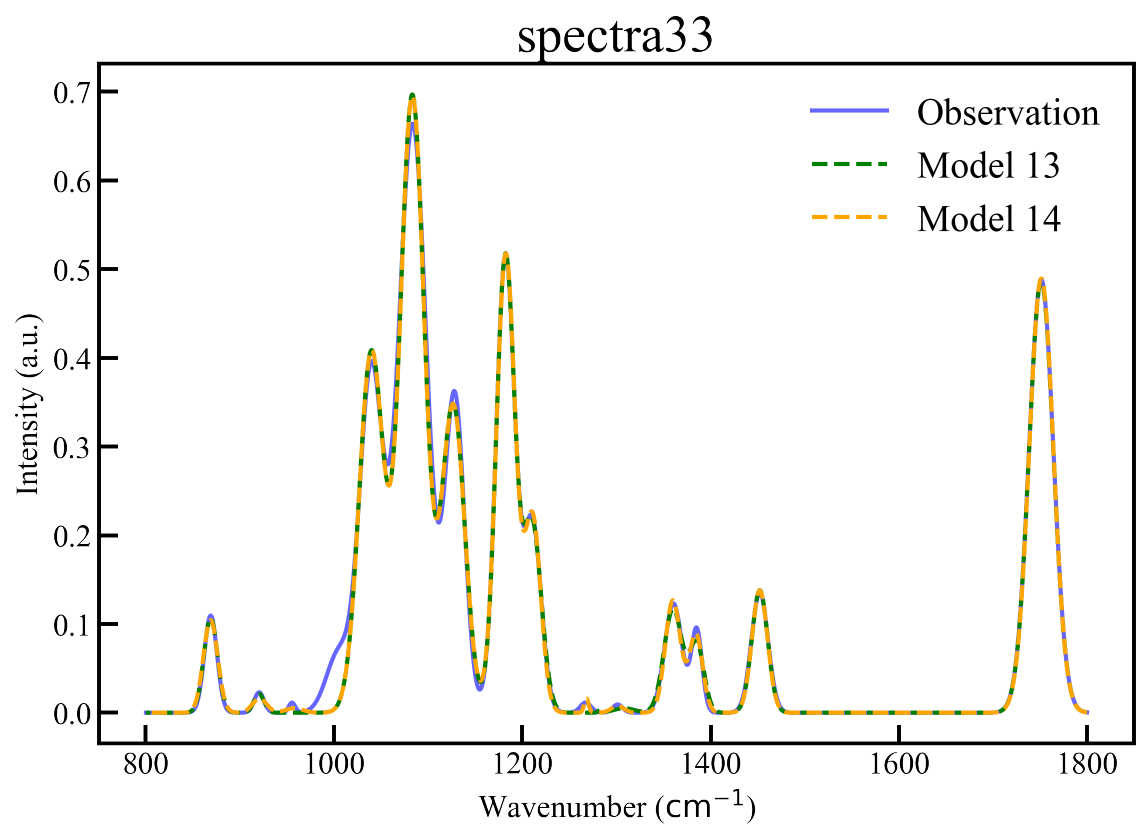} \caption{Spectrum 33} \end{subfigure}\hfill
    \begin{subfigure}[b]{0.45\linewidth} \centering \includegraphics[width=\linewidth]{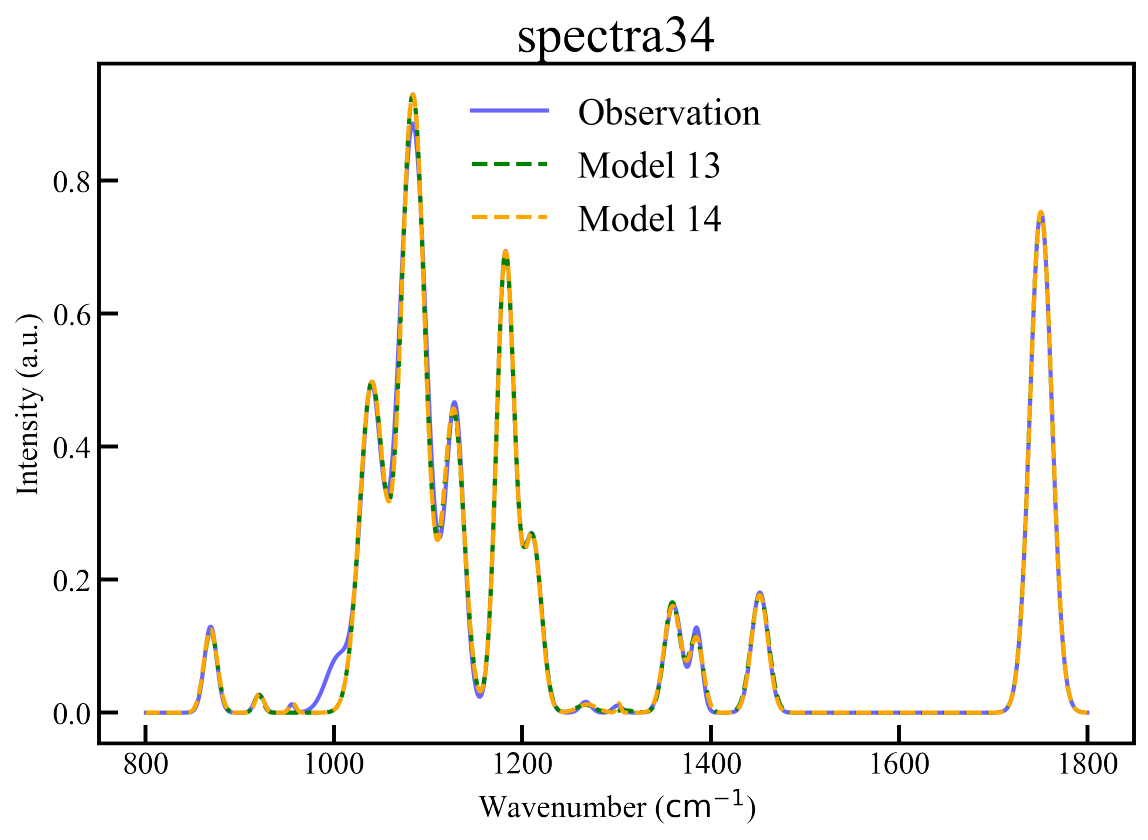} \caption{Spectrum 34} \end{subfigure}
    \vspace{2em}
    \begin{subfigure}[b]{0.45\linewidth} \centering \includegraphics[width=\linewidth]{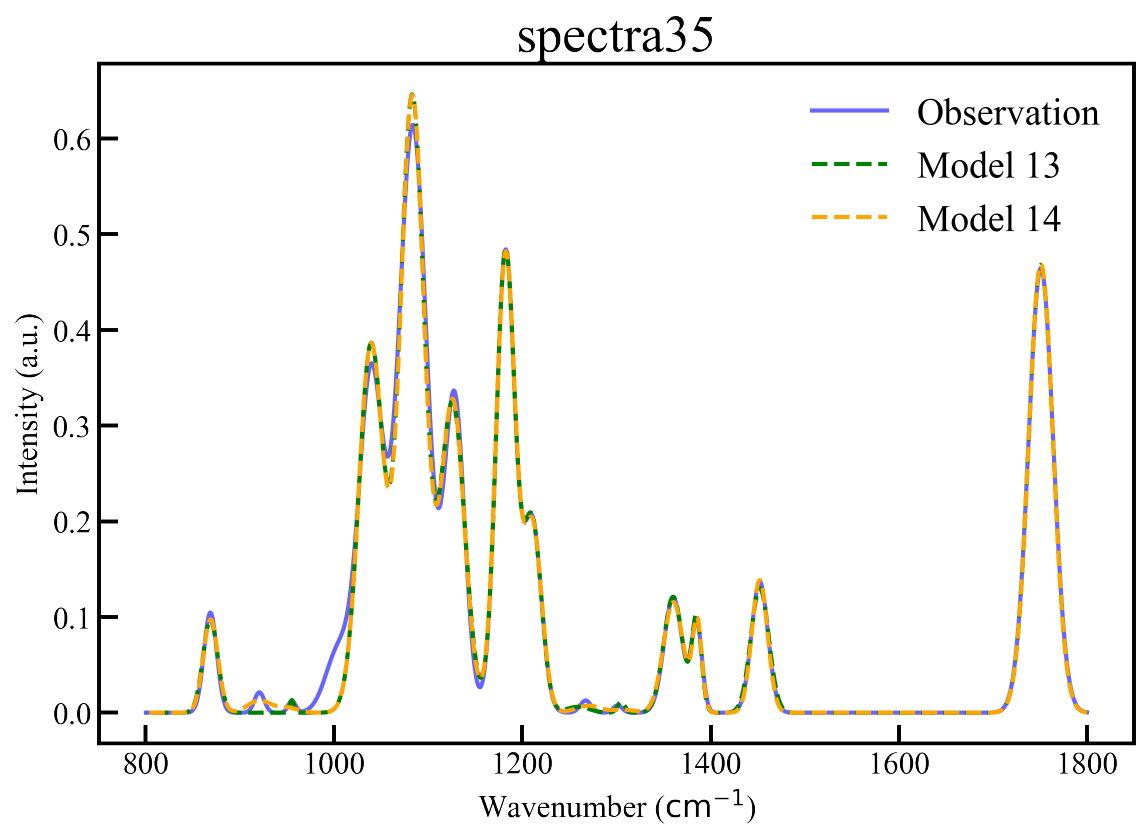} \caption{Spectrum 35} \end{subfigure}\hfill
    \begin{subfigure}[b]{0.45\linewidth} \centering \includegraphics[width=\linewidth]{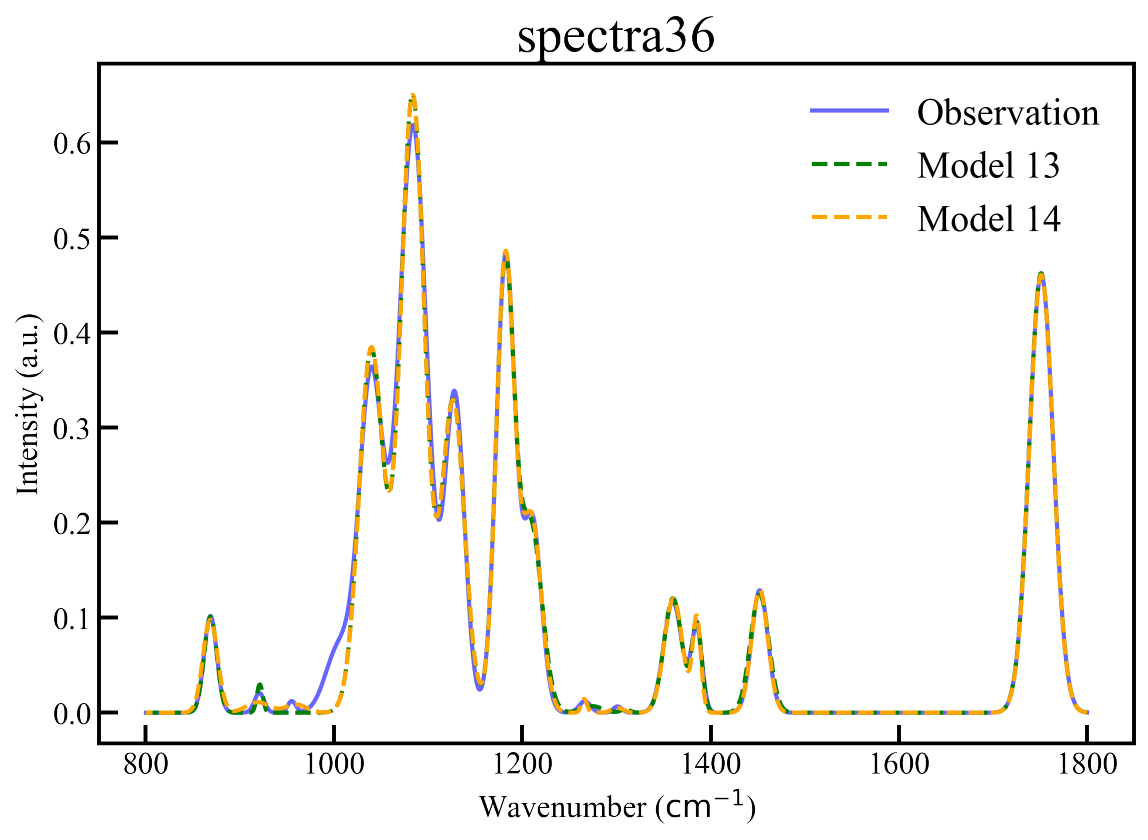} \caption{Spectrum 36} \end{subfigure}
    \caption{IR fitting results (Spectra 31--36). The blue solid line represents the observed spectral data, and the orange (green) dashed line represents the fitting curve for the 14-peak (13-peak) model.}
\end{figure}

\clearpage
\begin{figure}[p]
    \centering
    \begin{subfigure}[b]{0.45\linewidth} \centering \includegraphics[width=\linewidth]{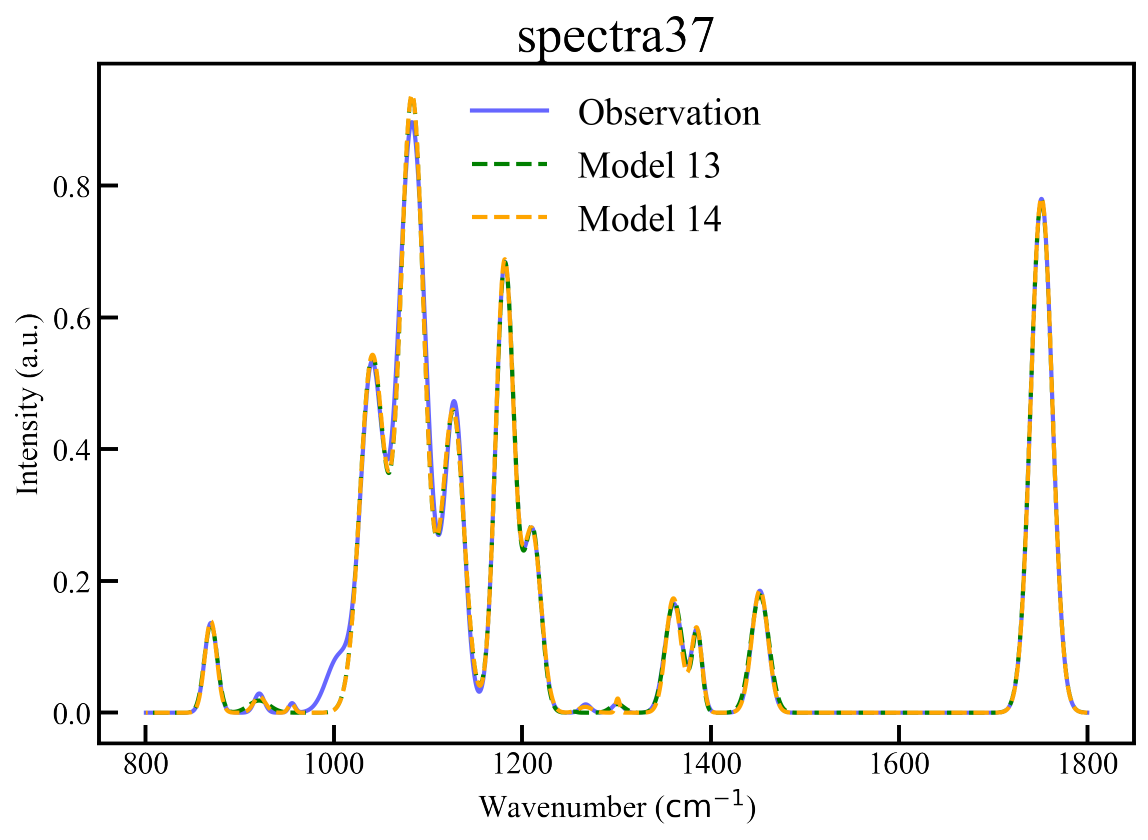} \caption{Spectrum 37} \end{subfigure}\hfill
    \begin{subfigure}[b]{0.45\linewidth} \centering \includegraphics[width=\linewidth]{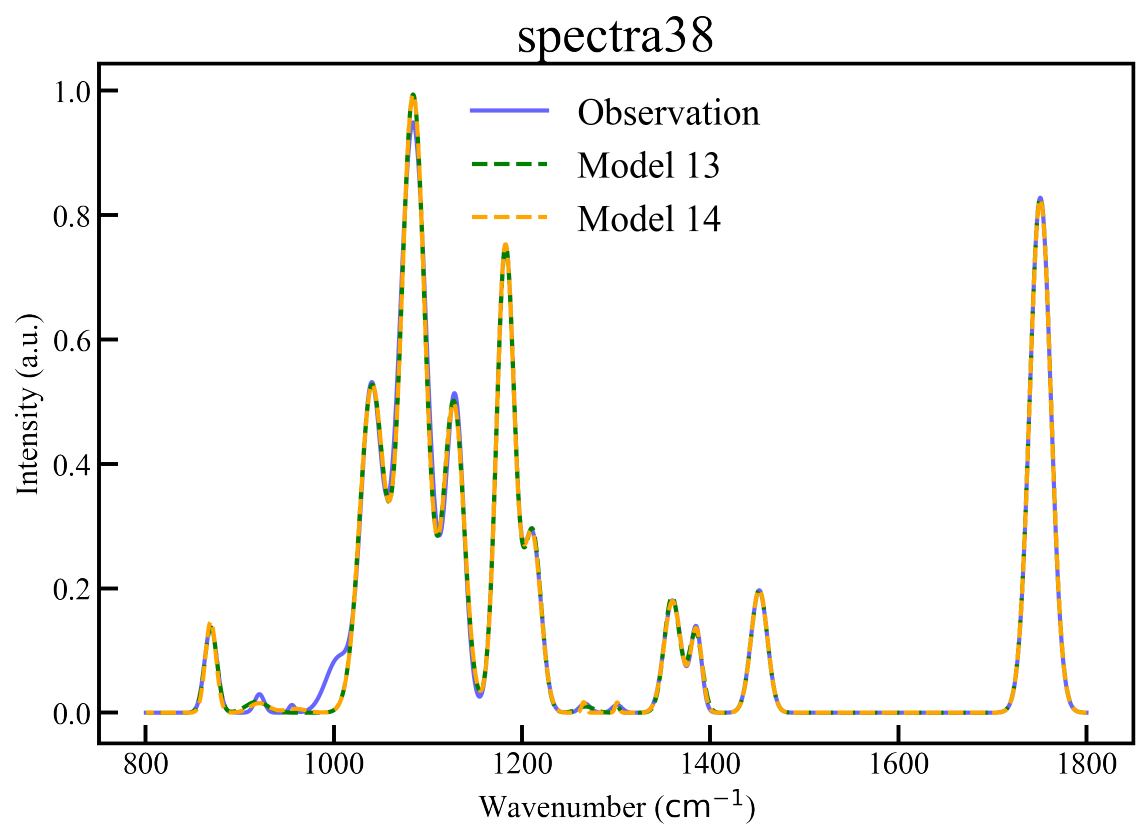} \caption{Spectrum 38} \end{subfigure}
    \vspace{2em}
    \begin{subfigure}[b]{0.45\linewidth} \centering \includegraphics[width=\linewidth]{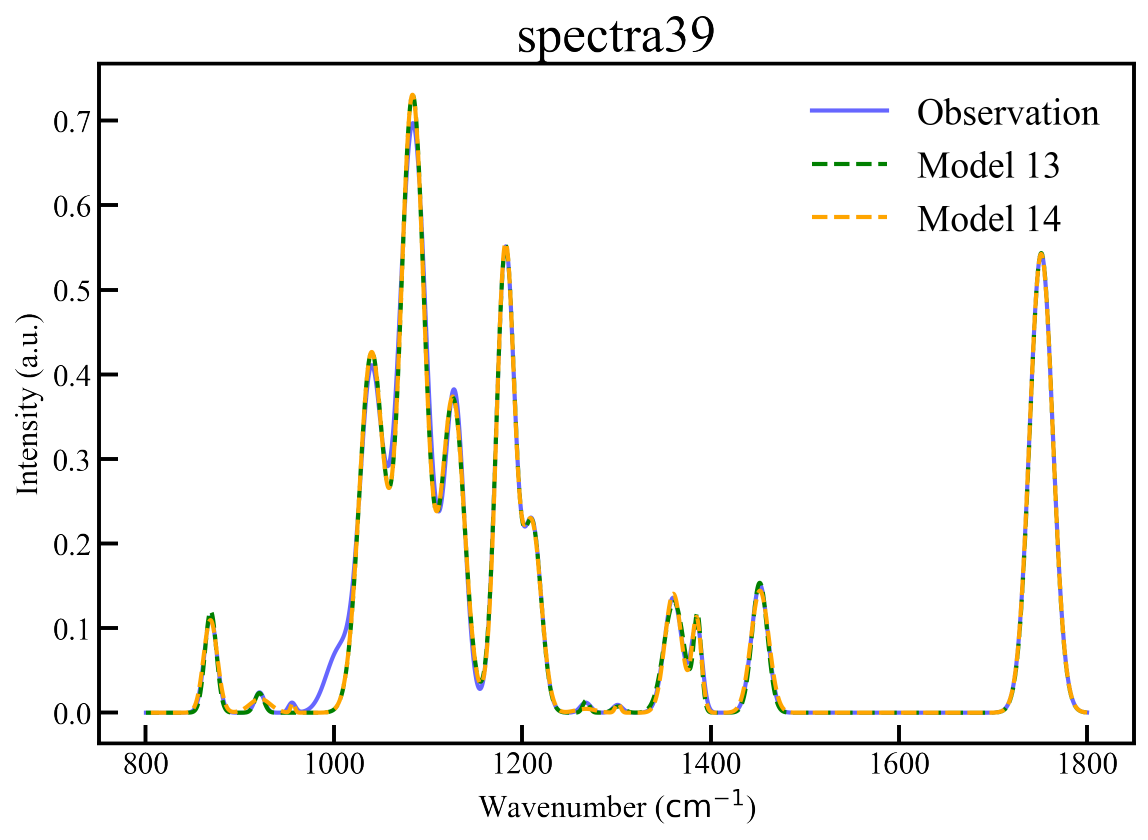} \caption{Spectrum 39} \end{subfigure}\hfill
    \begin{subfigure}[b]{0.45\linewidth} \centering \includegraphics[width=\linewidth]{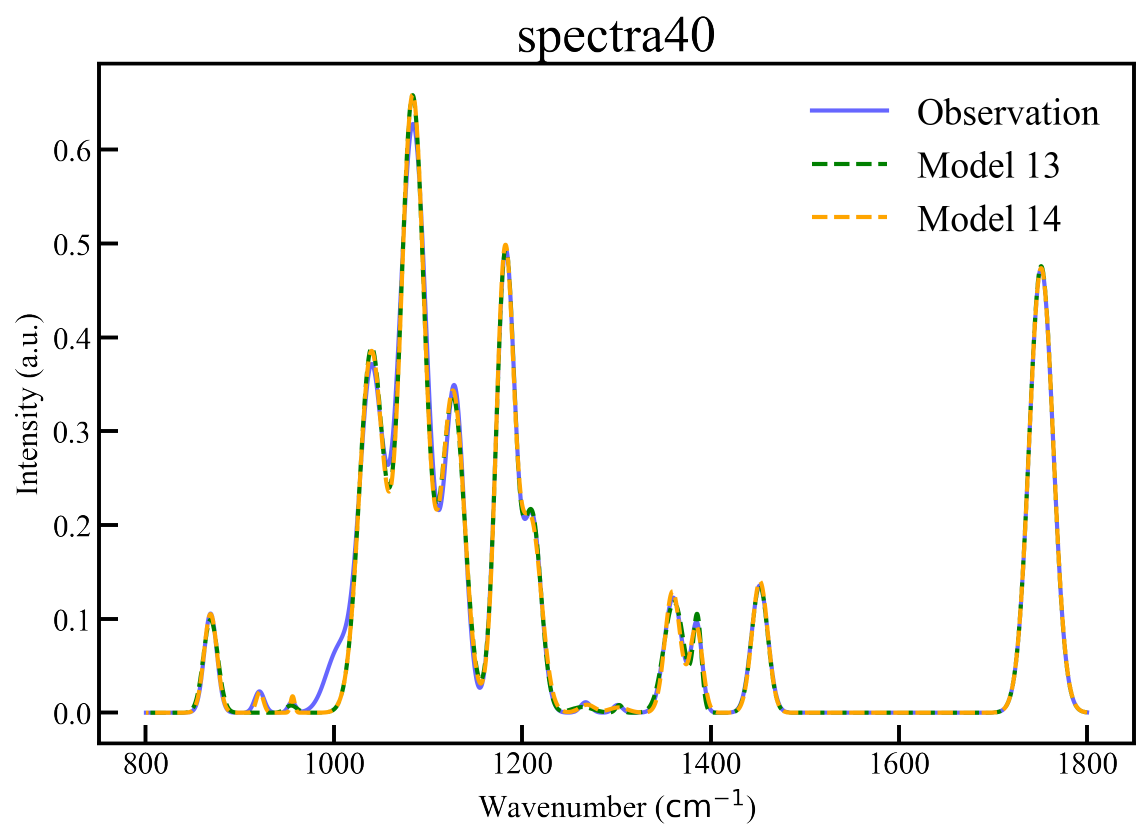} \caption{Spectrum 40} \end{subfigure}
    \vspace{2em}
    \begin{subfigure}[b]{0.45\linewidth} \centering \includegraphics[width=\linewidth]{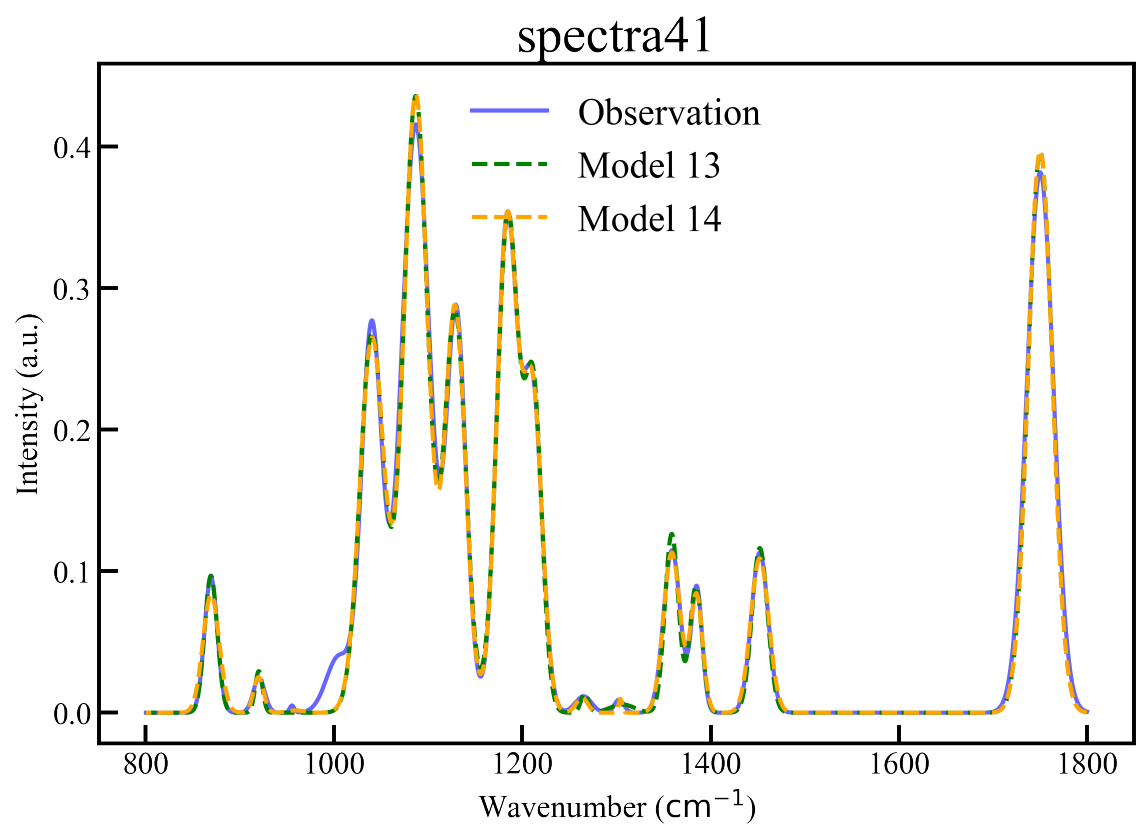} \caption{Spectrum 41} \end{subfigure}\hfill
    \begin{subfigure}[b]{0.45\linewidth} \centering \includegraphics[width=\linewidth]{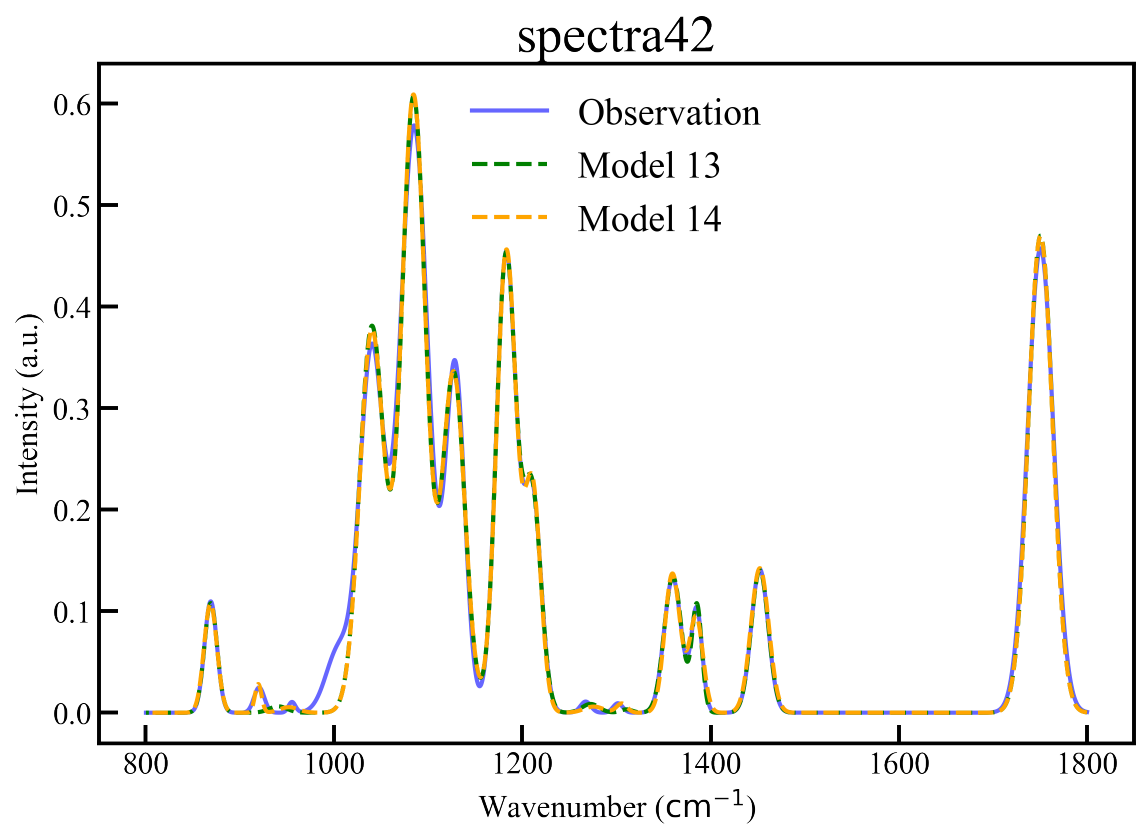} \caption{Spectrum 42} \end{subfigure}
    \caption{IR fitting results (Spectra 37--42). The blue solid line represents the observed spectral data, and the orange (green) dashed line represents the fitting curve for the 14-peak (13-peak) model.}
\end{figure}

\clearpage
\begin{figure}[p]
    \centering
    \begin{subfigure}[b]{0.45\linewidth} \centering \includegraphics[width=\linewidth]{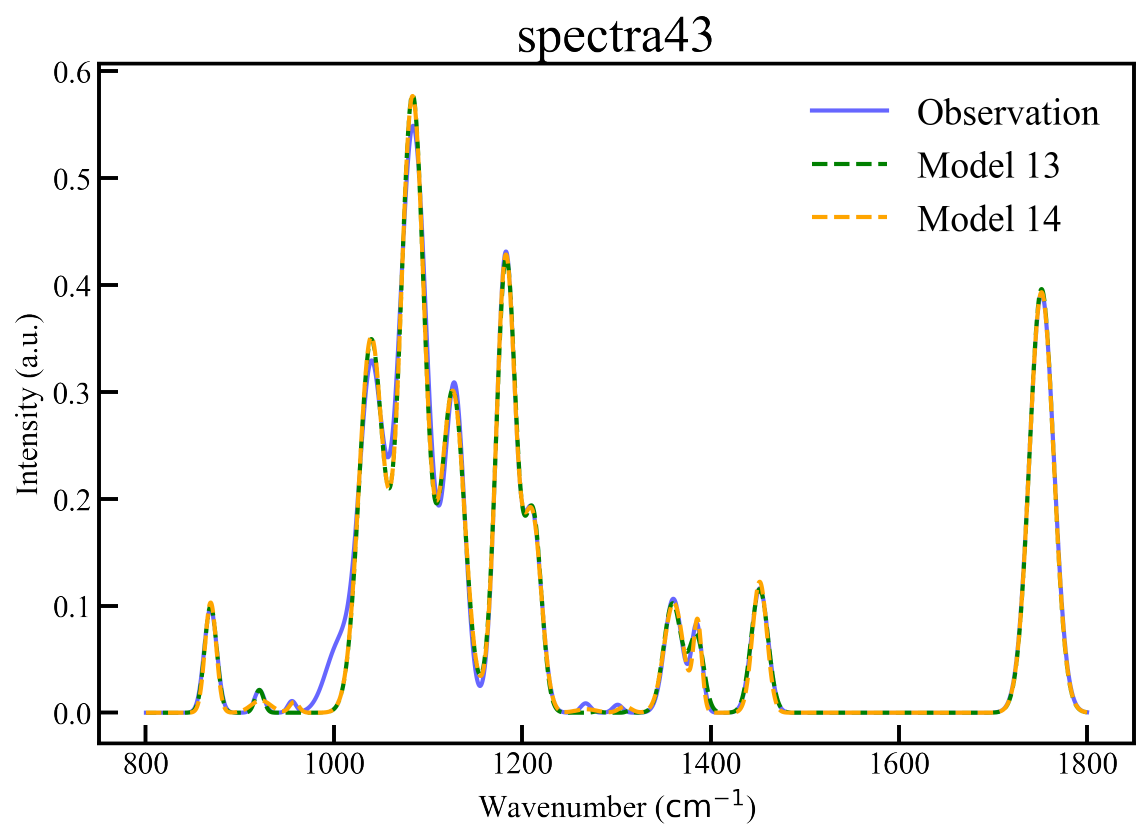} \caption{Spectrum 43} \end{subfigure}\hfill
    \begin{subfigure}[b]{0.45\linewidth} \centering \includegraphics[width=\linewidth]{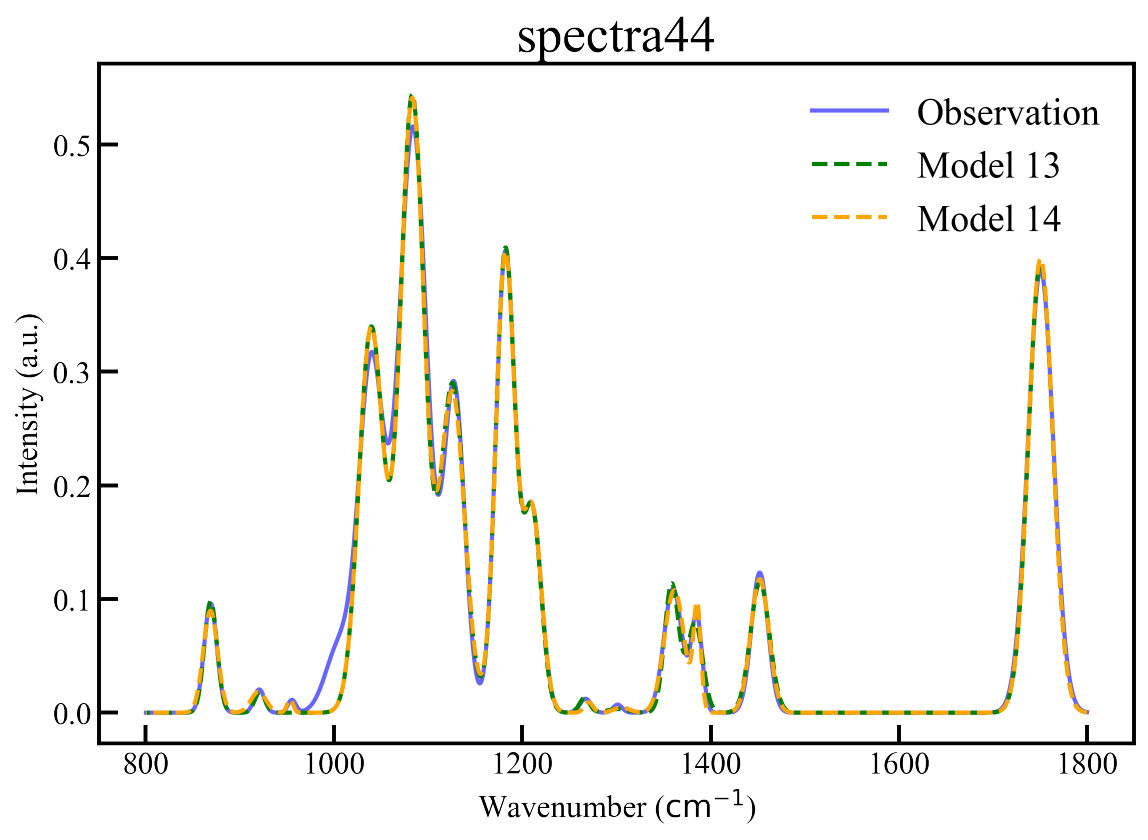} \caption{Spectrum 44} \end{subfigure}
    \vspace{2em}
    \begin{subfigure}[b]{0.45\linewidth} \centering \includegraphics[width=\linewidth]{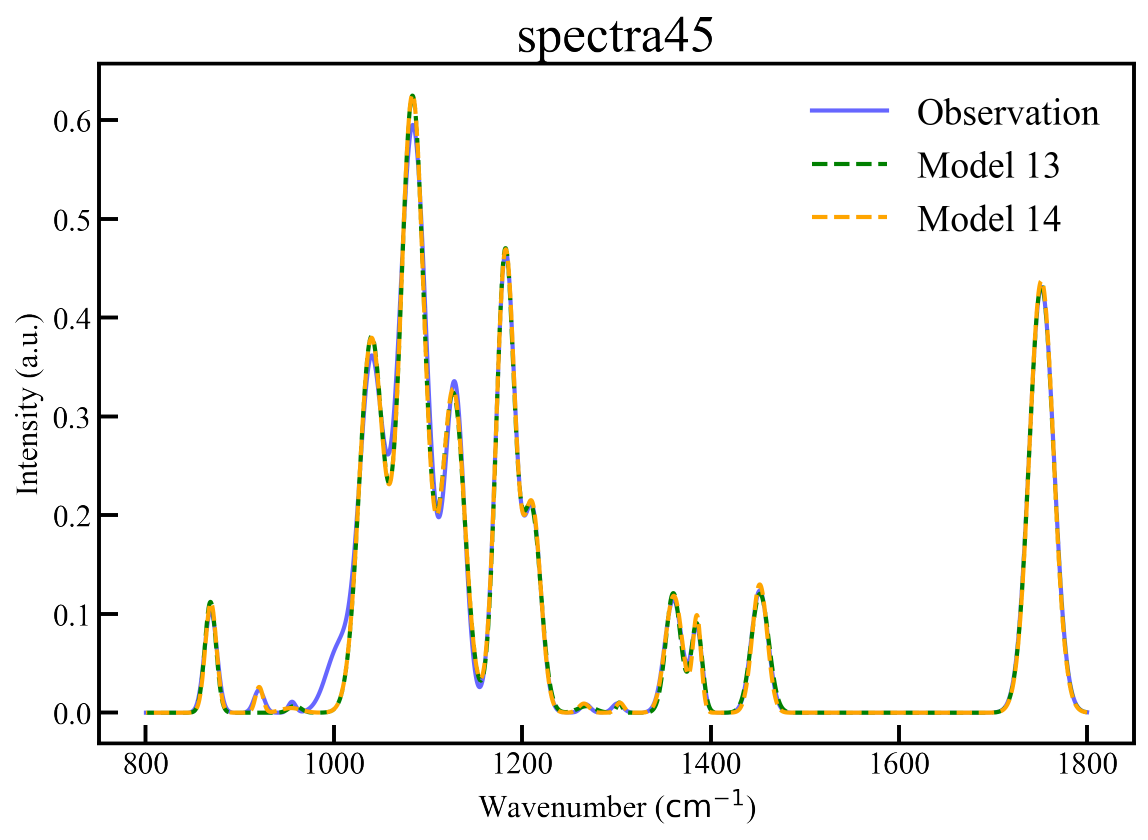} \caption{Spectrum 45} \end{subfigure}\hfill
    \begin{subfigure}[b]{0.45\linewidth} \centering \includegraphics[width=\linewidth]{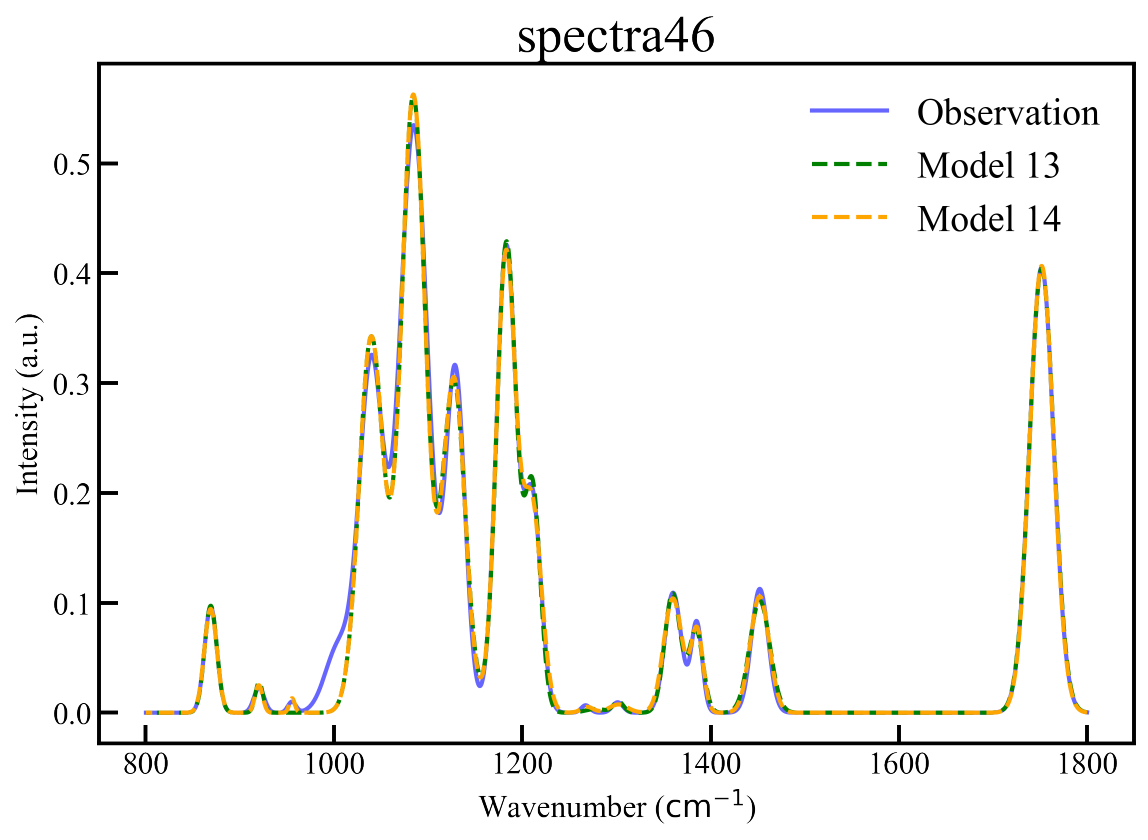} \caption{Spectrum 46} \end{subfigure}
    \vspace{2em}
    \begin{subfigure}[b]{0.45\linewidth} \centering \includegraphics[width=\linewidth]{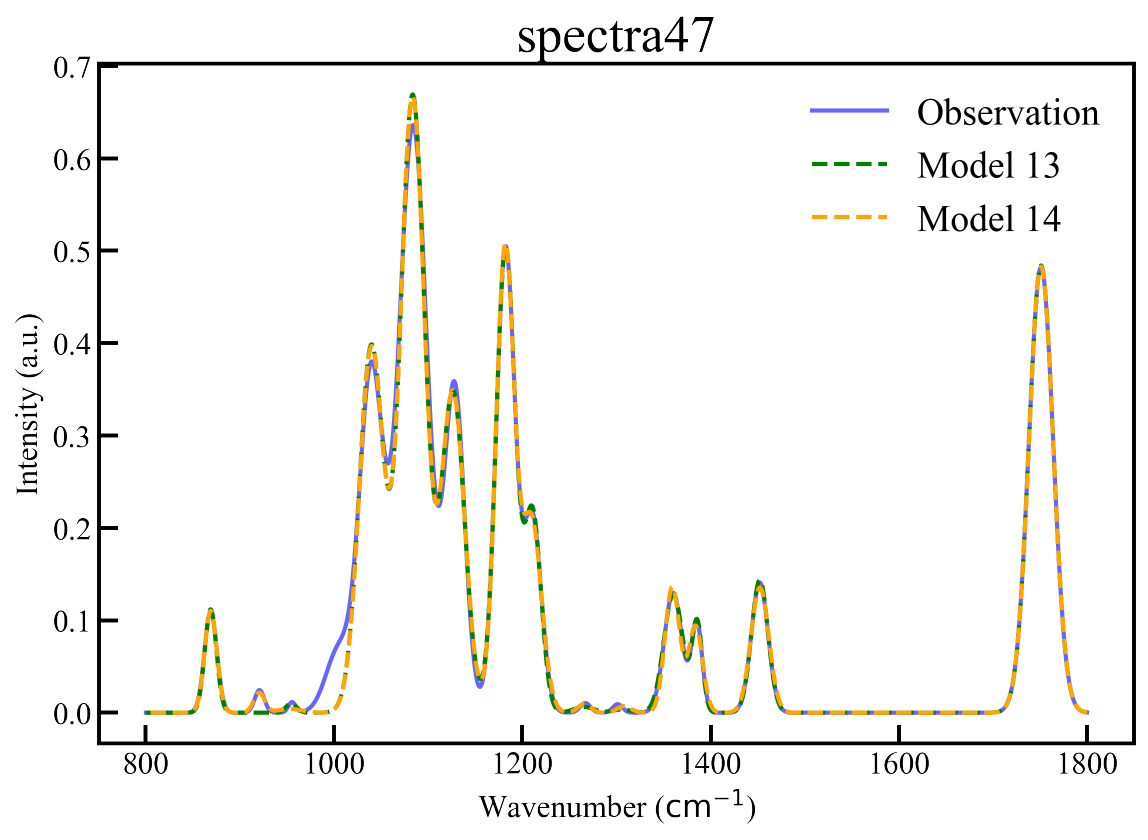} \caption{Spectrum 47} \end{subfigure}\hfill
    \begin{subfigure}[b]{0.45\linewidth} \centering \includegraphics[width=\linewidth]{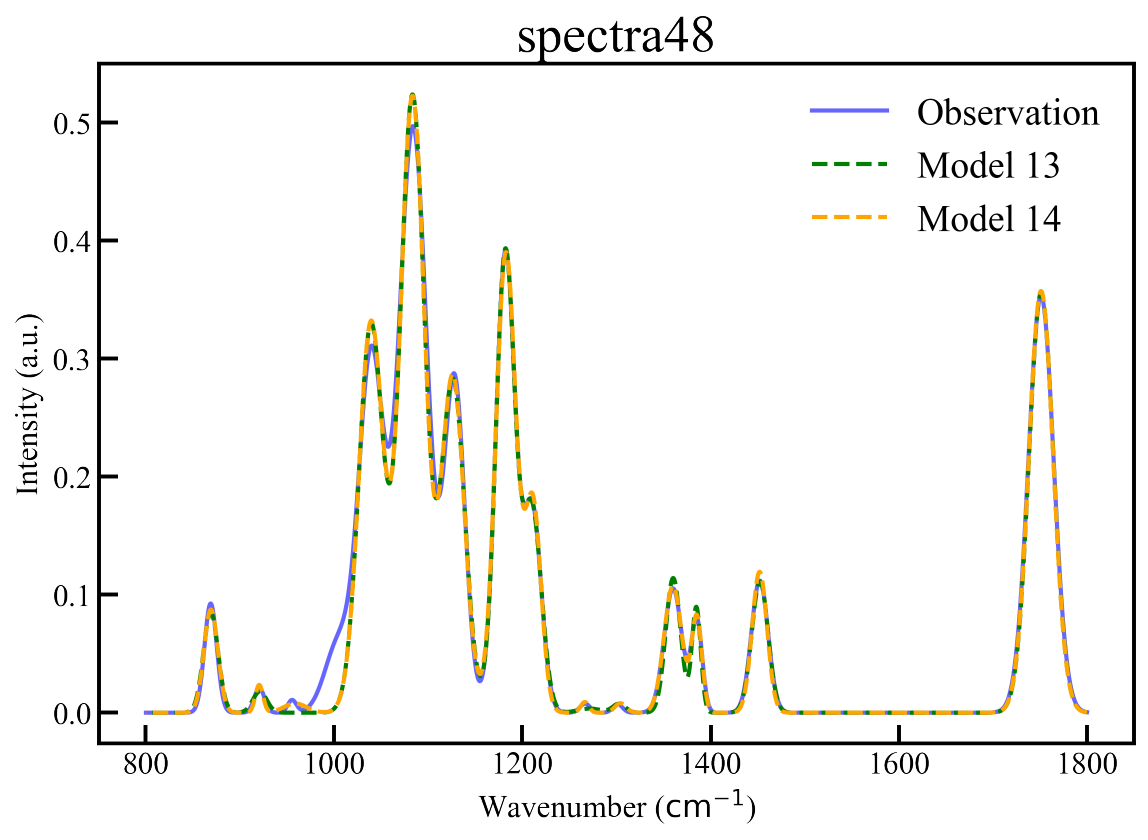} \caption{Spectrum 48} \end{subfigure}
    \caption{IR fitting results (Spectra 43--48). The blue solid line represents the observed spectral data, and the orange (green) dashed line represents the fitting curve for the 14-peak (13-peak) model.}
\end{figure}

\clearpage
\begin{figure}[p]
    \centering
    \begin{subfigure}[b]{0.45\linewidth} \centering \includegraphics[width=\linewidth]{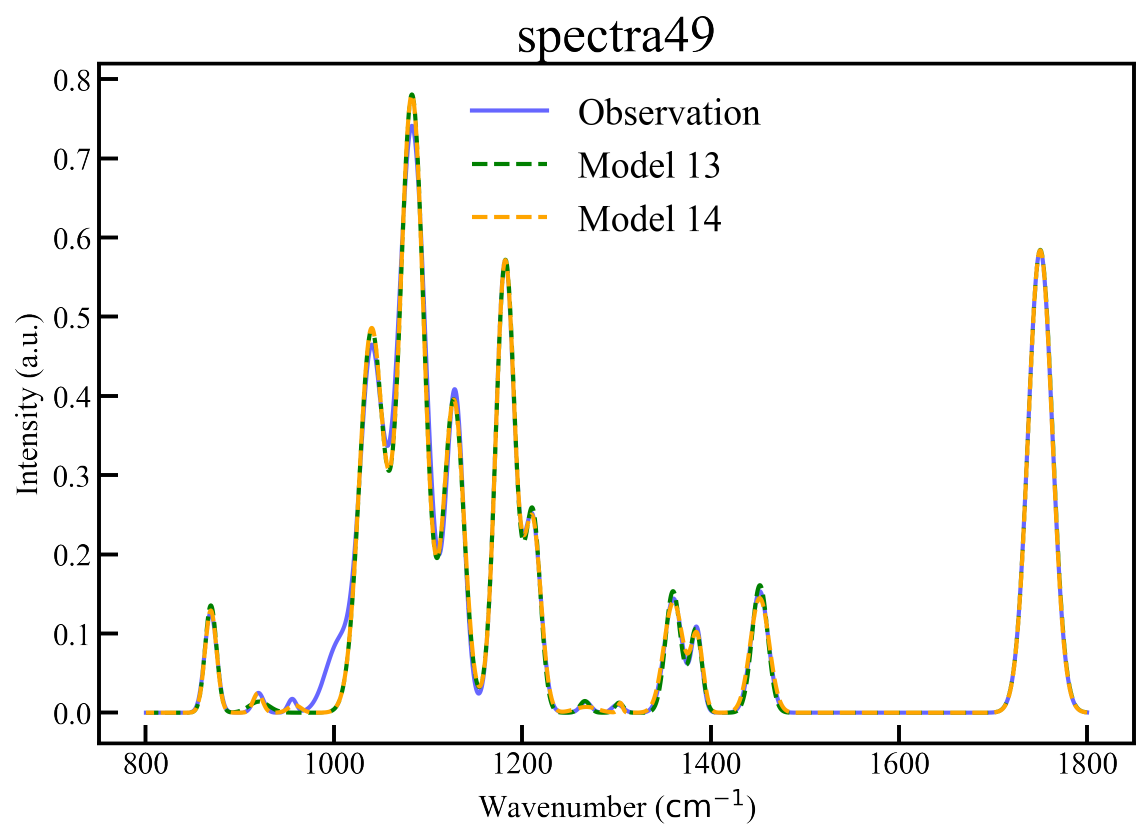} \caption{Spectrum 49} \end{subfigure}\hfill
    \begin{subfigure}[b]{0.45\linewidth} \centering \includegraphics[width=\linewidth]{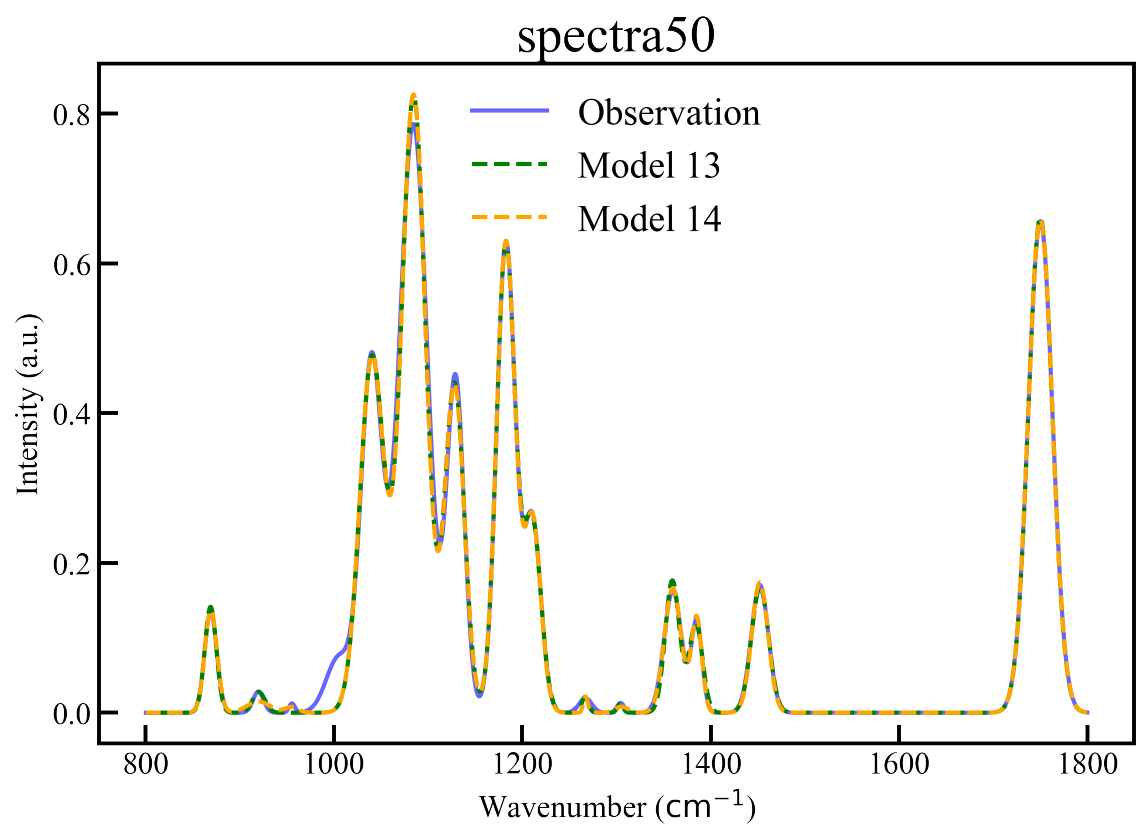} \caption{Spectrum 50} \end{subfigure}
    \vspace{2em}
    \begin{subfigure}[b]{0.45\linewidth} \centering \includegraphics[width=\linewidth]{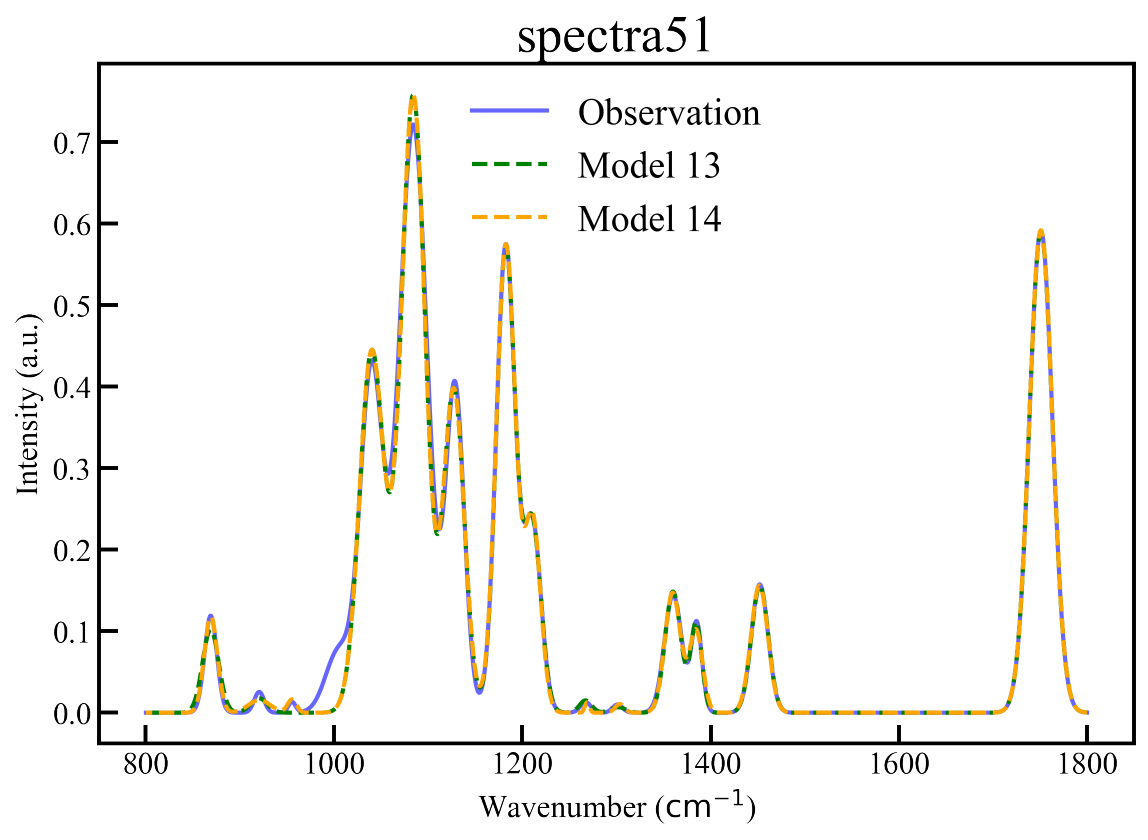} \caption{Spectrum 51} \end{subfigure}\hfill
    \begin{subfigure}[b]{0.45\linewidth} \centering \includegraphics[width=\linewidth]{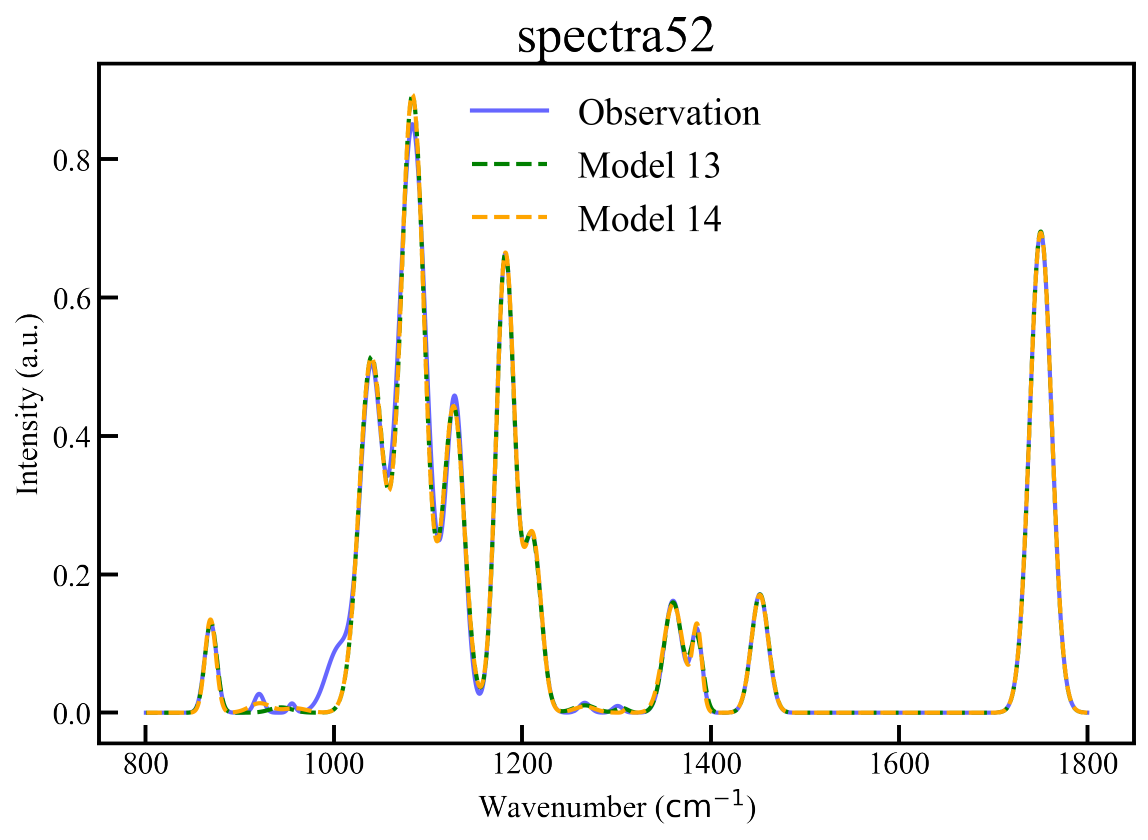} \caption{Spectrum 52} \end{subfigure}
    \vspace{2em}
    \begin{subfigure}[b]{0.45\linewidth} \centering \includegraphics[width=\linewidth]{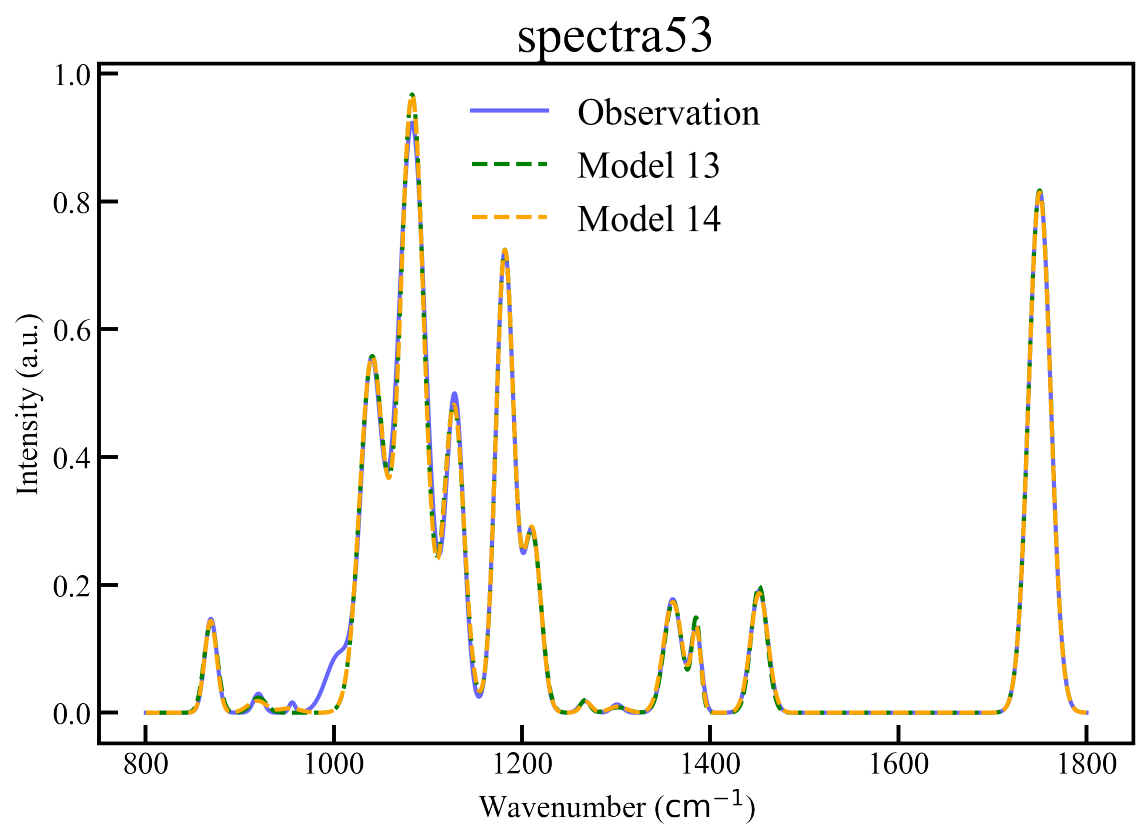} \caption{Spectrum 53} \end{subfigure}\hfill
    \begin{subfigure}[b]{0.45\linewidth} \centering \includegraphics[width=\linewidth]{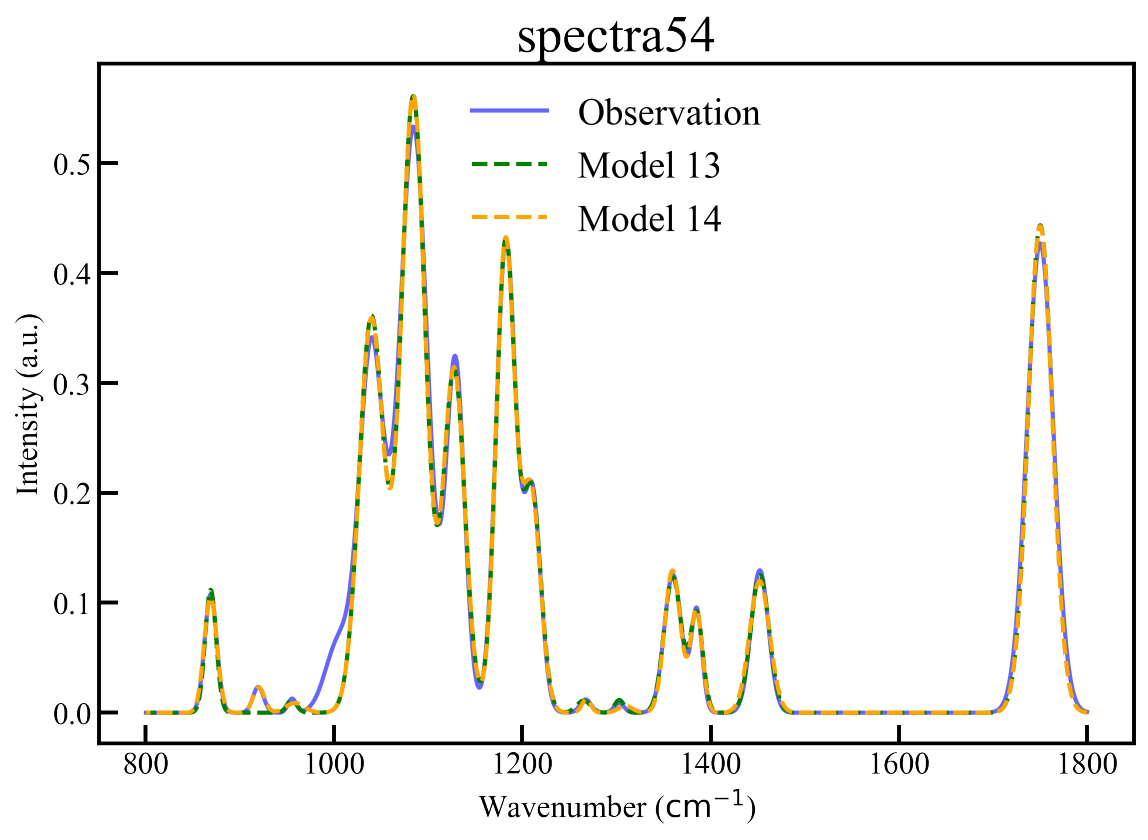} \caption{Spectrum 54} \end{subfigure}
    \caption{IR fitting results (Spectra 49--54). The blue solid line represents the observed spectral data, and the orange (green) dashed line represents the fitting curve for the 14-peak (13-peak) model.}
\end{figure}

\clearpage
\begin{figure}[p]
    \centering
    \begin{subfigure}[b]{0.45\linewidth} \centering \includegraphics[width=\linewidth]{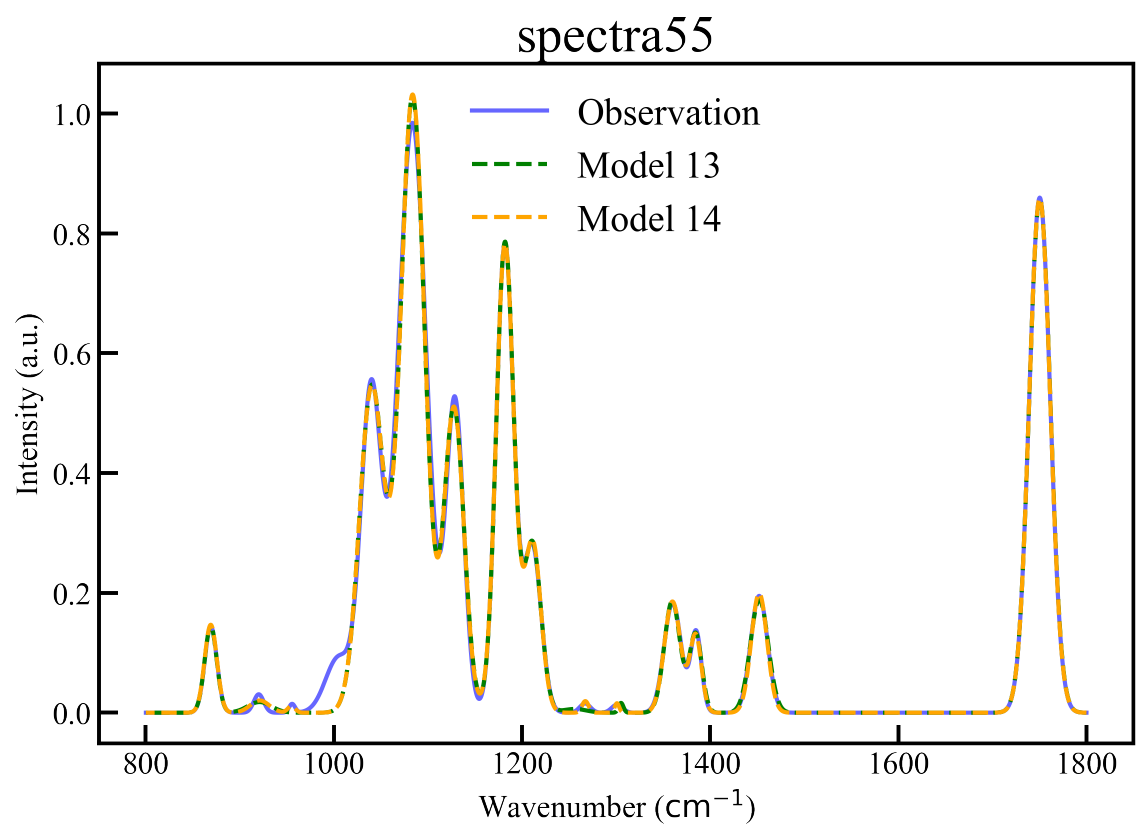} \caption{Spectrum 55} \end{subfigure}\hfill
    \begin{subfigure}[b]{0.45\linewidth} \centering \includegraphics[width=\linewidth]{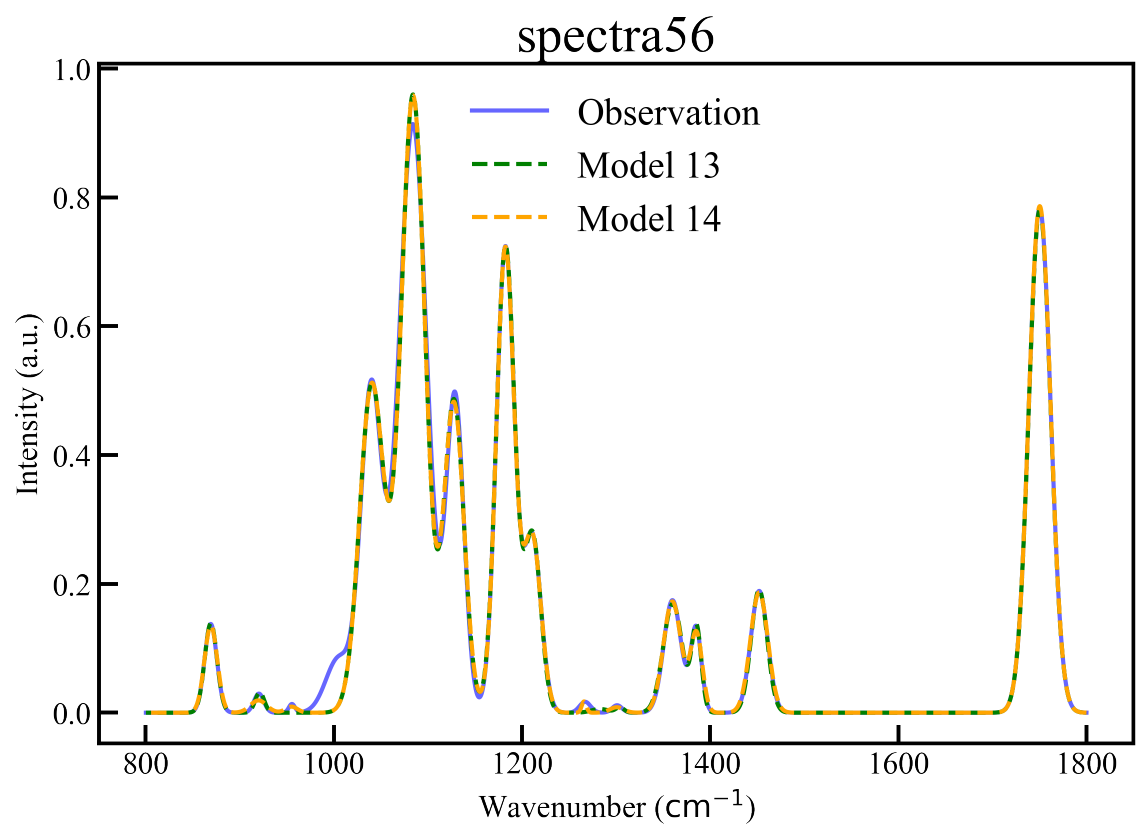} \caption{Spectrum 56} \end{subfigure}
    \vspace{2em}
    \begin{subfigure}[b]{0.45\linewidth} \centering \includegraphics[width=\linewidth]{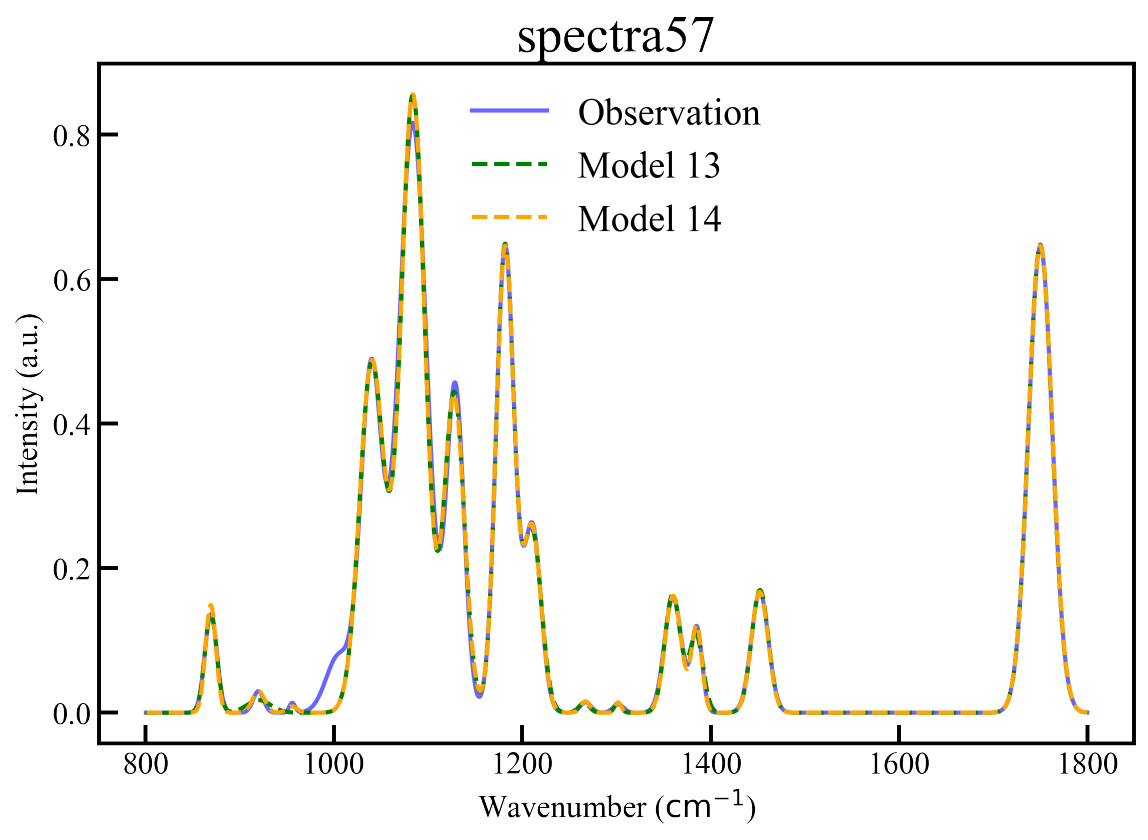} \caption{Spectrum 57} \end{subfigure}\hfill
    \begin{subfigure}[b]{0.45\linewidth} \centering \includegraphics[width=\linewidth]{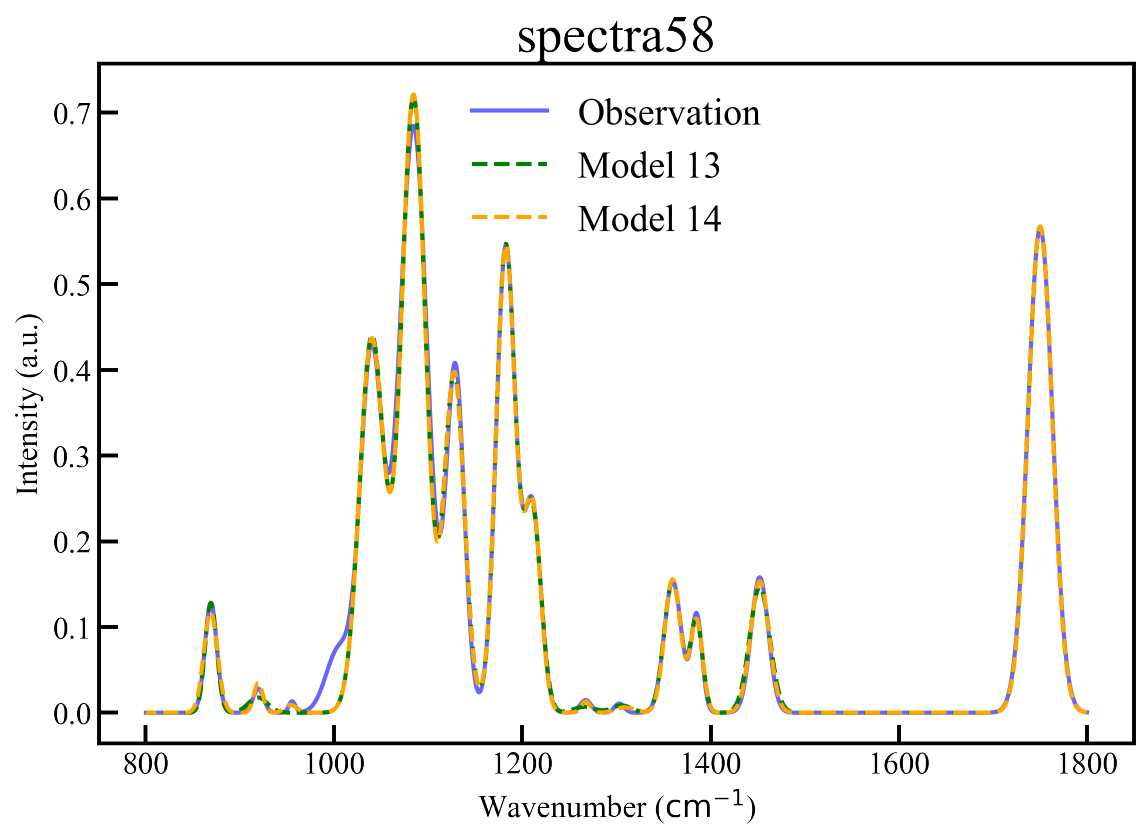} \caption{Spectrum 58} \end{subfigure}
    \vspace{2em}
    \begin{subfigure}[b]{0.45\linewidth} \centering \includegraphics[width=\linewidth]{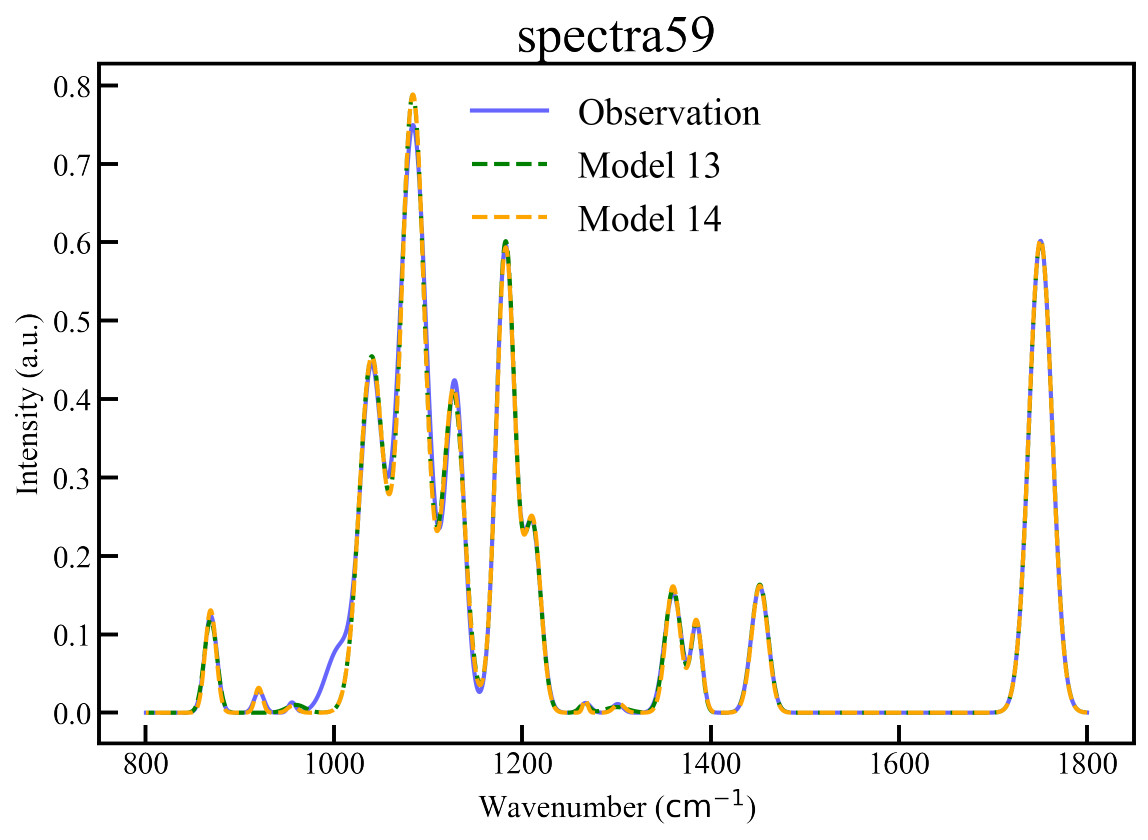} \caption{Spectrum 59} \end{subfigure}\hfill
    \begin{subfigure}[b]{0.45\linewidth} \centering \includegraphics[width=\linewidth]{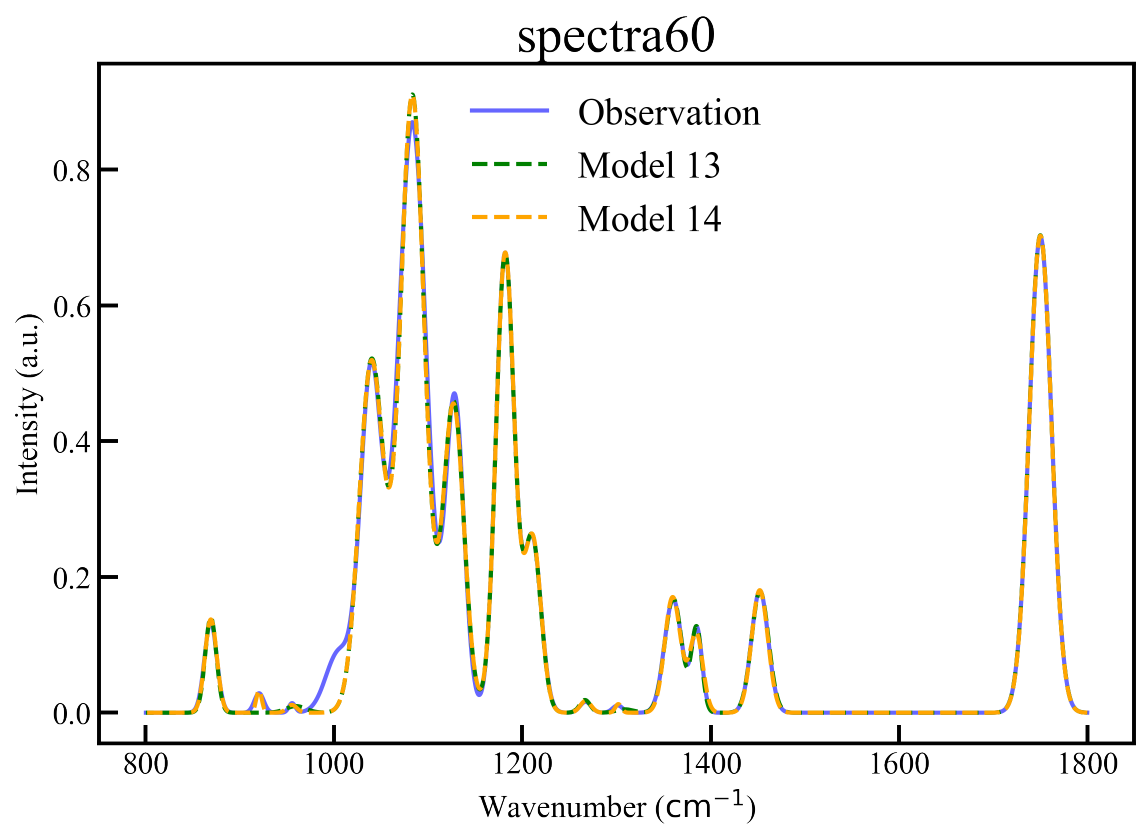} \caption{Spectrum 60} \end{subfigure}
    \caption{IR fitting results (Spectra 55--60). The blue solid line represents the observed spectral data, and the orange (green) dashed line represents the fitting curve for the 14-peak (13-peak) model.}
\end{figure}

\clearpage
\begin{figure}[p]
    \centering
    \begin{subfigure}[b]{0.45\linewidth} \centering \includegraphics[width=\linewidth]{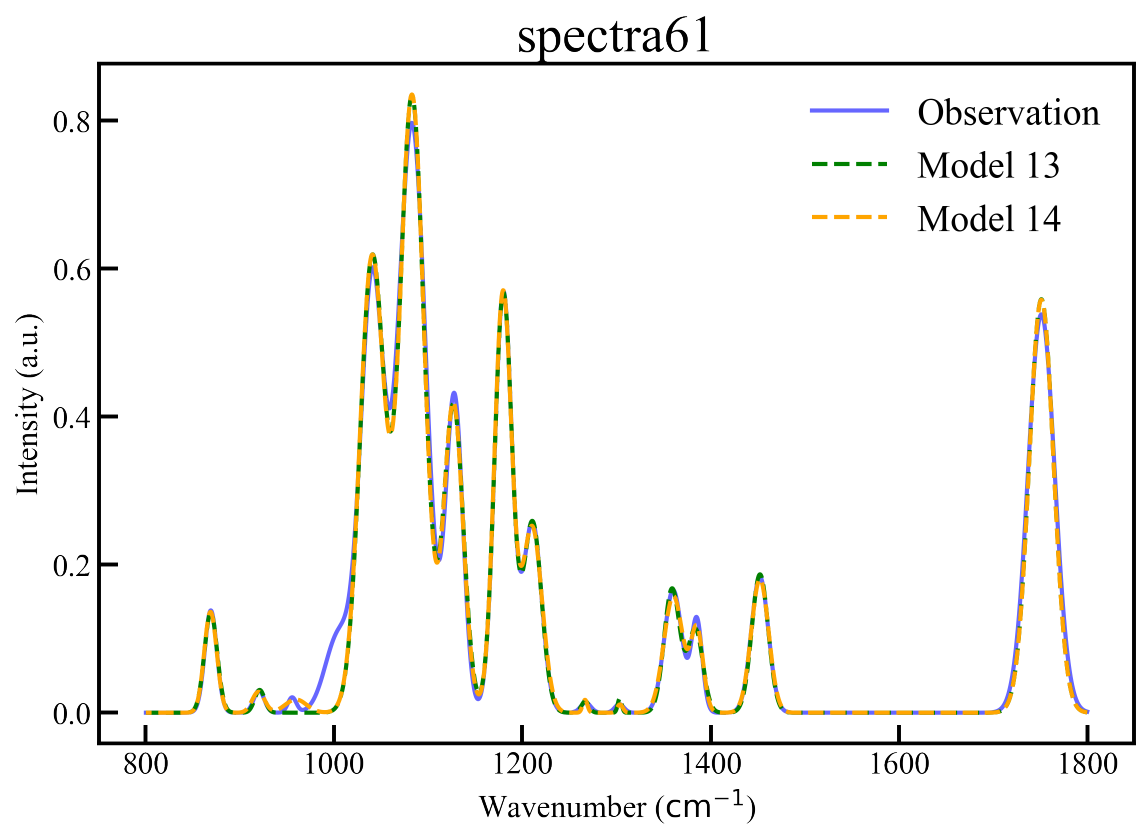} \caption{Spectrum 61} \end{subfigure}\hfill
    \begin{subfigure}[b]{0.45\linewidth} \centering \includegraphics[width=\linewidth]{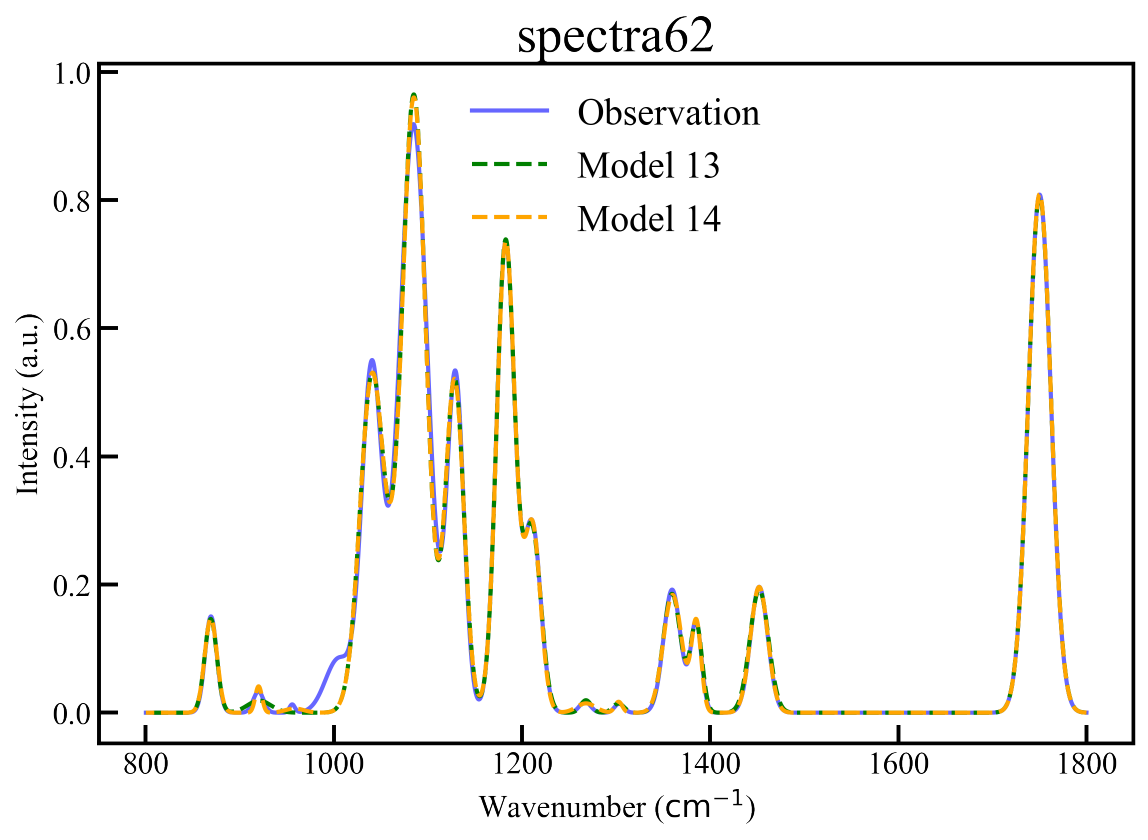} \caption{Spectrum 62} \end{subfigure}
    \vspace{2em}
    \begin{subfigure}[b]{0.45\linewidth} \centering \includegraphics[width=\linewidth]{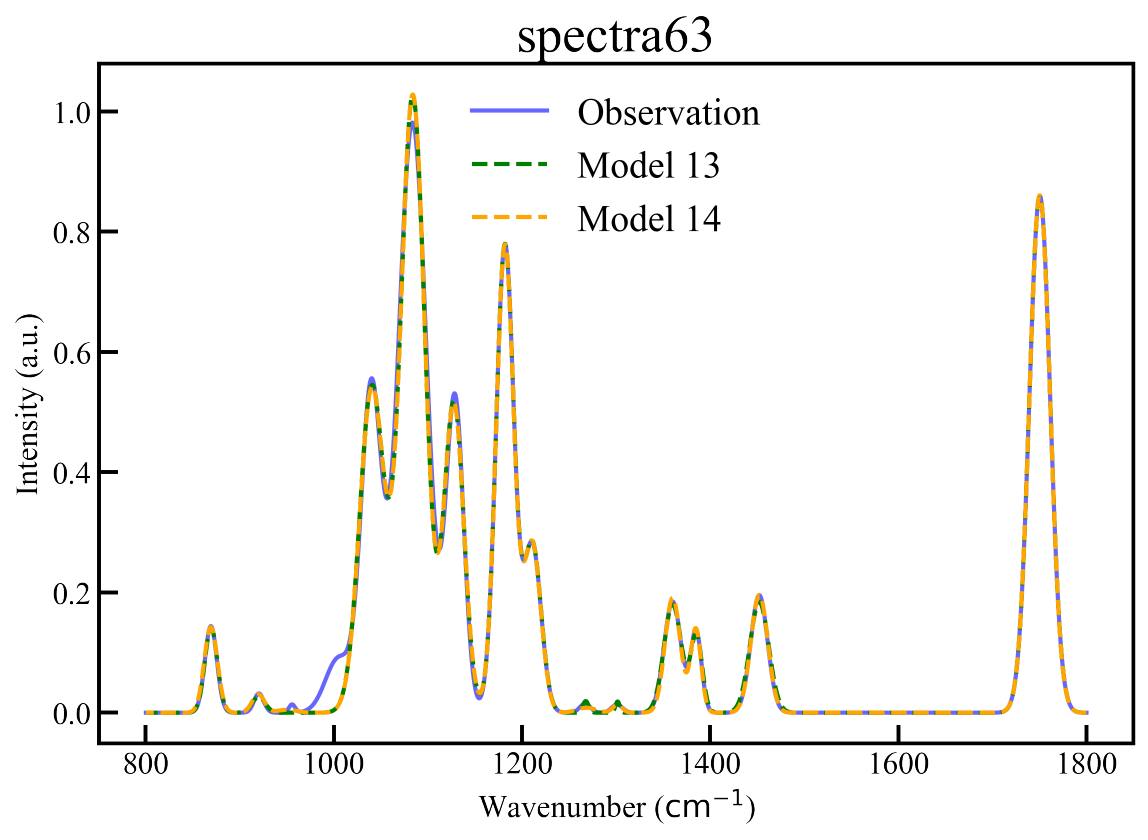} \caption{Spectrum 63} \end{subfigure}\hfill
    \begin{subfigure}[b]{0.45\linewidth} \centering \includegraphics[width=\linewidth]{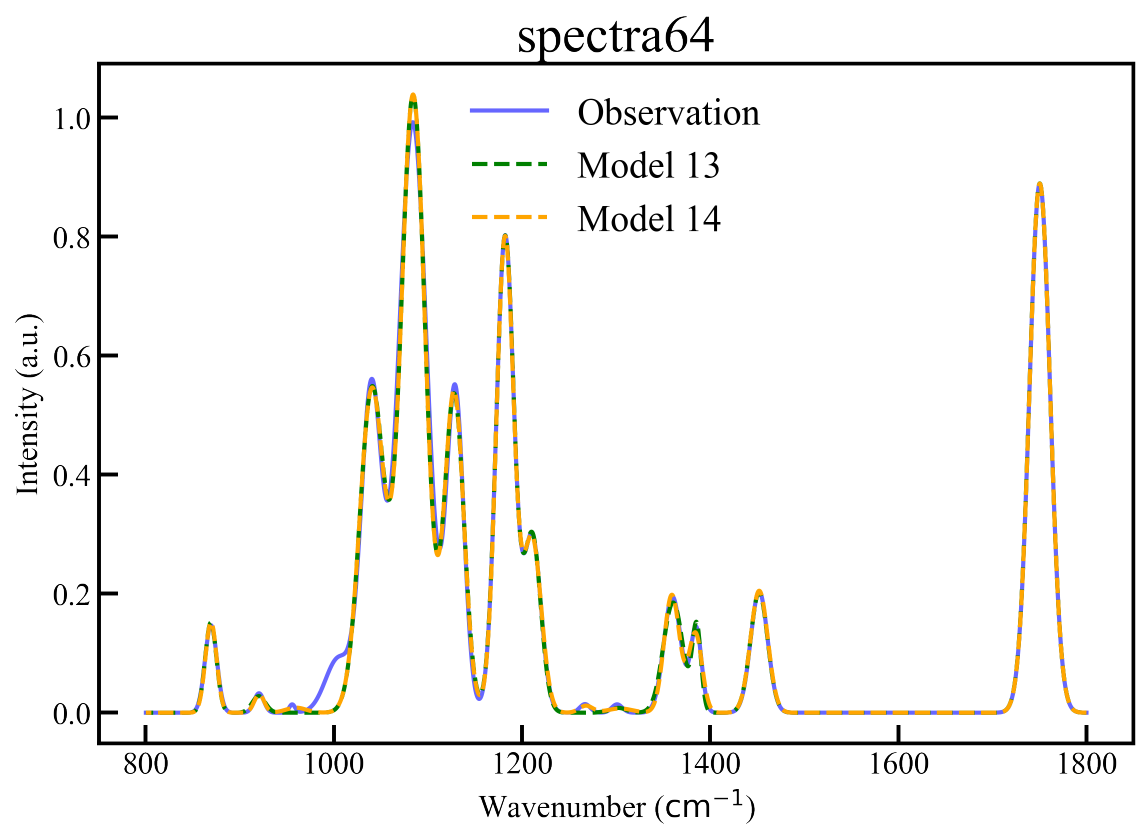} \caption{Spectrum 64} \end{subfigure}
    \vspace{2em}
    \begin{subfigure}[b]{0.45\linewidth} \centering \includegraphics[width=\linewidth]{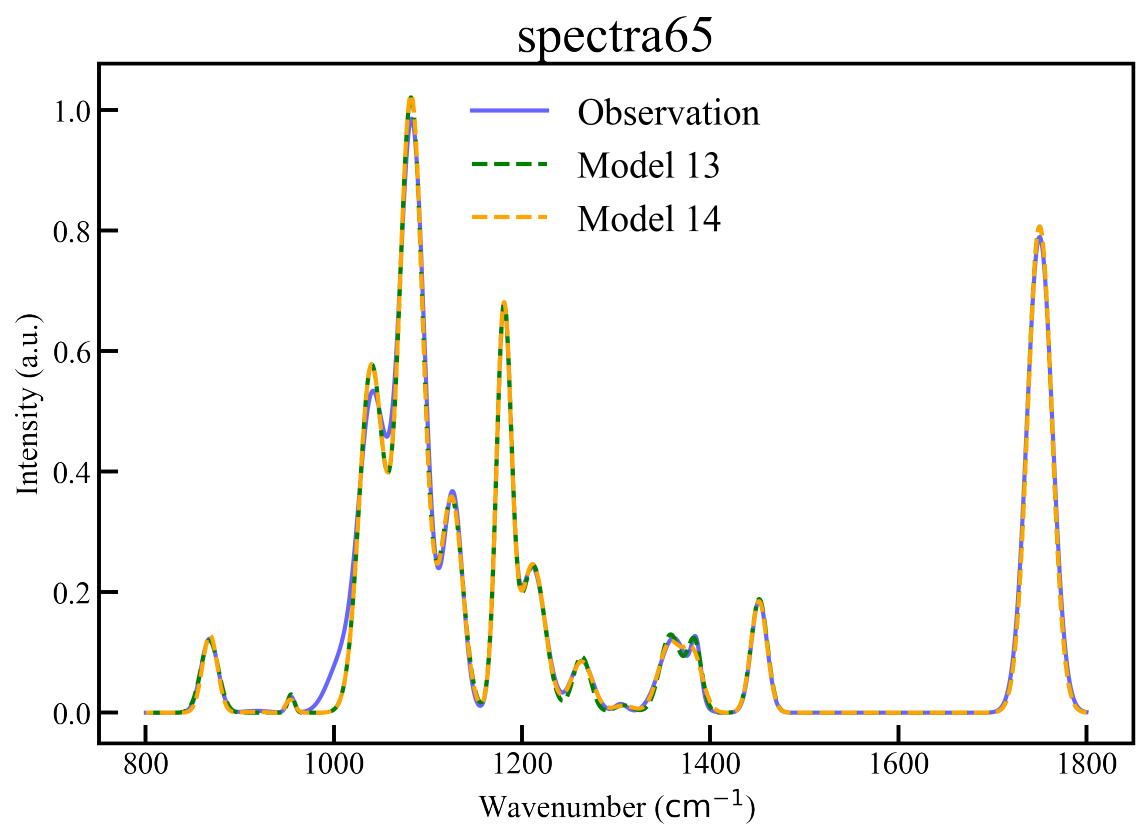} \caption{Spectrum 65} \end{subfigure}\hfill
    \begin{subfigure}[b]{0.45\linewidth} \centering \includegraphics[width=\linewidth]{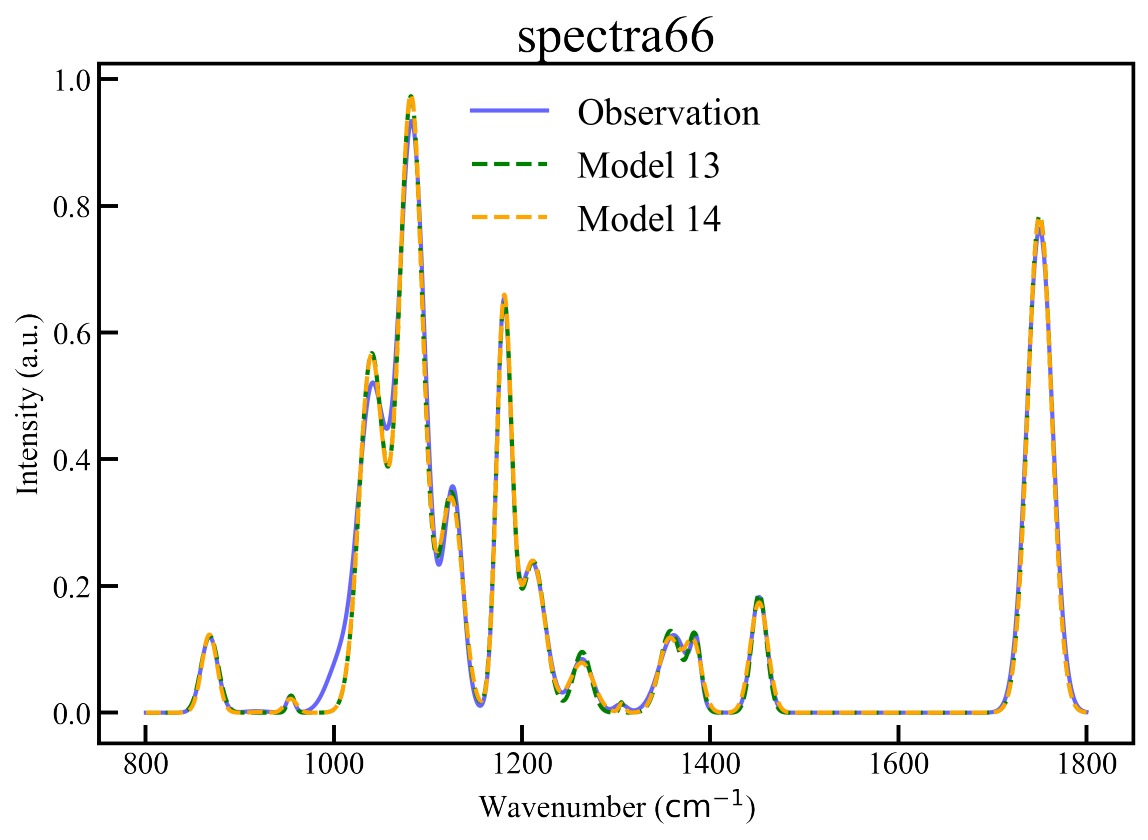} \caption{Spectrum 66} \end{subfigure}
    \caption{IR fitting results (Spectra 61--66). The blue solid line represents the observed spectral data, and the orange (green) dashed line represents the fitting curve for the 14-peak (13-peak) model.}
\end{figure}

\clearpage
\clearpage
\section{Demonstration Using XPS Spectra}\label{XPS_result}
This section presents the results obtained by applying the conventional Bayesian spectral deconvolution method and the proposed method for peak-number estimation in inner-shell XPS spectra.
\subsection{Verification Dataset}
\label{dateset}
We used a dataset containing artificially generated inner-shell XPS spectral data generated from an XPS spectroscopic model and the corresponding physical parameters related to the electronic levels specified in the data-generating process \cite{Mototake_2019}.
Here, we employed the spectroscopic model proposed by Kotani et al. \cite{KOTANI1985805} for 4f-orbital electrons associated with inner-shell electrons in the 3d orbital of rare-earth compounds.
Specifically, we generated the data from the parameter regions of the spectroscopic model corresponding to La$_2$O$_3$, which exhibits a two-peak structure in the region of interest, and CeO$_2$, which exhibits a three-peak structure.\par
The 4$f$ electrons have the following three eigenstates:
$|f^0\rangle$, $|f^1\rangle$, and $|f^2\rangle$.
In the $|f^0\rangle$ state, the 4$f$ orbital is unoccupied; in the $|f^1\rangle$ state, it is occupied by one electron; and in the $|f^2\rangle$ state, it is occupied by two electrons.
In the state space of these 4$f$ electrons, the effective Hamiltonian that reproduces the XPS spectroscopic process is given as follows.
\begin{align}
H &= \varepsilon_L \sum_{\nu} a_{L\nu}^\dagger a_{L\nu} + \varepsilon_f^0 \sum_{\nu} a_{f\nu}^\dagger a_{f\nu} + \varepsilon_c a_c^\dagger a_c \notag\\
&+ \frac{V}{\sqrt{N_f}} \sum_{\nu} (a_{f\nu}^\dagger a_{f\nu} + a_{f\nu} a_{f\nu}^\dagger) \notag\\
&+ U_{ff} \sum_{\nu \neq \nu'} a_{f\nu}^\dagger a_{f\nu} a_{f\nu'}^\dagger a_{f\nu'} \notag\\
&- U_{fc} \sum_{\nu} a_{f\nu}^\dagger a_{f\nu} (1 - a_c^\dagger a_c). %\tag{1}
\end{align}
Here, $\varepsilon_L$, $\varepsilon_f^0$, and $\varepsilon_c$ denote the energies of the conduction electrons ($5d$ and $6s$ electrons), $4f$ electrons, and core electrons, respectively, in $4f$ rare-earth metals.
The index $\nu$ ($\nu = 1,\dots,N_f$, $N_f = 14$) represents the spin and orbital quantum number of the $f$ orbital.
The parameters $V$, $U_{ff}$, and $-U_{fc}$ denote the energies of the hybridization interaction between the $4f$ and conduction electrons, Coulomb interaction between $4f$ electrons, and Coulomb potential from the core hole acting on the $4f$ electrons, respectively.
In both the initial and final states, the energy of the $|f^0\rangle$ state is set to zero as the reference energy level.
With this setting, the number of parameters in the effective Hamiltonian $H$ is reduced to the following:
$\Delta \,(= \varepsilon_f^0 - \varepsilon_L)$, $V$, $U_{ff}$, $U_{fc}$, and $\Gamma$.
The parameter set of the effective Hamiltonian $H$ is defined as follows.
\begin{align}
\theta = \{\Delta, V, U_{ff}, U_{fc},  \Gamma\}
\end{align}
The initial state refers to the system before X-ray irradiation, with all electrons in their ground states. The final state refers to the system after X-ray irradiation, in which a core hole has been created by the removal of a core electron.
The minimum-energy eigenstate of the initial state is $|G\rangle$. The three energy levels of the final state are $E_j(\theta)$ ($j = 0,1,2$), with the eigenstate $|F_j\rangle$.\par
Fermi's golden rule provides the transition probability between these states.
\begin{align}
F(\omega;\theta) = \sum_{j=0}^2 |\langle F_j | a_c | G \rangle|^2 \delta(\omega - E_j(\theta) + E_G(\theta)). %\tag{2}
\end{align}
By convolving $F(\omega; \theta)$ with the Lorentz function and adding Gaussian noise $\varepsilon$ with variance $\sigma^2$, the spectral data $Y$ are obtained as follows.
\begin{align}
y(\omega) = \sum_{j=0}^2 \frac{|\langle F_j | a_c | G \rangle|^2 \Gamma / \pi}{(\omega - (E_j(\theta) - E_G(\theta)))^2 + \Gamma^2} + \varepsilon%\tag{3}
\end{align}
$\Gamma$ is the half-width of the Lorentz function.
\begin{table}[b]
\centering
\caption{
XPS Spectrum Reproduction Parameters for La$_2$O$_3$ and CeO$_2$}
\begin{tabular}{lccccc}
\hline
Compound & $\Delta$ & $V$ & $U_{ff}$ & $U_{fc}$ & $\Gamma$ \\
\hline
La$_2$O$_3$ & 12.5 & 0.57 & 10.5 & 12.7 & 0.5 \\
CeO$_2$ & 1.60 & 0.76 & 10.5 & 12.5 & 0.7 \\
\hline
\end{tabular}
\label{tab1}
\end{table}\par
\begin{figure}[t]
\centering
\includegraphics[width=\linewidth]{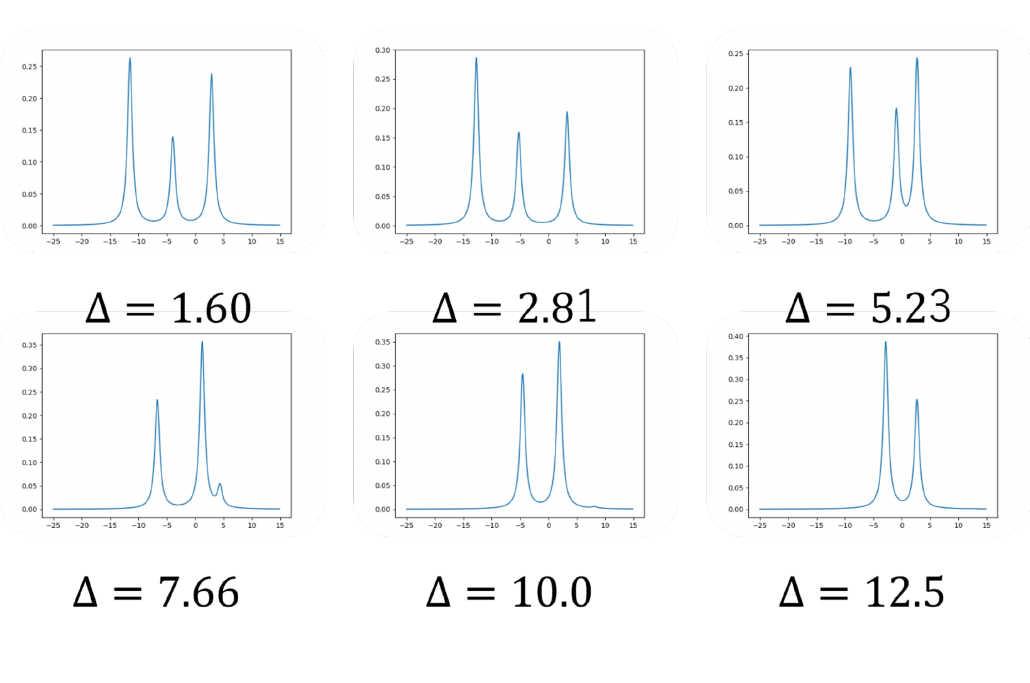} 
\caption{
Spectra generated for different values of $\Delta$.}
\label{data}
\end{figure}

The parameters corresponding to La$_2$O$_3$ and CeO$_2$ used to construct this dataset are listed in Table~\ref{tab1}.
Building on this, the table shows that the most prominent difference between the parameter sets for La$_2$O$_3$ and CeO$_2$ lies in $\Delta$.
Given this distinction, we treat $\Delta$ as the physical quantity $Z$ hereinafter.
With this definition in place, the following trends were observed as $\Delta$ varied.
For a small $\Delta$, multiple peaks appear in the spectrum, and each peak has a pronounced intensity.
For a large $\Delta$, the intensity of the peak on the high-energy side decreases, yielding a shape close to that of a two-peak structure.
The generated data are shown in Fig.~\ref{data}.
For the spectrum at $\Delta = 12.5$, the peak intensity corresponding to $E_2$ becomes nearly zero.
This is because the interaction between the $|f^2\rangle$ and $|f^0\rangle$ states becomes weaker as the energy of the $|f^2\rangle$ state increases.
This suggests that $\Delta$ plays an essential role in changing the peak-number properties.
Therefore, we generated $N' = 6$ spectral datasets $\mathcal{Y}$ at equal intervals from $\Delta = 1.6$ to $\Delta = 12.5$.
Each spectrum $Y$ consisted of 400 equally spaced points, that is, $Y = (y_1, y_2, \dots, y_{400})$.
By setting the standard deviation $\sigma$ of the noise added to the spectra to $0.01$, we generated a set of spectra $\mathcal{Y} \in \mathbb{R}^{400 \times 6}$ and physical quantities $Z = (z_1 = \Delta_1, z_2 = \Delta_2, \dots, z_{N'} = \Delta_{N'}) \in \mathbb{R}^{6}$.\par
To investigate the properties of the generated dataset, we applied Bayesian spectral deconvolution, as described in Section~\ref{spectrum_con}, to a dataset composed of spectral models with the noise level $\sigma = 0$.
As the spectral model, we used a linear sum of Gaussian functions and estimated the number of peaks $M$.
The selection of $M$ based on the Bayesian FE confirmed that only the spectrum at $\Delta = 12.5$ was estimated to have two peaks, whereas all the others were estimated to have three peaks (Table~\ref{tbl_fe}).
As noted above, only the spectrum at $\Delta = 12.5$ shows a nearly zero peak intensity for $E_2$, resulting in a two-peak structure.
These results confirm that Bayesian spectral deconvolution correctly estimates the number of peaks according to the spectral shape.\par
\begin{figure}[t]
\centering
\includegraphics[width=0.8\linewidth]{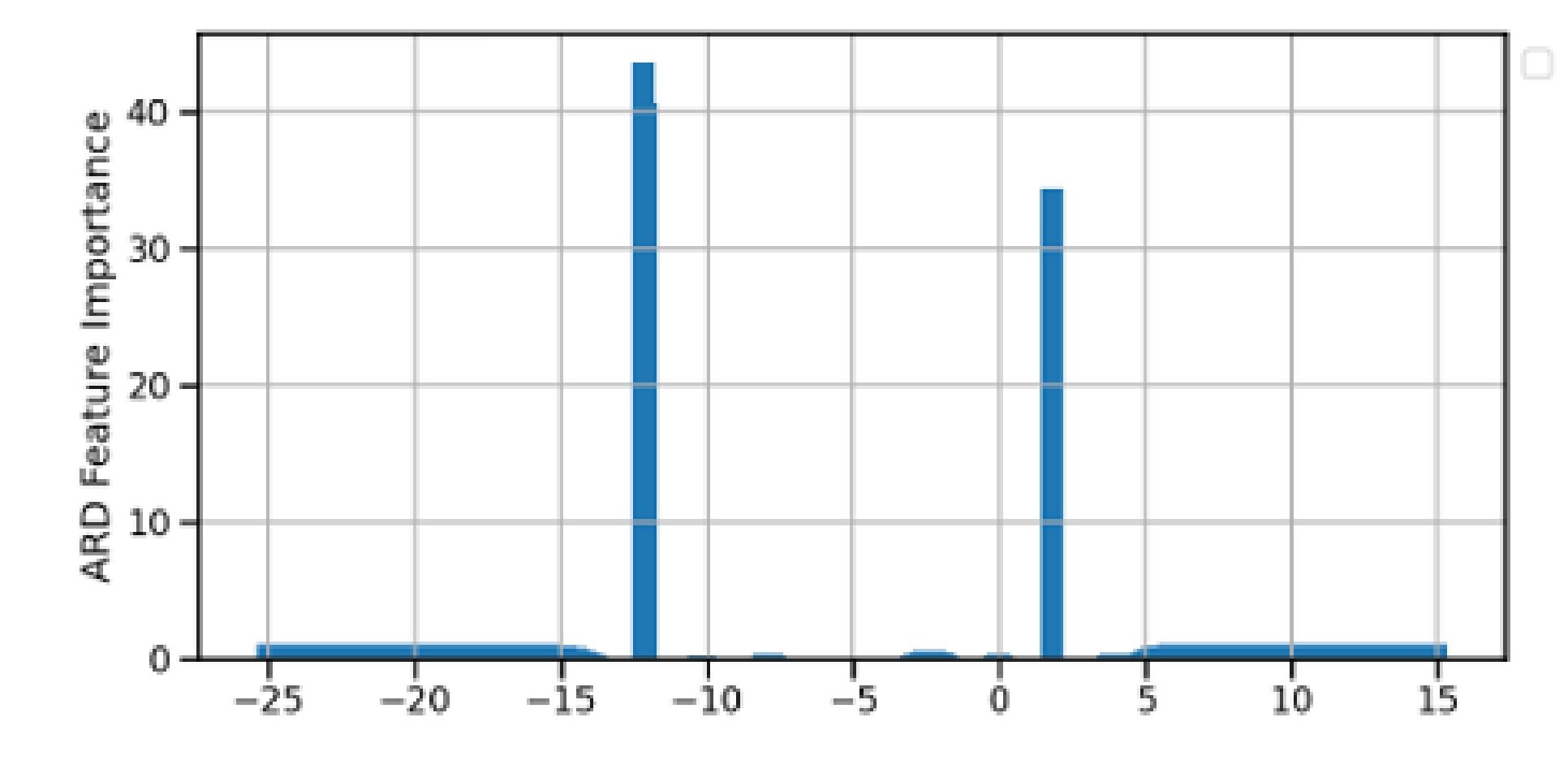}
\caption{
Feature importance obtained by ARD.}
\label{ARD}
\end{figure}

Next, to investigate the relationship between the physical quantity $Z$ and the spectrum $\mathcal{Y}$, we regressed \(Z\) on \(Y\) using an ARD model, which enables Gaussian process regression together with evaluation of the importance of the explanatory variables.
Figure~\ref{ARD} shows the importance of the explanatory variables obtained from ARD regression.
In Fig.~\ref{ARD}, the horizontal axis represents the energy coordinate \(\omega\), and the vertical axis represents the importance.
This result suggests that in the regression of the physical quantity $Z$, only two of the three peaks in the spectrum are important.
From the viewpoint of correspondence with the physical property, this implies that the two-peak structure is important.
Accordingly, when the proposed framework is applied to this dataset, a model that focuses only on these two peaks is expected to be selected as the spectral model.


\begin{table}[t]
\centering
\caption{
Bayesian FE for each value of $\Delta$.}
\label{tab:energy_results}

\begin{tabular}{ccc} % 枠で囲む設定
\hline
\(\Delta\) & \(K=2\) & \(K=3\) \\
\hline
1.60         & 3.325e+03 & \textbf{2.664e+03} \\
2.81  & 2.878e+03 & \textbf{2.440e+03} \\
5.23  & 3.478e+03 & \textbf{1.887e+03} \\
7.65  & 2.169e+03 & \textbf{1.785e+03} \\
10.0  & 1.915e+03 & \textbf{1.696e+03} \\
12.5        & \textbf{1.832e+03} & 1.858e+03 \\
\hline
\end{tabular}
\label{tbl_fe}
\end{table}
\subsection{Results of Applying the Proposed Method}
\label{spectrum_XPS}
The parameters for each method in the proposed framework were set as described below to ensure clarity and consistency.
A linear sum of the Gaussian functions was used as the spectral model for Bayesian spectral deconvolution.
The exchange Monte Carlo parameters in the spectral deconvolution model were defined as follows.
The number of temperature points \( L \) was set to 40, and based on the study by Nagata et al. \cite{exrate}, \(\beta_l\) was set according to the following configuration.
\begin{equation}
\beta_l =
\begin{cases} 
0.0 & \quad \text{for } l = 1, \\
d^{{l-L}} & \quad \text{for } l = 2, 3, \dots, L.\end{cases}
\label{eq:temperature}
\end{equation}
The burn-in period for the sampling process was set to 4000 steps, and the total number of samples was set to 10,000 steps.
From the resulting sequence of $\theta$ samples, we extracted 16 values of $\theta$ at 5-step intervals and used them to estimate \(p(Y|Y^{\mathrm{obs}},M)\).
\begin{figure}[t]
\centering
    \centering
    \includegraphics[width=1\columnwidth]{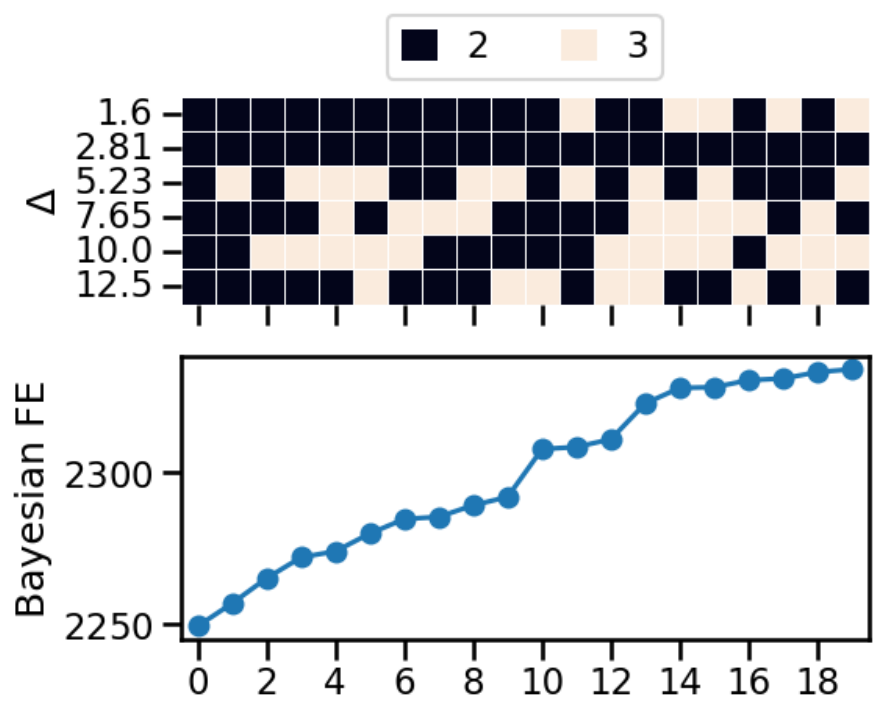}
    \caption{
    Top-ranked models and the $\widetilde{\rm FE}$ at $\sigma = 0.01$.}
    \label{fig:a}
\end{figure}
The hyperparameters of the Gaussian process regression module were determined as follows.
A rational quadratic kernel was adopted as the kernel function.
To compute the Gaussian process regression module $p(Z|\mathcal{Y})$, 30 samples were drawn from \(p(Y|Y^{\mathrm{obs}},M)\), and Gaussian process regression was applied to each sampled dataset $\{\mathcal{Y}^{\rm sample}_{i}\}_{i=1}^{30}$ to estimate
\[
p(Z\mid\mathcal{Y}) \sim \frac{1}{30}\sum_{i=1}^{30} p(Z\mid\mathcal{Y}^{\rm sample}_{i}).
\]
The kernel hyperparameters were optimized by an empirical Bayes method as the values that maximize $p(Z\mid\mathcal{Y}^{\rm sample}_{i})$.
\par
We applied the proposed method to the dataset $\{\mathcal{Y}, Z\}$ and evaluated whether a two-peak or three-peak model was more plausible for each physical quantity $Z$.
Applying the proposed framework yielded the results shown in Fig.~\ref{fig:a}.
The upper panel of Fig.~\ref{fig:a} shows the selected spectral models, ordered from left to right by increasing Bayesian FE; black pixels correspond to two-peak models, whereas beige pixels correspond to three-peak models.
The lower panel shows the $\widetilde{\rm FE}$ of the spectral models shown in the upper panel.
Figure~\ref{fig:a} confirms that a two-peak model is selected for many regions of the spectral data that exhibit a three-peak structure.
This result is consistent with the physical prior knowledge that emphasizes only two peaks in estimating the physical quantity $Z$ (see Section~\ref{dateset}).\par

\end{document}